\documentclass[12pt, a4paper, twoside, openright]{book}
\usepackage{latexsym}
\usepackage{amsmath}
\usepackage{verbatim}
\usepackage{mathrsfs}
\usepackage{amssymb,amsmath}
\usepackage{amsthm}
\usepackage{amscd}
\usepackage{amssymb}
\usepackage{amsfonts}
\usepackage{url}
\usepackage{slashed}
\usepackage[latin1]{inputenc}
\usepackage[english]{babel}

\usepackage{eso-pic}
\usepackage[dvips]{graphicx}
\usepackage{fancyheadings}

\DeclareMathAlphabet{\mathpzc}{OT1}{pzc}{m}{it}

 \csname
@addtoreset\endcsname{equation}{section}

\def\gz0{\gamma^{0}}

 \def\det{{\rm det\,}}

\def\ket#1{|#1\rangle}
\def\scs#1{\section{\sc #1}}
\def\scss#1{\subsection{\sc #1}}

\def\a{\alpha}
\def\b{\beta}
\def\g{\gamma}
\def\G{\Gamma}
\def\d{\delta}

\def\e{\epsilon}

\def\z{\zeta}

\def\l{\lambda}
\def\L{\Lambda}
\def\m{\mu}
\def\n{\nu}
\def\x{\xi}

\def\r{\rho}

\def\s{\sigma}

\def\t{\tau}

\def\o{\omega}
\def\O{\Omega}

\def\cA{{\cal A}}

\def\cC{{\cal C}}
\def\cD{{\cal D}}

\def\cF{{\cal F}}
\def\cG{{\cal G}}

\def\cI{{\cal I}}
\def\cJ{{\cal J}}

\def\cL{{\cal L}}

\def\cP{{\cal P}}

\def\cR{{\cal R}}

\def\cT{{\cal T}}

\def\cV{{\cal V}}

\def\be{\begin{equation}}
\def\ee{\end{equation}}
\def\bea{\begin{eqnarray}}
\def\eea{\end{eqnarray}}
\def\ba{\begin{array}}
\def\ea{\end{array}}
\def\bec{\begin{center}}
\def\ec{\end{center}}
\def\ba{\begin{align}}
\def\ena{\end{align}}

\def\12{\frac{1}{2}}

\def\bra{\langle \,}
\def\ket{\, \rangle}

\def\ra{\rightarrow}

\def\lra{\leftrightarrow}

\begin{document}

\addtolength{\hoffset}{23pt}
\begin{titlepage}
\begin{center}
   	\large{\textsc{Facolt\`a di Scienze Matematiche, Fisiche e Naturali}}\\
   	\small{{Corso di Laurea Specialistica in \textsc{Scienze Fisiche}}}\\
		\rule{5cm}{1pt}\\
	
			\makebox[\textwidth]{\rule{0pt}{.22\textheight}}\\
	\LARGE{\textsc{Higher Spins}}\\
	\LARGE{\textsc{and}}\\
	\LARGE{\textsc{String Interactions}}\\
	
	\bigskip \bigskip
	\large{Tesi di Laurea Specialistica}\\
        \makebox[.2\textwidth]{\rule{0pt}{.1\textheight}}\\
            Defended at the University of Pisa on July 21st, 2009
\end{center}

\vfill
\begin{small}
	\begin{center}
	\rule{3cm}{1pt}\\
	\LARGE{Massimo Taronna}\\		
	\end{center}
\end{small}

\end{titlepage}
\pagestyle{empty}
\cleardoublepage
\addtolength{\hoffset}{-23pt}


\addtolength{\hoffset}{23pt}
\begin{titlepage}\begin{center}
   	\large{\textsc{Facolt\`a di Scienze Matematiche, Fisiche e Naturali}}\\
		\rule{5cm}{1pt}\\
	{\small{Corso di Laurea Specialistica in \textsc{Scienze Fisiche}}}\\
		\makebox[\textwidth]{\rule{0pt}{.07\textheight}}\\
	\LARGE{\textsc{Higher Spins \\ and \\ String Interactions}}\\
		\makebox[\textwidth]{\rule{0pt}{.07\textheight}}\\
	\normalsize{Elaborato di:}\\
	\large{Massimo Taronna}\\
		\makebox[.2\textwidth]{\rule{0pt}{.1\textheight}}\\
\end{center}
\begin{small}
\makebox{\parbox[b]{.8\textwidth}{
	\flushleft{\textbf{Relatore:}}
		\flushright{\begin{tabular}{l c}
		Prof. Augusto Sagnotti &\makebox[.4\textwidth]{\dotfill}\\
		\end{tabular}}


	\flushleft{\textbf{Candidato:}}
		\flushright{\begin{tabular}{l c}
		Massimo Taronna &\makebox[.4\textwidth]{\dotfill}\\
		\end{tabular}}
}}
\vfill
\makebox[\textwidth]{\rule{0pt}{.02\textheight}}\\
	\begin{center}
	\rule{3cm}{1pt}\\
	Anno Accademico 2008-2009\\
	\end{center}
\end{small}

\end{titlepage}
\pagestyle{empty}
\cleardoublepage
\addtolength{\hoffset}{-23pt}

\pagestyle{fancy}
\addtolength{\headwidth}{0.7cm}

\renewcommand{\chaptermark}[1]{\markboth{\thechapter.\ #1}{}}
\renewcommand{\sectionmark}[1]{\markright{#1\ \thesection}}
\lhead[\fancyplain{}{\textbf{\footnotesize{\leftmark}}}]{}
\chead{}
\rhead[]{\fancyplain{}{\textbf{\footnotesize{\rightmark}}}}


\pagestyle{plain}
\pagenumbering{roman}
\null
\vspace{\stretch{1}}
\begin{flushright}
\emph{Alla mia famiglia, e ad Aurora}
\end{flushright}
\vspace{\stretch{2}}
\null
\cleardoublepage

\chapter*{Acknowledgments}
This Thesis is the result of the undergraduate training that I have received during the
last five years at the "E. Fermi" Physics Department of the University of
Pisa and at Scuola Normale Superiore, where my this research activity was done.
I also spent some time at the Galileo Galilei Institute in Florence, that I would like to
thank for the kind hospitality, with financial support from INFN and from the MIUR-PRIN
contract 5ATT78 during the year 2008/2009.

First of all, my gratitude goes to my advisor Augusto Sagnotti, who
introduced me to this wonderful and fascinating research field and supported my activities,
I am greatly indebted to him for having conveyed to me part of his
enthusiasm for one of the most intriguing research fields of Physics. I would like
to thank him also for many key suggestions, for the many discussions that we
had and for giving me the opportunity to meet other people and to discuss with
them in a very nice atmosphere.
Among those, Nicolas Boulanger, Andrea Campoleoni, Dario Francia,
Carlo Iazeolla and Per Sundell, to whom I am indebted for their help and for
the time spent together talking about Physics. It is also a pleasure to thank
Carlo Angelantonj for having invited me to Torino and for having been, since we first met,
very nice and helpful to me.
I would also like to thank, for their warm friendship and for all the adventures we had together,
my friends Alberto, Emeline, Federico, Giuseppe, Marco, Mariana, Mathilde, Michele,
Rayco, Sebastiano, Sebastien, Simone, Yaiza, and many others.
Above all, my love goes to my family for having always encouraged me and for having been always
close to me; and to Aurora, for the wonderful time that we have spent together.

\tableofcontents

\cleardoublepage

\pagestyle{fancy}
\pagenumbering{arabic}
\chapter{Introduction}
Higher-Spin Fields have been regarded as a very interesting subject since the role of space-time symmetries in Quantum Field Theory (QFT) was first recognized \cite{Majorana:1932rj,Dirac:1936tg,Fierz:1939ix,Rarita:1941mf,Wigner:1939cj,Bargmann:1948ck}. From the very beginning, the fundamental degrees of freedom of any QFT were organized via the solutions of Poincar\'e invariant equations, and it was soon understood that these are associated to unitary representations of the Poincar\'e group, or of its higher-dimensional generalizations for $D\,>\,4$. From first principles one is thus led to classify all such representations, and possibly also to construct a theory that combines all of them. By a relativistic field of higher spin we mean a generalization of the electromagnetic potential $A_{\,\m}$, or of the metric fluctuation $h_{\m\n}$, that transforms under an arbitrary representation of the Poincar\'e group. In four dimensions and up to dualities this set is exhausted by symmetric Bose fields of the form $\phi_{\,\m_1\cdots\m_s}$, together with their fermionic counterparts, symmetric spinor-tensors of the form $\psi_{\,\m_1\cdots\m_s}$, while in higher dimensions one generally needs to consider tensors of mixed symmetry that can be classified via Young tableaux.

Unfortunately, increasing difficulties have essentially limited our knowledge of Field Theory to spins not exceeding two. At the same time, the description of the fundamental particles encoded in the Standard Model involves fields whose spin is at most one, while the description of gravity, macroscopically at least, is related to a spin-$2$ field, and in supersymmetric cases also to spin-$3/2$ fields \cite{Weinberg:2000cr}.
However, although the current understanding of lower-spin gauge theories gives the possibility to explain huge amounts of phenomena, it is still an open problem how to build a quantum description of gravity that is consistent at short distances.
Here one faces also the issue of unification, that has long been the underlying philosophy of Physics: formerly distinct subjects can often be put on the same ground within a single conceptual framework.

The challenge today is to find such a well-defined conceptual framework combining a consistent quantum description of gravity with the other fundamental interactions.

String Theory \cite{Sagnotti:2002yc,stringtheory,Goddard:1973qh,Friedan:1985ge} is, up to now, a most promising candidate to be both a finite quantum theory of gravity and a framework in which all fundamental interactions can be unified. Basically, the crucial change of perspective is to postulate as fundamental constituents of the world one-dimensional objects, rather then point-like particles. In this way interactions are softened, since the string length $l_s\sim 10^{-33}cm$ acts as a natural UV cut-off, and fields of different spins arise as vibrational modes in an enticing unified picture. This picture has the virtue of reproducing, in principle, the known fundamental interactions at the quantum level, but an infinite tower of massive modes must be taken into account for the consistency of the theory. In fact, a vibrating relativistic string displays, already at the classical level, an infinite set of harmonics of increasing frequency and with prescribed transformation properties under the Lorentz group. The quantum counterpart of this classical picture is a plethora of massive states of arbitrary spin that dispose themselves on linear Regge trajectories on the mass-squared \textit{vs.} spin plane. For the first Regge trajectory the link between the mass squared $m^2$ and the spin $s$ is
\begin{gather}
m^{\,2}\sim\frac{s-a}{\a'}\ ,
\end{gather}
where $\a'$ is called Regge slope, and is related to the string tension by $T\,=\,\frac{1}{2\pi\a'}$, while $a$ depends on what kind of String Theory is being considered, so that its value is $1$ for open strings and $2$ for closed strings.

The typical choice of a string tension of the order of the Planck energy, $M_P \sim 10^{19}$ GeV, has long limited the analysis to the massless particles describing long-range interactions and, in supersymmetric cases, also to the spin-$3/2$ ``gravitini'' of Supergravity. However, it is conceivable that leaving out all higher-spin modes is potentially precluding the possibility to really go beyond the on-shell first quantized picture of String Theory and to understand its true quantum degrees of freedom. From this viewpoint String Theory itself might well prove to be the main motivation to study higher-spin theories. In fact, in the regime of extremely high energies, the higher-spin modes play an important role and it is necessary to treat all string excitations on the same footing. Such a unifying framework makes it imperative to seriously consider all higher-spin fields.
From this point of view, clarifying higher-spin dynamics will help to better understand String Theory and, vice versa, a closer look at String Theory at high energies or, equivalently, in the $\a'\ra\infty$ limit, can provide important clues on higher-spin dynamics.

Actually, although string spectra contain massive higher spins, the great achievements attained studying massless lower-spin theories have been a most stringent motivation to concentrate to a large extent the attention on massless higher-spin theories. In fact, it is well known that local symmetries place very strong restrictions on the structure of theories. Basically, two different approaches have been developed over the years, as we shall see better in the following. One of them exploits the frame-like formalism while the other tries to follow more closely the lessons of the metric-like formulation of gravity. In addition, for the latter one ought to distinguish between constrained and unconstrained forms. Still, only recently a complete ``metric'' Lagrangian formulation of free higher-spin theories has been attained and only in a flat background, while the problem of interactions remains largely unclear. For a review or a detailed discussion see \cite{Singh:1974rc,Fronsdal:1978rb,Fang:1978wz,Buchbinder:2001bs,Bouatta:2004kk,Francia:2005bu, Francia:2007qt,Campoleoni:2008jq}. For a long time no-go results have discouraged a systematic study of this subject, but String Theory and, more recently, the Vasiliev equations \cite{Vasiliev:1988sa} (for a review see \cite{Bekaert:2005vh}) have provided very interesting frameworks in which consistent higher-spin interactions manifest themselves.

The most stringent known no-go results are the Coleman-Mandula Theorem and its supersymmetric generalization \cite{Coleman:1967ad}, Weinberg's Theorem \cite{Weinberg:1964ew}, the Weinberg-Witten Theorem \cite{Weinberg:1980kq} and more recently the extension of the Weinberg-Witten Theorem developed by Porrati \cite{Porrati:2008rm}. The Coleman-Mandula Theorem and its supersymmetric generalization rule out, under the hypotheses of finiteness of the spectrum and non-triviality of the $S$-matrix, a consistent nontrivial embedding of the lower-spin symmetries into some bigger symmetry algebra that mixes fields of different spins. Weinberg's Theorem rules out the possibility that particles with spin greater than $2$ mediate long-distance interactions in flat space. At the same time, the Weinberg-Witten Theorem and its completion developed by Porrati forbid the possibility that a consistent higher-spin gravitational coupling for particles with spin greater than $2$ around flat space be a minimal coupling.
Still, none of these results suffices to forbid the possibility to construct a consistent higher-spin dynamics because, in one way or another, their statements are related to subtle assumptions that may well prove too restrictive, like finiteness of the spectrum, minimal couplings, existence of asymptotic states and more generally to the possibility of defining an $S$-matrix.

From this viewpoint the choice made in this Thesis is to focus on the available yes-go results in order to try to open a path toward the systematics of higher-spin dynamics.

In this direction there are the recent yes-go results of \cite{Boulanger:2008tg,Porrati:2009bs,Zinoviev:2008jz}, where some non minimal couplings involving higher-spin particles have been analyzed and deformations of the gauge symmetry that are consistent to lowest order have been explicitly uncovered. More in detail \cite{Boulanger:2008tg} addresses the uniqueness of the coupling between higher-spin fields and gravity, while \cite{Porrati:2009bs,Zinoviev:2008jz} construct consistent Lagrangians for massive spin-$3/2$ and massive spin-$2$ fields carrying $U(1)$ charges, to lowest order in the former case and to all orders in a constant electromagnetic field in the latter. 
In particular, \cite{Porrati:2009bs} shows how, pushing forward what may be regarded as a variant of the original observation of Fradkin and Vasiliev \cite{Fradkin:1986qy} and adding suitable non-minimal couplings to the action, it is possible to enforce to all orders the transversality and $\g$-trace constraints on the spin $3/2$-field:
\begin{gather*}
D\cdot\psi\,=\,0\ ,\\
\gamma^\a\psi_\a\,=\,0\ ,
\end{gather*}
where with $D$ we mean in general a deformed covariant derivative, providing also, in this fashion, a solution to the long-standing Velo-Zwanziger causality problem \cite{Velo:1970ur}. Similarly, \cite{Zinoviev:2008jz} is devoted to a consistent non-minimal coupling to lowest order that enforces transversality and tracelessness on a spin-$2$ field coupled to electromagnetism. In both cases transversality and tracelessness, or their generalizations to non-flat backgrounds, appear crucial for the consistency of the theory. Let us stress that this hint on the correct requirement that must be imposed on the system originates from an observation that was made long ago in String Theory in \cite{Argyres:1989cu}, where a non minimal Lagrangian for a charged massive spin-$2$ field coming from Open String Field Theory was presented. The crucial observation is that, even if the Lagrangian is very complicated, its equations of motion impose exactly the Fierz-Pauli constraints to all orders so that, after using them, one recovers a good hyperbolic system.

In some sense, the usual way of constructing massive interacting theories via the Stuckelberg completion of the massive free Lagrangian is very powerful, but appears less fundamental because it does not take into account the possible non minimal deformations of the gauge symmetry. Instead, the crucial requirement to demand is the exact validity of the Fierz-Pauli constraints determining the correct number of propagating degrees of freedom. String Theory from this perspective appears to be a powerful arena to understand how some long-standing problems can be solved.

Summarizing, it is by now well appreciated that an effective description of higher-spin massive particles in flat space requires that minimal coupling be properly overcome, as suggested, for instance by the recent analysis of \cite{Boulanger:2008tg}, and an interesting possibility that may well prove not too remote is to try and extend this type of analysis to all higher-spin fields.

The problem of higher-spin interactions, as already mentioned, is currently studied from two different but complementary perspectives. The first, developed in the frame-like formalism, could be called an ``algebraic approach'', was developed essentially by Vasiliev, and is a powerful generalization of the vielbein formulation of gravity. Through a reformulation of the free dynamics of spin-$s$ fields in terms of a set of one-form connections, mimicking the description of the spin-$2$ field in terms of vielbein $e^{\,a}$ and spin connection $\o^{\,ab}$, it is possible to grasp crucial information on the structure of the non-abelian higher-spin algebras in order to define higher-spin interactions \textit{\`a} la Yang-Mills. The basic idea is to generalize the correspondence between the one-forms $(e^{\,a},\o^{\,ab})$ and the generators of the Poincar\'e algebra $(P_{\,a},M_{\,ab})$ encoded in the non abelian connection $\O\,=\,-i(e^{\,a} P_{\,a}+\frac{1}{2}\,\o^{\,ab}M_{\,ab})$. This is attained choosing as a generalized connection $\Omega$ a generating function containing connections for all spin-$s$ fields, in such a way that the role of $P_{\,a}$ and $M_{\,ab}$ is played by an infinite-dimensional algebra. As a result, one faces a situation that is similar, in some respects, to the one for gravity in the Cartan-Weyl form.
The other perspective starts from what is, possibly, a more ``geometric'' point of view, since is reminiscent of the metric formulation for gravity and tries to extend the geometric intuition that we have acquired for spin two to higher-spins. However, to date this approach is far less successful in characterizing higher-spin interactions.
This direction draws its origin from the works of Dirac \cite{Dirac:1936tg}, Fierz and Pauli \cite{Fierz:1939ix}, Wigner \cite{Wigner:1939cj} and Bargmann and Wigner \cite{Bargmann:1948ck}, where it became clear that the physical requirement of positivity of the energy could be replaced by the condition that one-particle states carry unitary representations of the Poincar\'e group. Explicitly, for massive fields of integer and half-integer spin, represented respectively by totally symmetric tensors $\phi_{\,\m_1\cdots \m_s}$ and by totally symmetric tensor-spinors $\psi_{\,\m_1\cdots\m_s}$, such a requirement can be encoded in the Dirac-Fierz-Pauli conditions
\begin{align}
(\square-M^{\,2})\,\phi_{\,\m_1\cdots\m_s} &\,=\,0\ ,& (i\slashed{\partial}\,-\,M)\,\psi_{\,\m_1\cdots\m_s}&\,=\,0\ ,\\
\partial^{\,\m_1}\,\phi_{\,\m_1\cdots\m_s} &\,=\,0\ ,& \partial^{\,\m_1}\,\psi_{\,\m_1\cdots\m_s} &\,=\,0\ ,
\end{align}
where irreducibility for massive fields can be imposed via the $(\g-)$trace constraints
\begin{align}
\eta^{\,\m_1\m_2}\,\phi_{\,\m_1\cdots\m_s}&\,=\,0\ ,& \g^{\,\m_1}\,\psi_{\,\m_1\cdots\m_s}&\,=\,0\ .
\end{align}
As we anticipated, generalizations of these equations are known for mixed-symmetry fields where, in order to achieve irreducibility, one must also specify a given Young-projection.

As stressed long ago in \cite{Fierz:1939ix}, a Lagrangian formulation of interacting higher-spin fields is fundamental even classically, in order to avoid algebraic inconsistencies in the equations of motion, and was first obtained by Singh and Hagen in \cite{Singh:1974rc} for free \textit{massive} totally symmetric fields, and then by Fronsdal \cite{Fronsdal:1978rb} for free \textit{massless} totally symmetric bosonic fields, and by Fang and Fronsdal in \cite{Fang:1978wz} for free \textit{massless} totally symmetric tensor-spinors. The Fronsdal equations are a direct generalization of the Maxwell equations for spin-$1$
\begin{gather}
\square A_{\,\m}-\partial_{\m} \partial\cdot A\,=\,0\ ,
\end{gather}
with abelian gauge symmetry
\begin{gather}
\delta A_{\,\m}\,=\,\partial_{\m}\,\L\ ,
\end{gather}
or of the linearized Einstein's equations for spin-$2$
\begin{gather}
\square\, h_{\m\n}-\partial_{\m}\,\partial\cdot h_\n-\partial_{\n}\,\partial\cdot h_\m+\partial_{\m}\partial_{\n}\, h'\,=\,0\ ,
\end{gather}
where $h'\,=\,h^\a_\a$, with the abelian gauge symmetry
\begin{gather}
\delta h_{\m\n}\,=\,\partial_\m\L_\n+\partial_\n\L_\m\ .
\end{gather}
For example, the spin-$3$ equation reads
\begin{multline}
\square \phi_{\m\n\r}-(\partial_\m\,\partial\cdot \phi_{\n\r}+\partial_\n\,\partial\cdot \phi_{\r\m}+\partial_\r\,\partial\cdot \phi_{\m\n})\\+\partial_{\m}\,\partial_\n\,\phi'_\r+\partial_{\n}\,\partial_\r\,\phi'_\m+
\partial_{\r}\,\partial_\m\,\phi'_\n\,=\,0\ ,\label{Fro}
\end{multline}
but now, in order to maintain the abelian gauge invariance
\begin{gather}
\delta \phi_{\m\n\r}\,=\,\partial_\m\L_{\n\r}+\partial_\n\L_{\r\m}+\partial_\r\L_{\m\n}\ ,
\end{gather}
it is necessary to impose the additional constraint
\begin{gather}
\L'\,=\,0\ .
\end{gather}
Moreover, for spin $s\geq 4$ local gauge invariant Lagrangians for $\phi_{\m_1\cdots\m_s}$ alone require the additional constraint
\begin{gather}
\phi''\,=\,0\ .
\end{gather}

Recently the Fronsdal constraints have been overcome in a particular simple framework by Francia and Sagnotti in \cite{Francia:2002aa} and, for mixed symmetry fields, in \cite{Bekaert:2002dt}. These works stress the potential interest of moving from the original Fronsdal formulation for massless higher-spins \cite{Fronsdal:1978rb,Fang:1978wz} to an unconstrained formulation that appears manifestly connected with a geometric description codified by the curvature tensors first constructed by de Wit and Freedman \cite{de Wit:1979pe}.
The basic step leading from the constrained Fronsdal formulation to an unconstrained formulation is the introduction of a spin-$(s-3)$ totally symmetric compensator field $\a$ with a Stuckelberg-like gauge invariance
\begin{gather}
\delta \a_{\,\m_1\cdots\m_{s-3}}\,=\,\L'_{\,\m_1\cdots\m_{s-3}}\ .
\end{gather}
In this fashion, for instance, for spin $s\,=\,3$ the equation $(\ref{Fro})$ becomes
\begin{multline}
\square\, \phi_{\m\n\r}-(\partial_\m\,\partial\cdot \phi_{\n\r}+\partial_\n\,\partial\cdot \phi_{\r\m}+\partial_\r\,\partial\cdot \phi_{\m\n})\\+\partial_{\m}\,\partial_\n\,\phi'_\r+\partial_{\n}\,\partial_\r\,\phi'_\m+
\partial_{\r}\,\partial_\m\,\phi'_\n\,=\,3\,\partial_\m\,\partial_\n\,\partial_\r\, \a\ .
\end{multline}
With this modification gauge invariance is no more constrained, while eliminating the compensator field it is possible to cast the resulting equations in the non-local geometric form \cite{Francia:2002aa}
\begin{gather}
\frac{1}{\square^{\,n}\!\!}\ \partial\cdot \cR^{[n]}_{;\,\m_1\cdots\m_{2n+1}}\,=\,0\ ,
\end{gather}
for odd spin $s\,=\,2n+1$, and
\begin{gather}
\frac{1}{\square^{\,n-1}\!\!}\ \cR^{[n]}_{;\,\m_1\cdots\m_{2n}}\,=\,0\ ,
\end{gather}
for even spin $s\,=\,2n$, while similar equations can be written for Fermi fields. Here the $[n]$ indicates the $n$-th trace in the first group of indices of the tensor.\\
These linearized versions of geometric quantities associated to Higher-Spin Fields are exactly the counterparts of the linearized Einstein tensor for spin-$2$ and of the electromagnetic tensor for spin-$1$. 
Moreover, the compensator $\a$ and an additional Lagrange multiplier $\b$ suffice to build the so called minimal unconstrained theory \cite{Francia:2005bu} for totally symmetric fields. This minimal Lagrangian form is actually a partial gauge fixing of the BRST construction of Pashnev, Tsulaia, Buchbinder and others \cite{Buchbinder:2001bs}, that was the first concrete positive result going beyond the constrained Fronsdal construction. A similar, although far more involved, Lagrangian construction is now available for all mixed-symmetry fields \cite{Campoleoni:2008jq}.

The next step, after the completion of the higher-spin program for the free theory, it is to study interactions in a systematic way.
Again, String Theory can prove a convenient starting point from which one can investigate particular dynamical regimes were higher-spin effects become important.

In this direction, a beautiful idea behind this program, that however was never fully quantified, it is to regard String Theory as a broken phase of a Higher-Spin Gauge Theory where masses originate from a generalized Higgs effect. This enormous process, called by some authors ``La Grande Bouffe'' \cite{Sezgin:2002rt,Bianchi:2003wx}, would produce a non-vanishing tension capable of giving rise to the known string spectrum. From this perspective a better understanding of higher-spin dynamics might shed light on the very structure of String Theory itself.

Higher-spin interactions can be formulated in different ways. A useful perspective, that dates back to the work of Berends, Burgers and Van Dam \cite{Berends:1985xx,Berends:1984wp,Berends:1984rq}, is further elaborated in many subsequent works, and very recently in \cite{Bekaert:2007mi}. It starts with a list of all Poincar\'e-invariant local deformations of the form
\begin{gather}
S[h]\,=\,S^{(0)}[h]\,+\,\e S^{(1)}[h]\,+\,O(\e)\ ,
\end{gather}
including at least one field of spin $s>2$, such that
\begin{gather}
S^{(0)}[h]\,=\,\sum_s S^{(0)}[h_{\m_1\cdots\m_s}]\ ,
\end{gather}
is the sum of all free quadratic actions and such that the deformed local gauge symmetries
\begin{gather}
\d_{\x} h\,=\,\d^{(0)}_{\x} h+\e\d^{(1)}_{\x} h+O(\e^2)\ ,
\end{gather}
are non abelian to first order in the deformation parameter $\e$ and do not arise from local field redefinitions
\begin{gather}
h\,\ra\, h\,+\,\e\,\phi(h)\,+\,O(\e^2)\ ,\\
\xi\,\ra\, \xi\,+\,\e\,\z(h,\xi)\,+\,O(\e^2)\ ,
\end{gather}
of the gauge fields and parameters. This approach has the advantage of being formally perturbative, and can be used to deal with interactions order by order.

With this perspective in mind, this Thesis is devoted to the study of open-string tree level scattering amplitudes with three and four external states in the first Regge trajectory of the bosonic string. We thus obtain, for the first time, explicit forms of three-point amplitudes involving higher-spin modes and their currents. These amplitudes are analyzed in detail and are turned into cubic couplings for massive and massless higher-spin excitations. Most of the calculations are done using the formalism of generating functions, that makes it possible to compute, in a simple and manifestly projective invariant fashion, the generating functions of three and four-point amplitudes. In fact, grouping together all totally symmetric tensors in generating functions of the type
\begin{gather}
\tilde{H}_i(k_{\,i},p_{\,i})\,=\,\sum_{n\,=\,0}^{\infty}\,\tilde{H}_{i\,\m_1\cdots\m_n}(k_{\,i})\ p_{\,i}^{\,\m_1}\cdots p_{\,i}^{\,\m_n}\ ,
\end{gather}
one can compute \textit{any} tree-level three-point or four-point scattering amplitude with open-string external states in the first Regge trajectory. 
Thus, the explicit result for the three-point amplitude can be cast in the rather compact form
\begin{multline}
\cA\,=\,i\,\frac{g_o}{\a'}\ (2\pi)^{d}\delta^{(d)}(k_{\,1}+k_{\,2}+k_{\,3})
\left\{\cA_{\,+}\ Tr[\L^{a_1}\L^{a_2}\L^{a_3}]\right.\\\left.\,+\,\cA_{\,-}\ Tr[\L^{a_2}\L^{a_1}\L^{a_3}]\right\}\ ,
\end{multline}
with
\begin{multline}
\cA_{\pm}\,=\,\tilde{H}_1\left(k_{\,1}\,,\,p\,+\,\frac{\partial}{\partial p}\,\pm\,\sqrt{\frac{\a'}{2}}\ k_{\,23}\right)\tilde{H}_2\left(k_{\,2}\,,\,\frac{\partial}{\partial p}\,\pm\,\sqrt{\frac{\a'}{2}}\ k_{\,31}\right)\\
\times\tilde{H}_3\left(k_{\,3}\,,\,p\,\pm\,\sqrt{\frac{\a'}{2}}\ k_{\,12}\right)\Bigg|_{p\,=\,0}\ ,
\end{multline}
where we have also taken into account the Chan-Paton factors \cite{Paton:1969je} and where $\partial_p$ acts as a contraction operator among polarization tensors.\\
From this formula, one can then compute explicitly the current generating function, given by
\begin{gather}
\cJ(x,k_{\,1}')\,=\,i\,\frac{g_o}{\a'}\,\left\{J_{\,+}\ Tr[\ \cdot\ \L^{a_2}\L^{a_3}]\,+\,J_{\,-}\ Tr[\ \cdot\ \L^{a_3}\L^{a_2}]\right\}\ ,
\end{gather}
where
\begin{multline}
J_{\pm}\,=\,{H}_2\left(x\,\mp\,i\sqrt{\frac{\a'}{2}}\ k_{\,1}',\ \frac{\partial}{\partial p}\,+\,k_{\,1}'\,\mp\, i\sqrt{2\a'}\ \partial_{3}\right)\\\times{H}_3\left(x\,\pm\, i\sqrt{\frac{\a'}{2}}\ k_{\,1}',\ p\,+\,k_{\,1}'\,\pm\, i\sqrt{2\a'}\ \partial_{2}\right)\Bigg|_{p\,=\,0}\ ,
\end{multline}
in coordinate space, and
\begin{multline}
\tilde{J}_\pm\,=\,\exp\left\{\pm \sqrt{\frac{\a'}{2}}\ k_{\,1}'\cdot k_{\,23}\right\}\\\times\tilde{H}_2\left(k_{\,2}\,,\,\partial_p\,+\,
k_{\,1}'\,\pm\,\sqrt{\frac{\a'}{2}}\ k_{\,31}\right)\tilde{H}_3\left(k_{\,3}\,,\,p\,+\,
k_{\,1}'\,\pm\,\sqrt{\frac{\a'}{2}}\ k_{\,12}\right)\Bigg|_{p\,=\,0}\ ,
\end{multline}
in momentum space.

The couplings extracted from String Theory are analyzed in a number of cases in order to identify the off-shell conserved currents that play a key role in string interactions. The main obstacle in this respect is that the currents emerge in a form that uses crucially the on-shell equations for the external states. Nonetheless, in some simple cases one can recognize such terms and extract from the original expansion truly off-shell couplings. For string amplitudes of the types $s-0-0$ and $s-1-1$, it turns out that the off-shell couplings have a consistent massless limit and coincide with the corresponding currents of Berends, Burgers and van Dam \cite{Berends:1985xx}. These results are to be regarded as a first concrete indication on how all higher-spin string couplings are realized off-shell by suitable conserved currents. Notice that gauge invariance is broken, at the cubic order, only by the massive equations of motion for the external states.

The conserved currents that we have identified give consistent cubic couplings and correspond to Noether currents associated with global symmetries of the free spin-$s$ Lagrangian. From them, computing the generating function of propagators for totally symmetric fields one can also compute on the field theory side some interesting tree-level four-point scattering amplitudes involving exchanges of infinite numbers of higher-spin particles to study their high-energy behavior, generalizing the recent work of Bekaert, Mourad and Joung \cite{Bekaert:2009ud}.

Summarizing, this Thesis rests on the explicit computation of all tree-level three-point or four-point scattering amplitudes involving open string states in the first Regge trajectory of the open bosonic string. Chapter $2$ is devoted to a brief review of the basic ingredients that are needed to construct the string $S$-matrix. To this end we summarize some basic properties of the path integral quantization by which the $S$-matrix is defined and we explain briefly how to construct vertex operators describing incoming and outgoing states. In Chapter $3$ we take as our starting point the usual CFT on the sphere and, considering the Quantum Field Theory generating function with a judicious choice of the external current, we construct a generating function for all vertex operator correlation functions. Moreover, restricting the attention to states in the first Regge trajectory, the result is cast in a manifestly projective invariant fashion for both three-point and four-point amplitudes. In Chapter $4$, grouping all totally symmetric polarization tensors in a generating function, we translate the tree-level scattering amplitudes into Lagrangian couplings of a compact and useful form. Actually, the result can be cast in various useful forms, extracting the current generating function from the amplitudes and also computing its Fourier transform. In Chapter $5$ we present some interesting applications of the couplings obtained from String Theory, computing on the field theory side some scattering amplitudes involving infinitely many higher-spin exchanges, and study their behavior in some interesting limits. Moreover, we construct an expression for the current exchanges for any spin-$s$ fields extending the result previously obtained in \cite{Francia:2007qt} to the general case of mixed symmetry fields and of non-transverse external currents. The last Chapter contains a brief summary of the results obtained and some indications on possible directions for further research.

\chapter{Functional Integral and String Amplitudes}
In this chapter we review some basic facts concerning String Theory scattering amplitudes that will be used extensively in this Thesis. In particular, we review the formalism by which $S$-matrix amplitudes can be defined and also some issues related to the first quantization of String Theory, with special reference to the bosonic case.


\scs{First Quantization and the $S$-matrix}
String Theory can be defined as a first quantized theory of one dimensional objects moving in a $D$-dimensional space time. A convenient starting point is the Euclidean Polyakov action
\begin{gather}
S_P[X,\g]\,=\,-\frac{1}{4\pi\a'}\int_M d\t d\s\ \sqrt{\g}\,\g^{ab}\,\partial_a X^\m \, \partial_b X_\m\,+\,\l\,\chi\ ,\label{Polyakov}
\end{gather}
where $X^{\m}(\s,\t)$ are the string coordinates, $\g_{ab}$ is the world-sheet metric considered as an independent field and $\chi$ is the Euler characteristic of the Riemann surface spanned by $\xi^a\equiv(\s,\t)$. From this action one can construct the Euclidean path integral describing, in a first quantized language, transition amplitudes between an initial configuration $\psi_i$ and a final configuration $\psi_f$
\begin{gather}
\int_{\psi_i}^{\psi_f} \frac{\cD \g\,\cD X^\m}{Vol_{\,\text{diff x Weyl}}}\ \exp\left[-\frac{1}{4\pi\a'}\int_M d\t d\s\ \sqrt{\g}\,\g^{ab}\,\partial_a X^\m \, \partial_b X_\m\,+\,\l\chi\right]\ .
\end{gather}
Here we have divided the path integral measure by the volume of the group of local symmetries of the action, diffeomorphisms and Weyl transformations, to stress that one has to factor out the contribution associated to the gauge orbits, leaving the relevant integral over some gauge slice to avoid an overcounting of gauge equivalent configurations.

This setting is most natural, after a Wick rotation, to construct an $S$-matrix, where initial and final states are taken to be infinitely far apart from each other. To this end, it is sufficient to describe asymptotic states in the string spectrum.
For String Theory this problem can be solved using the conformal symmetry of the action. In fact, in Conformal Field Theory (CFT) asymptotic free states are mapped, via the operator-state correspondence, to local operators defined on punctures. This leads to the definition of ``Vertex Operators'' $\cV(k_{\,i},\x_i)$, codifying initial and final states of ingoing momentum $k_{\,i}$ on the punctures $\x_i$, and to the definition
\begin{multline}
S_{j_1\cdots j_n}(k_{\,1},\cdots k_n)\,=\,\sum_{\text{compact topologies}}\int\frac{\cD \g\,\cD X^\m}{Vol_{\text{diff x Weyl}}}\\\times\exp\left[-\frac{1}{4\pi\a'}\int_M d\t d\s\ \sqrt{\g}\,\g^{ab}\,\partial_a X^\m \, \partial_b X_\m\,+\,\l\,\chi\right]\\\prod_{i\,=\,1}^n\int d\x_i \sqrt{\g(\x_i)}\ \cV_{j_i}(k_{\,i},\x_i)\ ,
\end{multline}
for the $S$-matrix where, in order to maintain diffeomorphism invariance, we have explicitly integrated over the positions of the world-sheet punctures. For surfaces of non-negative $\chi$ one must actually fix some punctures to account for global symmetries and conformal Killing vectors (CKV). For a more detailed account see \cite{stringtheory,Friedan:1985ge}.


\scs{Light-Cone Quantization}
Let us now consider in more detail the quantization of bosonic strings. As said before, a convenient starting point is the Polyakov action $(\ref{Polyakov})$ that in this section is considered with Minkowski world-sheet signature $(+,-)$.\\
Solving for the equations of motion one obtains
\begin{gather}
\partial_a(\sqrt{-\g}\,\g^{ab}\,\partial_b X^{\m})\,=\,0\ ,\label{Constr1}\\
T_{ab}\,=\,\partial_a X^\m\,\partial_b X_\m\,-\,\frac{1}{2}\,\g_{ab}\,\partial^{\,c} X^\m\,\partial_c X_\m\,=\,0\ ,\label{Constr2}
\end{gather}
where the second is to be regarded as a constraint relating longitudinal and transverse string coordinates.

If a suitable choice for the world-sheet coordinates $\x^a$ is made, turning the metric into the diagonal form
\begin{gather}
\g_{ab}\,=\,e^{\,\phi(\t,\s)}\,\eta_{ab}\ ,\label{gauge}
\end{gather}
eq.$\ (\ref{Constr1})$ can be reduced to the two-dimensional wave equation
\begin{gather}
\left(\partial_{\t}^{\,2}-\partial_{\s}^{\,2}\right)\ X^{\m}\,=\,0\ ,
\end{gather}
whose general solution is a linear combination of left-moving and right-moving modes. 
At this point there are two key options. The first, associated to closed strings, requires the identification $\s\sim\s+2\pi$ with periodic boundary conditions, so that the general solution is of the form
\begin{gather}
X^\m(\t,\s)\,=\,x^\m\,+\,2\,\a' p^{\,\m} \t+i\sqrt{\frac{\a'}{2}}\ \sum_{n\,\neq\, 0}\left(\frac{\a_{\,n}^{\,\m}}{n}\ e^{-2in(\t-\s)}+\frac{\tilde{\a}_{\,n}^{\,\m}}{n}\ e^{-2in(\t+\s)}\right)\ .\label{closed}
\end{gather}
The second, associated to open strings with the simplest choice of Neumann boundary conditions $\partial_\s X\,=\,0$ at $\s\,=\,0$ and $\s\,=\,\pi$, is described by the mode expansion
\begin{gather}
X^\m\,=\,x^\m\,+\,2\,\a' p^{\,\m} \t\,+\,i\sqrt{2\a'}\ \sum_{n\,\neq\, 0}\frac{\a_{\,n}^{\,\m}}{n}\ e^{-in\t}\,\cos(n\s)\ .
\end{gather}
Having fixed the gauge according to $(\ref{gauge})$ does not use up completely the gauge freedom. Rather, it leaves aside a residual symmetry that can be used to choose the light-cone gauge for the string coordinates. Defining
\begin{gather}
X^{\pm}\,=\,\frac{X^0\pm X^{D-1}\!\!}{\sqrt{2}}\ ,
\end{gather}
one can set
\begin{gather}
X^+\,=\,x^++\,2\,\a' p^{\,+}\tau\ ,
\end{gather}
and then, thanks to the constraint equation $(\ref{Constr2})$, the $X^-$ oscillators can be completely determined in terms of the transverse modes $X^i$, with the end result \cite{Goddard:1973qh}
\begin{gather}
\a_{\,n}^{\,-}\,=\,\frac{1}{p^{\,+}}\ L_n\ ,\label{out1}\\
\tilde{\a}_{\,n}^{\,-}\,=\,\frac{1}{p^{\,+}}\ \tilde{L}_n\ ,\label{out2}
\end{gather}
where $L_n$ and $\bar{L_n}$ are the transverse Virasoro operators, defined as
\begin{gather}
L_n\,=\,\frac{1}{2}\,\sum_{m\,=\,-\,\infty}^{\infty}\,\a_{\,n-m}^{\,i}\,\a_{\,m}^{\,i}\ ,\\
\bar{L}_n\,=\,\frac{1}{2}\,\sum_{m\,=\,-\,\infty}^{\infty}\,\tilde{\a}_{\,n-m}^{\,i}\,\tilde{\a}_{\,m}^{\,i}\ .
\end{gather}
The quantization of the system can be performed in the usual way, replacing Poisson brackets with commutators and choosing the normal ordering prescription in order to avoid ordering ambiguities.
In this respect, the only ambiguous Virasoro operators are $L_0$ and $\bar{L}_0$, and one has to choose the normal ordering constant compatibly with Lorentz invariance. Furthermore, the $L_m$ and the $\bar{L}_m$ build two commuting copies of the Virasoro algebra with central charge $c\,=\,D-2$, so that
\begin{gather}
[L_m,L_n]\,=\,(m-n)\,L_{m+n}\,+\,\frac{D-2}{12}\ n(n^2-1)\,\delta_{n+m,0}\ .
\end{gather}
The spectrum can be easily recovered considering the zero-mode part of the constraint $(\ref{Constr2})$, with the correct normal ordering constant for $L_0$ and $\bar{L}_0$, so that the eqs$\ (\ref{out1})$ and $(\ref{out2})$ for $\a_0^-$ and $\tilde{\a}_0^-$ yield
\begin{gather}
M^{\,2}\,=\,\frac{2}{\a'}\,\left(N\,+\,\tilde{N}\,-\,\frac{D-2}{12}\right)\ ,
\end{gather}
for the closed string, and
\begin{gather}
M^{\,2}\,=\,\frac{1}{\a'}\,\left(N\,-\,\frac{D-2}{24}\right)\ ,
\end{gather}
for the open string. As a result, in the critical dimension $D\,=\,26$, the mass spectra of closed and open bosonic strings are described by
\begin{align}
M^{\,2}\,=\,&\frac{2}{\a'}\,\left(N\,+\,\tilde{N}\,-\,2\right)\ ,& M^{\,2}\,=\,&\frac{1}{\a'}\,\left(N\,-\,1\right)\ .&
\end{align}
For later convenience, it is useful to reformulate these simple results using a Euclidean world-sheet signature and complex coordinates \cite{Friedan:1985ge}.
Starting from the closed string mode expansion $(\ref{closed})$, after a Wick rotation to Euclidean signature $(+,+)$, one ends up with a sum of holomorphic and anti-holomorphic functions that can be nicely expressed in terms of the complex coordinates $w\,=\,\t+i\s$ and $\bar{w}\,=\,\t-i\s$. After a conformal transformation of the form $z\,=\,e^w$, that maps the cylinder to the complex plane, the expansion $(\ref{closed})$ turns into
\begin{gather}
X^{\m}(z,\bar{z})\,=\,x^\m\,-\,i\,\frac{\a'}{2}\, p^{\,\m}\,\ln|z|^{\,2}\,+\,i\,\sqrt{\frac{\a'}{2}}\,\sum_{n\,\neq\, 0}\,\frac{1}{n}\,\left(\frac{\a_{\,n}^{\,\m}}{z^{\,n}}\,+\,\frac{\tilde{\a}_{\,n}^{\,\m}}{\bar{z}^{\,n}}\right)\ .
\end{gather}
In this notation all properties of the quantum theory are encoded in the analytic structure of the Operator Product Expansion (OPE) of fields, the short distance expansion of the product of pairs of fields, and thus, eventually, in key properties of analytic functions. All the usual operations have their counterpart in this kind of framework so that, for example, constant time integrals expressing conserved charges are mapped into contour integrals, commutators of conserved charges are mapped into double contour integrals, and on. The counterpart of energy-momentum tensor conservation is holomorphicity or anti-holomorphicity of the corresponding components, while their Laurent expansions encode the Virasoro operators as
\begin{gather}
T_{zz}(z)\,=\,\sum_{n\,=\,-\infty}^{\infty}\frac{L_n}{z^{\,n+2}}\ ,\\
\bar{T}_{\bar{z}\bar{z}}(\bar{z})\,=\,\sum_{n\,=\,-\infty}^{\infty}\frac{\bar{L}_n}{\bar{z}^{\,n+2}}\ .
\end{gather}
The explicit form of $T$, in complex notation, is
\begin{gather}
T(z)\,=\,-\frac{1}{\a'}:\partial X\cdot\partial X:(z)\ ,\\
\bar{T}(\bar{z})\,=\,-\frac{1}{\a'}:\bar{\partial} X\cdot \bar{\partial} X:(\bar{z})\ .
\end{gather}
The open string case can be obtained from the closed one starting from a theory defined on the upper-half plane and extending it over all the complex plane by the ``doubling trick''. This procedure will naturally enforce Neumann boundary conditions imposing, on the real axis, the identification $\partial X^\m\,=\,\bar{\partial} X^\m$, that implies $\a_{\,n}^{\,\m}\,=\,\tilde{\a}_{\,n}^{\,\m}$.


\scs{Vertex Operators}
A vertex operator is the local operator associated with some asymptotic state of the string spectrum via the state-operator isomorphism. For the closed bosonic string in conformal gauge the key local operators associated with string modes are, in complex notation, the weight $(1,0)$ and $(0,1)$ operators
\begin{gather}
\partial X^{\m}(z)\,=\,-i\sqrt{\frac{\a'}{2}}\sum_{m\,=\,-\,\infty}^{\infty}\a_{\,m}^{\,\m} z^{\,-m-1}\ ,\\
\bar{\partial} X^\m(\bar{z})\,=\,-i\sqrt{\frac{\a'}{2}}\sum_{m\,=\,-\,\infty}^{\infty}\tilde{\a}_{\,m}^{\,\m}\bar{z}^{\,-m-1}\ .
\end{gather}
The usual oscillators of String Theory can thus be recovered as conserved charges constructed from these local operators
\begin{gather}
\a_{-m}^{\,\m}\,=\,\sqrt{\frac{2}{\a'}}\oint \frac{dz}{2\pi}\ z^{-m}\partial X^\m(z)\ ,
\end{gather}
so that one is led to the correspondence
\begin{gather}
\a_{-m}^{\,\m}|\,0\ket\lra i\sqrt{\frac{2}{\a'}}\,\frac{1}{(m-1)!}\ \partial^{\,m} X^\m(0)\ ,\ \ m\geq1\ ,
\end{gather}
and similarly, for their conjugates,
\begin{gather}
\tilde{\a}_{-m}^{\,\m}|\,0\ket\lra i\sqrt{\frac{2}{\a'}}\,\frac{1}{(m-1)!}\ \bar{\partial}^{\,m} X^\m(0)\ ,\ \ m\,\geq\,1\ .
\end{gather}
In a similar fashion, the zero-modes $x^\m$ and $\a_{\,0}^{\,\m}\,=\,p^{\,\m}$ can be mapped, respectively, to the local operators $X^\m(0,0)$ and $:e^{ik\cdot X(0,0)}:$, giving rise to the correspondence
\begin{gather}
|\,0;k \ket \lra :e^{\,i\,k\,\cdot\, X(0\,,\,0)}:\ .
\end{gather}
The open string case, as already noticed in the previous paragraph, is obtained from the closed string one starting from a CFT defined on the upper half plane extended to the whole complex plane by the ``doubling trick''. In this case the extension is equivalent to imposing Neumann boundary conditions, so that $\partial X\,=\,\bar{\partial}X$ on the real axis, and consequently one has the identification $\a_{-n}^{\,\m}\,=\,\bar{\a}_{-m}^{\,\m}$.

It is important to stress that in the presence of boundaries there are really two types of vertex operators, those defined in the interior, that describe the emission of closed strings, and those defined on the boundaries, that describe the emission of open strings. These two types of vertex operators are different, because the second feels the presence of the image charges realizing a particular boundary condition. 

In order to attain a definite quantum meaning, every combination of local operators at the same point should be regularized. In a free theory case the regularization entails the operation of normal ordering, that can be compactly expressed in terms of the Green function $\Delta(\s,\s')$, that codifies the short distance behavior, as
\begin{gather}
:\cF:\,=\,\exp\left(-\frac{1}{2}\,\int d^{\,2}\s_1\, d^{\,2}\s_2\ \Delta(\s_1,\s_2)\,\frac{\delta}{\delta X^\m(\s_1)}\,\frac{\delta}{\delta X_\m(\s_2)}\right)\,\cF\ ,
\end{gather}
where, in the Polyakov approach,
\begin{gather}
\Delta(\s,\s')\,=\,-\frac{\a'}{2}\,\ln d^{\,2}(\s,\s')\ ,
\end{gather}
and $d(\s,\s')$ is the geodesic distance between the points $\s_1$ and $\s_2$. In the presence of a boundary, as will be the case for open strings, one has to account for the image charges, so that the Green function 
takes the form
\begin{gather}
\Delta_{\partial}(\s,\s')\,=\,\Delta(\s,\s')+\Delta(\s,\s^{'*})\ ,
\end{gather}
and, for real $\s$, reduces simply to
\begin{gather}
\Delta_{\partial}(y,y')\,=\,2\Delta(y,y')\ .
\end{gather}
Up to now, exploiting the state-operator correspondence we have recognized that the most general bosonic string vertex operator is a product of local operators of the form
\begin{gather}
:\partial X^{\m}\cdots\partial^{\,2} X^{\n}\cdots\partial^{\,n} X^{\r}\ e^{ik\cdot X}:\ .
\end{gather}
However, in order to be compatible with the conformal invariance of the action, that one would like to preserve also at the quantum level, physical vertex operators must satisfy a key constraint. 
They must be primary conformal fields, that basically amounts to not giving rise to poles of order larger than two in the OPE with the energy-momentum tensor (see \cite{Friedan:1985ge}).

Let us restrict our attention to open string states in the first Regge trajectory, for which the relevant vertex operators are of the form
\begin{gather}
\cV^{\,(s)}\,=\,g_o\left(\frac{-i}{\sqrt{2\a'}}\right)^s\oint H_{\m_1\cdots \m_{s}}\,\partial X^{\m_1}\cdots \partial X^{\m_s}\, e^{ik\cdot X}\ ,\label{vertex}
\end{gather}
where $H_{\m_1\cdots \m_s}$ is a fully symmetric polarization tensor for a spin $s$ excitation, $g_o\,=\,e^{\,\l/2}$ is the open string coupling constant, the factor $i^s$ is associated with the number of time derivatives in the vertex consistently with the Wick rotation, the factors of $\a'$ balance the dimension of the derivatives and the integral is on the real axis.
The conformal invariance of the vertex can be shown to be precisely equivalent to the Dirac-Fierz-Pauli conditions for the physical fields
\begin{gather}
\begin{split}
p\cdot H_{\m_2\cdots \m_s}&\,=\,0\ ,\\
H^{\m}_{\m\m_3\cdots \m_s}&\,=\,0\ ,\\
-p^{\,2}H_{\m_1\cdots \m_s}&\,=\,\frac{s\,-\,1}{\a'}\ .
\end{split}
\end{gather}


\scs{Generating Functions}
In the following I would like to describe the computation of tree level scattering amplitudes for states in the first Regge trajectory of both closed and open bosonic strings. In this case the relevant topology is the sphere or the disk, or equivalently the complex plane or the upper-half plane, and a very useful tool, that I will briefly introduce here, is the Euclidean Generating Function with external currents, defined by the functional integral
\begin{gather}
Z[J]\,=\,\int\cD X^\m\ \exp\left(-\frac{1}{4\pi \a'}\int_M d^{\,2}\s\ \partial X^\m \, \bar{\partial}X_\m\,+\,i\int_M d^{\,2}\s\ J(\s)\cdot X(\s)\right)\ .
\end{gather}
A standard way of performing the functional integration is to expand the field $X^\m(\s)$ in terms of a complete set of normalized orthonormal eigenfunctions $X_I(\s)$ of the Laplacian $\partial\,\bar{\partial}$ as
\begin{gather}
X^\m(\s)\,=\,\sum_I\, x_I^\m\, X_I(\s)\ ,\label{modes}
\end{gather}
where
\begin{gather}
\begin{split}
\partial\,\bar{\partial}\, X_I&\,=\,-\omega_I^{\,2}\, X_I\ ,
\end{split}
\end{gather}
and
\begin{gather}
\int d^{\,2}\s\sqrt{\g}\ X_I\, X_J\,=\,\delta_{IJ}\ .\label{orto}
\end{gather}
Then, substituting in the functional integral the mode expansion $(\ref{modes})$ and using $(\ref{orto})$, yields
\begin{gather}
Z[J]\,=\,\prod_{I,\m}\int dx_I^{\,\m} \exp\left(-\frac{1}{4\pi\a'}\ \omega_I^{\,2}\, x_I^{\,\m} x_{I\m}\,+\,i\,x_I^{\,\m}\, J_{I\m}\right)\ ,
\end{gather}
that is a Gaussian integral, apart from the constant zero mode whose integral gives a $\d$-function. Finally, completing the square in the exponent and performing the Gaussian integration yields the useful result
\begin{multline}
Z[J]\,=\,i(2\pi)^d\delta^{(d)}(J_{\,0})\left(\det'\frac{-\partial\,\bar{\partial}}{4\pi^2\a'}\right)^{d/2}\\
\times\exp\left(-\frac{1}{2}\int d^{\,2}\s\, d^{\,2}\s'\ J(\s)\cdot J(\s')\,G'(\s,\s')\right)\ ,\label{Gen}
\end{multline}
where $G'(\s,\s')$ is the Green function, defined excluding the zero mode contribution, and thus satisfying the equation
\begin{gather}
-\frac{1}{2\pi\a'}\ \partial\,\bar{\partial}\,G'(\s,\s')\,=\,\delta^{\,2}(\s_1-\s_2)\,-\,X_0^2\ .\label{Green}
\end{gather}
The overall $\d$-function in $(\ref{Gen})$ implies that one can actually forget the $X_0^2$ term in $(\ref{Green})$. The solution is the Green function on the complex plane or on the upper-half plane considered in the previous section, where the geodesic distance is given by the square modulus. Hence, in our convention
\begin{gather}
G'(\s_1,\s_2)\,=\,-\frac{\a'}{2}\ln|z_{12}|^{\,2}\label{Green2}
\end{gather}
for the complex plane, and
\begin{gather}
G'(\s_1,\s_2)\,=\,-\a'\ln|y_{12}|^{\,2}\label{Green3}
\end{gather}
on the real axis for the upper-half plane.

\chapter{Generating Function for tree-level string amplitudes}
In this chapter we shall deal with the issue of gauge fixing the path integral in order to obtain a well defined set of tree-level $S$-matrix amplitudes. We shall then derive a generating function for the tree-level scattering amplitudes of open bosonic-string states, that will be presented in a manifestly M\"obius invariant fashion. A similar result was previously obtained with a different technique, albeit not in a manifestly M\"{o}bius invariant form, by Moeller and West \cite{Moeller:2005ez}, who also extended it to higher loops. Our formalism, as an intermediate step, reproduces exactly their result at tree level in a different form, and can be similarly generalized to one loop as well as to higher loops. Moreover, in the high energy limit at fixed angle\footnote{The high-energy symmetries proposed by Gross in \cite{Gross:1988ue} where corrected by Moeller and West \cite{Moeller:2005ez}. Further works on these issues can be found in \cite{Amati:1987wq} and in \cite{Taiwan}.} this type of result can be used to extract important information on the high-energy behavior of String Theory.

\scs{Tree-level $S$-matrix amplitudes}
In this section we shall complete, for the simple case of tree-level amplitudes, the brief review of the string $S$-matrix initiated in the previous chapter. The general expression for the $S$-matrix that we considered in the previous chapter is
\begin{multline}
S_{j_1\cdots j_n}(k_{\,1},\cdots k_n)\,=\,\!\!\!\!\sum_{\text{compact topologies}}\int\frac{\cD\g\,\cD X^\m}{Vol_{\,\text{diff x Weyl}}}\\\times\exp\Big[-\frac{1}{4\pi\a'}\,\int_M d\t d\s\ \sqrt{\g}\,\g^{ab}\,\partial_a X^\m\, \partial_b X_\m\,+\,\l\,\chi\Big]\\\prod_{i\,=\,1}^n\int d\s_i \sqrt{\g(\s_i)}\ \cV_{j_i}(k_{\,i},\s_i)\ .
\end{multline}
At tree-level, in the closed string case the unique compact topology to consider is the sphere. This implies that the metric integral can be gauge fixed completely. However, this leaves aside globally defined transformations that do not change the metric at all, associated with Conformal Killing Vectors (CKV). On the sphere these transformations generate the group $SL(2,\mathbb{C})$, and this redundancy can be fixed factoring out the volume of this group by fixing the positions of three vertex operators. This yields, for $n\geq 3$
\begin{multline}
S_{j_1\cdots j_n}(k_{\,1},\cdots k_n)\,=\,\int d^{\,2} z_4\cdots d^{\,2} z_n\ |z_{12}z_{13}z_{23}|^2\\\times\bra \cV_{j_{\,1}}(\hat{z}_1,k_{\,1})\cV_{j_{\,2}}(\hat{z}_2,k_{\,2})\cV_{j_{\,3}}(\hat{z}_3,k_{\,3})\cdots\cV_{j_{\,n}}(z_n,k_n)\ket\ ,
\end{multline}
where we have added a ``hat'' on the coordinates that are not integrated, and
\begin{gather}
z_{ij}\,=\,z_{i}-z_j\ .
\end{gather}
The measure factor $|z_{12}z_{13}z_{23}|^2$, draws its origin from volume factorization, and can be determined in a number of ways imposing that the final result for the amplitude be invariant under $SL(2,\mathbb{C})$.

For open strings, that will be of primary interest for us, the vertex operators are defined on the real axis while the world sheet is the upper-half plane. In this case the group generated by the CKV is the M\"obius group $SL(2,\mathbb{R})$ and the $S$-matrix is given, leaving aside for the moment the Chan-Paton factors \cite{Paton:1969je}, by $(n\geq3)$
\begin{multline}
S_{j_1\cdots j_n}(k_{\,1},\cdots k_n)\,=\,\int_{\mathbb{R}^{n-3}} d y_4\cdots d y_n\ |y_{12}y_{13}y_{23}|\\\times\bra \cV_{j_{\,1}}(\hat{y}_1,k_{\,1})\cV_{j_{\,2}}(\hat{y}_2,k_{\,2})\cV_{j_{\,3}}(\hat{y}_3,k_{\,3})
\cdots\cV_{j_{\,n}}(y_n,k_n)\ket+(1\lra 2)\ ,
\end{multline}
where
\begin{gather}
y_{ij}\,=\,y_i-y_j\ ,
\end{gather}
and where we have added a ``hat'' on the coordinates that are not integrated. Notice that a distinct contribution with a pair of vertex operators interchanged is needed because only cyclic permutations of the three fixed external legs are M\"obius equivalent. The factor coming from volume factorization is in this case the holomorphic part of the closed string one computed on the real axis. Moreover, the open string amplitude thus obtained can be generalized, since cyclic symmetry allows to dress it with Chan-Paton degrees of freedom \cite{Paton:1969je} that can be associated with the string end points. More explicitly, one can define a set of matrices $\L^a$ satisfying
\begin{gather}
\begin{split}
Tr(\L^a \L^b)&\,=\,\delta^{\,ab}\ ,\\
\sum_{a_I} Tr(A\L^{a_I})Tr (B\L^{a_I})&\,=\,Tr(AB)\ ,
\end{split}
\end{gather}
or their generalization, taking into account the flip symmetry of string amplitudes, given by
\begin{gather}
\begin{split}
Tr([\L_{\,o}^{a_1},\L_{\,o}^{a_2}][\L_{\,o}^{a_3},\L_{\,o}^{a_4}])&\,\sim\,\sum_{a_I}
Tr([\L_{\,o}^{a_1},\L_{\,o}^{a_2}]\L_{\,o}^{a_I})Tr(\L_{\,o}^{a_I}[\L_{\,o}^{a_3},\L_{\,o}^{a_4}])\ ,\\
Tr(\{\L_{\,o}^{a_1},\L_{\,o}^{a_2}\}\{\L_{\,o}^{a_3},\L_{\,o}^{a_4}\})&\,\sim\,\sum_{a_I}
Tr(\{\L_{\,o}^{a_1},\L_{\,o}^{a_2}\}\L_{\,e}^{a_I})Tr(\L_{\,e}^{a_I}\{\L_{\,o}^{a_3},\L_{\,o}^{a_4}\})\ ,\\
Tr([\L_{\,e}^{a_1},\L_{\,e}^{a_2}][\L_{\,e}^{a_3},\L_{\,e}^{a_4}])&\,\sim\,\sum_{a_I}
Tr([\L_{\,e}^{a_1},\L_{\,e}^{a_2}]\L_{\,o}^{a_I})Tr(\L_{\,o}^{a_I}[\L_{\,e}^{a_3},\L_{\,e}^{a_4}])\ ,\\
Tr(\{\L_{\,e}^{a_1},\L_{\,e}^{a_2}\}\{\L_{\,e}^{a_3},\L_{\,e}^{a_4}\})&\,\sim\,\sum_{a_I}
Tr(\{\L_{\,e}^{a_1},\L_{\,e}^{a_2}\}\L_{\,e}^{a_I})Tr(\L_{\,e}^{a_I}\{\L_{\,e}^{a_3},\L_{\,e}^{a_4}\})\ ,
\end{split}
\end{gather}
where the labels $e$ and $o$ are associated to the freedom, that arises thanks to the flip symmetry, of associating different Chan-Paton factors to even and odd mass levels of the open spectrum.
Then, up to these set of matrices it is possible to define consistently with unitarity dressed amplitudes given by
\begin{multline}
S_{j_1\cdots j_n}(k_{\,1},\cdots k_n)\,=\,\int_{\mathbb{R}^{n-3}} d y_4\cdots d y_n\ |y_{12}y_{13}y_{23}|\\\times\bra \cV_{j_{\,1}}(\hat{y}_1,k_{\,1})\cV_{j_{\,2}}(\hat{y}_2,k_{\,2})\cV_{j_{\,3}}(\hat{y}_3,k_{\,3})\cdots\cV_{j_{\,n}}(y_n,k_n)\ket \,Tr(\L^{a_1}\cdots \L^{a_n})+(1\lra 2)\ ,\label{Smatrix}
\end{multline}
where the ordering of the $\L^{a_i}$'s is supposed to reflect the ordering of the $y_i$'s in the various regions of integration.
This formula will be our starting point in the next sections.


\scs{Generating Function}
As one can see from $(\ref{Smatrix})$, scattering amplitudes in String Theory are basically integrals of vertex operator correlation functions. They can be computed very conveniently using the generating function $(\ref{Gen})$ obtained in the previous chapter.
To this end, the first step is to use a current of the form
\begin{multline}
J(\s)\,=\,\sum_{i\,=\,1}^n\left(\vphantom{k^{\,(n)}_{\,i}}k_{\,i}\,\delta^{\,2}(\s-\s_i)\,- \,k'_i\,\partial\,\delta^{\,2}(\s-\s_i)\,+\,\cdots\,+\,\right.\\\left.
+\,(-1)^{n}\,k^{\,(n)}_{\,i}\,\partial^{\,(n)}\delta^{\,2}(\s-\s_i)\,+\,\cdots\right)\ ,\label{curr}
\end{multline}
where in principle one should consider all possible derivatives of the $\delta$-function that, integrating by parts, give rise to sources for all terms of the form $\partial^{(n)}X^{\m}$, that are building blocks of any vertex operators.
Using this current the generating function $(\ref{Gen})$ turns into
\begin{multline}
Z\,=\,i(2\pi)^d\delta^{(d)}(J_{\,0})\ \cC\\\times\exp\left[-\frac{1}{2}\left(\sum_{i,j\,=\,1}^n\sum_{n,m\,=\,0}^{\infty}k_{\,i}^{(n)}\cdot k_{\,j}^{(m)}\ \partial_i^{\,n}\,\partial_j^{\,m}\, G'(\s_i,\s_j)\right)\right]\ ,
\end{multline}
where the label on the derivative indicates on which argument of the Green function the derivative acts upon and where the constant $\cC$ includes all constant factors coming from the functional integration.

The terms with $i\,=\,j$ are to be properly regulated and, in general, are proportional to the two-dimensional metric and its derivatives. With a convenient gauge choice, considering a flat metric, they only affect the overall constant $\cC$, so that one is finally led to the generating function
\begin{multline}
Z\,=\,i\,(2\pi)^d\,\delta^{\,(d)}(J_{\,0})\ \cC\\\times\exp\left[-\frac{1}{2}\left(\sum_{i\,\neq\, j}^n\sum_{n,m\,=\,0}^{\infty}k_{\,i}^{\,(n)}\cdot k_{\,j}^{\,(m)}\ \partial_i^{\,n}\,\partial_j^{\,m} \, G'(\s_i,\s_j)\right)\right]\ ,\label{Gen2}
\end{multline}
whose expansion coefficients encode precisely all correlation functions that are needed and where all constant factors are included in the constant $\cC$.
The most general correlation function of vertex operators can in fact be evaluated taking a number of derivatives with respect to the sources $k^{\,(n)}$ and letting at the end $k^{\,(n)}\,=\,0$.

In the following it will prove convenient to specialize the result $(\ref{Gen2})$ to describe correlation functions of vertex operators for states in the first Regge trajectory, letting $k^{\,(n)}\,=\,0$ for $n\,>\,1$. The end result is the generating function
\begin{multline}
Z\,=\,i\,(2\pi)^d\,\delta^{\,(d)}(J_{\,0})\,\cC\,\exp\left[-\frac{1}{2}\sum_{i\,\neq\, j}^n\left(k_{\,i}\cdot k_{\,j}\, G'(\s_i,\s_j)\right.\right.\\\left.\vphantom{-\frac{1}{2}\sum_{i\,\neq\, j}^n}\left.+\,2\,k_{\,i}'\cdot k_{\,j} \, \partial_i\, G'(\s_i,\s_j)\,+\,k_{\,i}'\cdot k_{\,j}'\, \partial_i\,\partial_j \, G'(\s_i,\s_j)\right)\right]\ .\label{Gen3}
\end{multline}
Until now our discussion has been quite general, so that in principle it also applies to the closed string case. From now on however, we specialize to the case of the open bosonic string, where the derivatives are supposed to be along the world sheet boundary.

Starting from the Green function formula $(\ref{Green3})$, one can then obtain the relations
\begin{gather}
\begin{split}
G'(\s_i,\s_j)\,=\,&-\,\a'\ln|y_i\,-\,y_j|^{\,2}\ ,\\
\partial_i\, G'(\s_i,\s_j)\,=\,&-\frac{2\a'}{y_i\,-\,y_j}\ ,\\
\partial_i\,\partial_j\, G'(\s_i,\s_j)\,=\,&-\frac{2\a'}{(y_{\,i}\,-\,y_{\,j})^{\,2}}\ ,
\end{split}
\end{gather}
and substituting in $(\ref{Gen3})$ gives
\begin{multline}
Z\,=\,i\,(2\pi)^d\,\delta^{\,(d)}\left(\sum_i k_{\,i}\right)\,\cC\,\\\exp\Bigg\{\a'\sum_{i\,\neq\, j}^n\Big[k_{\,i}\cdot k_{\,j}\ln|y_{ij}|\,-\,\frac{2\,k_{\,i}\cdot k_{\,j}'}{y_{ij}}+\frac{k_{\,i}'\cdot k_{\,j}'}{y_{ij}^2}\Big]\Bigg\}\ .\label{FirstRegge}
\end{multline}
As a consequence, in this formalism the vertex operator $(\ref{vertex})$ acquires the operatorial form
\begin{multline}
\cV_s\,=\,g_o\left(\frac{-i}{\sqrt{2\a'}}\right)^s H_{\m_1\cdots \m_s}\partial X^{\m_1}\cdots \partial X^{\m_s}\rightarrow\\\rightarrow g_o\left(\frac{-1}{\sqrt{2\a'}}\right)^s H_{\m_1\cdots \m_s}\frac{\partial}{\partial k_{\,\m_1}'}\cdots \frac{\partial}{\partial k_{\,\m_s}'}\ ,\label{Vertex2}
\end{multline}
while the effective theory for the first Regge trajectory maps naturally into an extended space with auxiliary coordinates $k_{\,i}'$. It is important to notice that one has the freedom to rescale all $k_{\,i}'$ by any non-vanishing complex number, provided the same rescaling is effected in the vertex operator formula $(\ref{Vertex2})$, without affecting the amplitude.


\scs{Generating Function for $3$-point Amplitudes}

In this section we shall take as our starting point the generating function $(\ref{FirstRegge})$ and, for the case of three-point functions, we shall write it in a manifestly M\"obius invariant fashion. To this end, it will be necessary to take a closer look at the kinematics of three-point scattering.
In our mostly-plus convention one can parameterize the masses as
\begin{gather}
\begin{split}
-k_{\,1}^{\,2}&\,=\,\frac{m\,-\,1}{\a'}\ ,\\
-k_{\,2}^{\,2}&\,=\,\frac{n\,-\,1}{\a'}\ ,\\
-k_{\,3}^{\,2}&\,=\,\frac{l\,-\,1}{\a'}\ .
\end{split}
\end{gather}
and momentum conservation is then equivalent to the conditions
\begin{gather}
\begin{split}
2\a'k_{\,1}\cdot k_{\,2}&\,=\,\a'(k_{\,3}^{\,2}\,-\,k_{\,1}^{\,2}\,-\,k_{\,2}^{\,2})\ ,\\
2\a'k_{\,1}\cdot k_{\,3}&\,=\,\a'(k_{\,2}^{\,2}\,-\,k_{\,1}^{\,2}\,-\,k_{\,3}^{\,2})\ ,\\
2\a'k_{\,2}\cdot k_{\,3}&\,=\,\a'(k_{\,1}^{\,2}\,-\,k_{\,2}^{\,2}\,-\,k_{\,3}^{\,2})\ .
\end{split}
\end{gather}
Moreover, let us recall that the Fierz-Pauli constraints on the polarization tensors are
\begin{gather}
\begin{split}
p\cdot H_{\m_2\cdots \m_s}&\,=\,0\ ,\\
H^{\m}_{\m\m_3\cdots \m_s}&\,=\,0\ ,\\
-p^{\,2}\,H_{\m_1\cdots \m_s}&\,=\,\frac{s\,-\,1}{\a'}\ .
\end{split}
\end{gather}
With these relations in mind, it is possible to write the purely momentum dependent part of the generating function in the form
\begin{gather}
|y_{12}\,y_{13}\,y_{23}|\exp\left[\a'\sum_{i\,\neq\, j}\,k_{\,i}\cdot k_{\,j}\,\ln|y_{ij}\,|\right]\,=\,\Big|\frac{y_{12}y_{23}}{y_{13}}\Big|^{\,m}
\Big|\frac{y_{12}y_{13}}{y_{23}}\Big|^{\,n}\Big|\frac{y_{13}y_{23}}{y_{12}}\Big|^{\,l}\ ,\label{aaa}
\end{gather}
where, for convenience, we have also included the measure factor $|y_{12}\,y_{13}\,y_{23}|$.
For three-point amplitudes there is no integration to perform and one must only apply the vertex operators to the generating function which yields
\begin{gather}
\cA_{(m,n,l)}^{(3)}\,=\,\hat{\cV}^{\,(1)}_m\hat{\cV}^{\,(2)}_n\hat{\cV}^{\,(3)}_l Z\ .
\end{gather}
For later convenience we shall put the dependence on all coefficients in the generating function redefining the vertex operators $(\ref{Vertex2})$ as
\begin{gather}
\hat{\cV}^{(s)}\,=\,H_{\m_1\cdots \m_s}\frac{\partial}{\partial k_{\,\m_1}'}\cdots \frac{\partial}{\partial k_{\,\m_s}'}\ .
\end{gather}
As a result the generating function of three-point amplitudes turns into
\begin{multline}
Z\,=\,i\,g_o^{\,3}\, \cC\, e^{-\l}\,(2\pi)^{d}\,\delta^{\,(d)}\left(\sum_i k_{\,i}\right)\left(\frac{-1}{\sqrt{2\a'}}\right)^{m\,+\,n\,+\,l}\Big|\frac{y_{12}y_{23}}{y_{13}}\Big|^{\,m}
\Big|\frac{y_{12}y_{13}}{y_{23}}\Big|^{\,n}\\\Big|\frac{y_{13}y_{23}}{y_{12}}\Big|^{\,l}
\times\exp\left\{\a'\sum_{i\,\neq\, j}^N\left[-\frac{2\,k_{\,i}\cdot k_{\,j}'}{y_{ij}}\,+\,\frac{k'_i\cdot k'_j}{y_{ij}^2}\right]\right\}\ ,\label{Gen3}
\end{multline}
where $m$, $n$ and $l$ depend on the vertex operators applied. This means that the term with exponent $m$, $n$ and $l$ is multiplied with the term in the Taylor expansion of the exponential with $m$ derivatives respect to $k_{\,1}'$, $n$ derivatives respect to $k_{\,2}'$ and $l$ derivatives respect to $k_{\,3}'$.

Here the constant $\cC$ can be easily determined without computing it explicitly. In fact, using unitarity in the simple case of tachyon scattering, it is possible to obtain the general relation
\begin{gather}
g_o^{\,2}\,\cC\, e^{-\l}\,=\,\frac{1}{\a'}\ ,
\end{gather}
and substituting in eq.$\ (\ref{Gen3})$ one finally obtains
\begin{multline}
Z\,=\,i\,\frac{g_o}{\a'}\,(2\pi)^{d}\,\delta^{\,(d)}\left(\sum_i k_{\,i}\right)\left(\frac{-1}{\sqrt{2\a'}}\right)^{m\,+\,n\,+\,l}\Big|\frac{y_{12}y_{23}}{y_{13}}\Big|^{\,m}
\Big|\frac{y_{12}y_{13}}{y_{23}}\Big|^{\,n}\Big|\frac{y_{13}y_{23}}{y_{12}}\Big|^{\,l}\\
\times\exp\left\{\a'\sum_{i\,\neq\, j}^N\left[-\frac{2\,k_{\,i}\cdot k_{\,j}'}{y_{ij}}\,+\,\frac{k'_{\,i}\cdot k'_{\,j}}{y_{ij}^2}\right]\right\}\ .\label{Gen4}
\end{multline}
This generating function is rather cumbersome because of the presence of the explicit factors depending on $m$, $n$ and $l$. Explicitly, however, $Z$ has the form:
\begin{gather}
Z\,=\,\sum_{n\,,\,m\,,\,l}\,A^{\,m} B^{\,n} C^{\,l} (k_{\,1}'\cdot\partial_{k_{\,1}'})^m(k_{\,2}'\cdot\partial_{k_{\,2}'})^n (k_{\,3}'\cdot\partial_{k_{\,3}'})^l\,E(0,0,0)\ ,
\end{gather}
where $A$, $B$ and $C$ are the $y$-dependent prefactors and $E\left(k_{\,1}',k_{\,2}',k_{\,3}'\right)$ is the exponential function in $(\ref{Gen3})$.
One can now sum this series noticing that it is the Taylor expansion of the function $E(Ak_{\,1}',Bk_{\,2}',Ck_{\,3}')$ around $k_{\,1}'\,=\,k_{\,2}'\,=\,k_{\,3}'\,=\,0$:
\begin{gather}
Z\,=\,E\left(Ak_{\,1}',Bk_{\,2}',Ck_{\,3}'\right)\ .
\end{gather}
Using the last identity, $Z$ takes a more elegant form. Taking into account transversality\footnote{The transversality of the polarization tensors translates into the transversality of the $k_{\,i}'$.} and momentum conservation, the exponent becomes
\begin{multline}
\a'\sum_{i\,\neq\, j}^N\left[-\frac{2\,k_{\,i}\cdot k_{\,j}'}{y_{ij}}\,+\,\frac{k'_i\cdot k'_j}{y_{ij}^2}\right]\,=\,\a' \frac{y_{12}}{y_{13}y_{23}}\ k_{\,12}\cdot k_{\,3}'\\\,+\,\a'\frac{y_{13}}{y_{12}y_{23}}\ k_{\,31}\cdot k_{\,2}'\,+\,\a' \frac{y_{23}}{y_{12}y_{13}}\ k_{\,23}\cdot k_{\,1}'\,+\,2\a'\left(\frac{k_{\,1}'\cdot k_{\,2}'}{y_{12}^2}\,+\,\frac{k_{\,1}'\cdot k_{\,3}'}{y_{13}^2}\,+\,\frac{k_{\,2}'\cdot k_{\,3}'}{y_{23}^2}\right)\ ,
\end{multline}
and, when it is properly combined with the prefactor, one finally obtains the $y_{\,i}$\,-independent generating function
\begin{multline}
Z\,=\,i\,g_o\,\frac{(2\pi)^{d}}{\a'}\,\delta^{\,(d)}\left(\sum_i k_{\,i}\right)\exp\left\{-\sqrt{\frac{\a'}{2}}\left(k_{\,1}'\cdot k_{\,23}\left\langle\frac{y_{23}}{y_{12}y_{13}}\right\rangle\right.\right.\\\left.\,+\,k_{\,2}'\cdot k_{\,31}\left\langle\frac{y_{13}}{y_{12}y_{23}}\right\rangle\,+\,k_{\,3}'\cdot k_{\,12}\left\langle\frac{y_{12}}{y_{13}y_{23}}\right\rangle\right)\\\left.+\,(k_{\,1}'\cdot k_{\,2}'\,+\,k_{\,1}'\cdot k_{\,3}'\,+\,k_{\,2}'\cdot k_{\,3}')\vphantom{\sqrt{\frac{\a'}{2}}}\right\}\ ,
\end{multline}
where the notation $\bra\cdots\ket$ indicates the sign of the expression within brackets. Taking also into account the Chan-Paton factors and considering separately the two non-equivalent orderings $(y_1,y_2,y_3)$ and $(y_2,y_1,y_3)$ yields the final result
\begin{gather}
\begin{split}
Z&\,=\,i\,g_o\,\frac{(2\pi)^{d}}{\a'}\,\delta^{\,(d)}\left(\sum_i k_{\,i}\right)\left[\vphantom{\sqrt{\frac{\a'}{2}}}\right.\\
&\exp\left\{+\,\sqrt{\frac{\a'}{2}}\left(k_{\,1}'\cdot k_{\,23}+k_{\,2}'\cdot k_{\,31}+k_{\,3}'\cdot k_{\,12}\right)+(k_{\,1}'\cdot k_{\,2}'+k_{\,1}'\cdot k_{\,3}'+k_{\,2}'\cdot k_{\,3}')\right\}\\\vphantom{\sqrt{\frac{\a'}{2}}}\times&Tr[\L^{a_1}\L^{a_2}\L^{a_3}]\\
+&\exp\left\{-\sqrt{\frac{\a'}{2}}\left(k_{\,1}'\cdot k_{\,23}+k_{\,2}'\cdot k_{\,31}+k_{\,3}'\cdot k_{\,12}\right)+(k_{\,1}'\cdot k_{\,2}'+k_{\,1}'\cdot k_{\,3}'+k_{\,2}'\cdot k_{\,3}')\right\}\\\times&Tr[\L^{a_2}\L^{a_1}\L^{a_3}]\vphantom{\sqrt{\frac{\a'}{2}}}\Bigg]\ .
\end{split}
\end{gather}
Any three-point amplitude can now be computed applying three of the operators
\begin{gather}
\hat{\cV}_s\,=\,H_{\m_1\cdots\m_s}\,\frac{\partial}{\partial k_{\,\m_1}'}\cdots \frac{\partial}{\partial k_{\,\m_s}'}\ ,
\end{gather}
to $Z$, as
\begin{gather}
\cA_{(m,n,l)}^{(3)}\,=\,\hat{\cV}^{\,(1)}_m\hat{\cV}^{\,(2)}_n\hat{\cV}^{\,(3)}_l Z\ .
\end{gather}


\scs{Generating Function for $4$-point Amplitudes}

We now turn to the generating function of four-particle amplitudes for states of the first Regge trajectory. The starting point is, as before, the generating function
\begin{multline}
Z\,=\,i\,(2\pi)^{d}\,\delta^{\,(d)}(k_{\,1}+k_{\,2}+k_{\,3}+k_{\,4})\ \cC\\\times\exp\left\{\a'\sum_{i\,\neq\, j}\left[k_{\,i}\cdot k_{\,j}\,\ln|y_{ij}|\,-\,\frac{2\,k_{\,i}\cdot k_{\,j}'}{y_{ij}}\,+\,\frac{k_{\,i}'\cdot k_{\,j}'}{y_{ij}^2}\,\right]\right\}\ ,
\end{multline}
with the vertex operators
\begin{gather}
\hat{\cV}_s\,=\,g_o\left(\frac{-1}{\sqrt{2\a'}}\right)^s H_{\m_1\cdots \m_s}\frac{\partial}{\partial k_{\,\m_1}'}\cdots \frac{\partial}{\partial k_{\,\m_s}'}\ .
\end{gather}
In order to simplify the purely momentum dependent term it is useful, as before, to take a closer look at the four-particle kinematics.
Parameterizing the masses as
\begin{align}
-k_{\,1}^{\,2}&\,=\,\frac{m-1}{\a'}\ ,& -k_{\,2}^{\,2}&\,=\,\frac{n-1}{\a'}\ ,& -k_{\,3}^{\,2}&\,=\,\frac{p-1}{\a'}\ ,& -k_{\,4}^{\,2}&\,=\,\frac{q-1}{\a'}\ ,
\end{align}
the four-particle kinematics can be nicely expressed in terms of the Mandelstam variables $s$, $t$ and $u$, with the constraint
\begin{gather}
\a'(s+t+u)\,=\,m+n+p+q-4\ ,
\end{gather}
while the terms of the form $k_{\,i}\cdot k_{\,j}$ take the form
\begin{align}
2\a'k_{\,1}\cdot k_{\,2}&\,=\,m+n-2-\a's\ ,& 2\a'k_{\,3}\cdot k_{\,4}&\,=\,p+q-2-\a's\ ,\\
2\a'k_{\,1}\cdot k_{\,3}&\,=\,m+p-2-\a't\ ,& 2\a'k_{\,2}\cdot k_{\,4}&\,=\,n+q-2-\a't\ ,\\
2\a'k_{\,1}\cdot k_{\,4}&\,=\,m+q-2-\a'u\ ,& 2\a'k_{\,2}\cdot k_{\,3}&\,=\,n+p-2-\a'u\ ,\label{Man}
\end{align}
so that the part of the generating function that is purely momentum dependent turns into\footnote{We already introduced at this stage the correct normalization for the measure $|y_{12}y_{13}y_{23}|$, while discarding three of the four vertex integrations.
}
\begin{multline}
|y_{12}\,y_{13}\,y_{23}|\exp\left[\a'\sum_{i\,\neq\, j}k_{\,i}\cdot k_{\,j}\,\ln|y_{ij}|\right]\,=\,\Big|\frac{y_{13}y_{24}}{y_{12}y_{34}}\frac{y_{23}}{y_{24}y_{34}}
\Big|\Big|\frac{y_{13}y_{14}}{y_{34}}\Big|^m \\\times\Big|\frac{y_{23}y_{24}}{y_{34}}\Big|^n\Big|\frac{y_{13}y_{23}}{y_{12}}\Big|^p
\Big|\frac{y_{14}y_{24}}{y_{12}}\Big|^q\Big|\frac{y_{13}y_{24}}{y_{12}y_{34}}
\Big|^{-\a't-2}\Big|\frac{y_{14}y_{23}}{y_{12}y_{34}}\Big|^{-\a'u-2}\ .\label{fac}
\end{multline}
As before, it is convenient to move all factors inside the generating function, so that the overall coefficient reduces to
\begin{gather}
i\,\frac{g_o}{\a'}\,(2\pi)^{d}\,\delta^{\,(d)}(k_{\,1}+k_{\,2}+k_{\,3}+k_{\,4})\,
\left(\frac{-1}{\sqrt{2\a'}}\right)^{m+n+p+q}\ ,
\end{gather}
while the vertex operators turn into
\begin{gather}
\begin{split}
\hat{\cV}_s\,=\,H_{\m_1\cdots \m_s}\frac{\partial}{\partial k_{\,\m_1}'}\cdots \frac{\partial}{\partial k_{\,\m_m}'}\ .
\end{split}
\end{gather}
The non purely momentum dependent part can now be cast in a more convenient form using momentum conservation and transversality $(k_{\,i}\cdot k_{\,i}'\,=\,0)$, so that
\begin{gather}
\begin{split}
k_{\,1}'\cdot\left(\frac{k_{\,2}}{y_{21}}+\frac{k_{\,3}}{y_{31}}+\frac{k_{\,4}}{y_{41}}\right)&
\,=\,+\frac{y_{34}}{y_{13}y_{14}}\left(k_{\,3}\frac{y_{14}y_{23}}{y_{12}y_{34}}
+k_{\,4}\frac{y_{13}y_{24}}{y_{12}y_{34}}\right)\cdot k_{\,1}'\ ,\\
k_{\,2}'\cdot\left(\frac{k_{\,1}}{y_{12}}+\frac{k_{\,3}}{y_{32}}+\frac{k_{\,4}}{y_{42}}\right)&
\,=\,-\frac{y_{34}}{y_{23}y_{24}}\left(k_{\,3}\frac{y_{13}y_{24}}{y_{12}y_{34}}
+k_{\,4}\frac{y_{14}y_{23}}{y_{12}y_{34}}\right)\cdot k_{\,2}'\ ,\\
k_{\,3}'\cdot\left(\frac{k_{\,1}}{y_{13}}+\frac{k_{\,2}}{y_{23}}
+\frac{k_{\,4}}{y_{43}}\right)&\,=\,+\frac{y_{12}}{y_{13}y_{23}}
\left(k_{\,1}\frac{y_{14}y_{23}}{y_{12}y_{34}}+k_{\,2}\frac{y_{13}y_{24}}{y_{12}y_{34}}\right)\cdot k_{\,3}'\ ,\\
k_{\,4}'\cdot\left(\frac{k_{\,1}}{y_{14}}+\frac{k_{\,2}}{y_{24}}
+\frac{k_{\,3}}{y_{34}}\right)&\,=\,-\frac{y_{12}}{y_{14}y_{24}}
\left(k_{\,1}\frac{y_{13}y_{24}}{y_{12}y_{34}}+k_{\,2}\frac{y_{14}y_{23}}{y_{12}y_{34}}\right)\cdot k_{\,4}'\ ,
\end{split}
\end{gather}
while the purely $k_{\,i}'$ dependent term is given by
\begin{gather}
\frac{k_{\,1}'\cdot k_{\,2}'}{y_{12}^2}+\frac{k_{\,1}'\cdot k_{\,3}'}{y_{13}^2}+\frac{k_{\,1}'\cdot k_{\,4}'}{y_{14}^2}+\frac{k_{\,2}'\cdot k_{\,3}'}{y_{23}^2}+\frac{k_{\,2}'\cdot k_{\,4}'}{y_{24}^2}+\frac{k_{\,3}'\cdot k_{\,4}'}{y_{34}^2}\ .
\end{gather}
Then, following the same steps as in the previous case, one can notice that the term with exponents $m$, $n$, $p$ and $q$ is to be multiplied with the term in the Taylor expansion of the exponential with the corresponding number of derivatives.
As before, it is thus possible to combine these terms, and at the end one is left with a manifestly M\"obius invariant measure integrated over the real axis,
\begin{small}
\begin{gather}
\begin{split}
Z=i\frac{g_o^2}{\a'}&(2\pi)^{d}\delta^{(d)}(k_1+k_2+k_3+k_4)\int_{-\infty}^{\infty} \Big|\frac{y_{13}y_{24}}{y_{12}y_{34}}\Big|\Big|\frac{y_{23}}{y_{24}y_{34}}\Big|d\l\ \exp\Bigg\{\sqrt{2\a'}\\\times\left[\left\langle\frac{y_{34}}{y_{13}y_{14}}\right\rangle\right.&
\Big(k_3\frac{y_{14}y_{23}}{y_{12}y_{34}}+k_4\frac{y_{13}y_{24}}{y_{12}y_{34}}\Big)\cdot k_1'-\left\langle\frac{y_{34}}{y_{23}y_{24}}\right\rangle
\Big(k_3\frac{y_{13}y_{24}}{y_{12}y_{34}}+k_4\frac{y_{14}y_{23}}{y_{12}y_{34}}\Big)\cdot k_2'\\+\left\langle\frac{y_{12}}{y_{13}y_{23}}\right\rangle&\left.
\Big(k_1\frac{y_{14}y_{23}}{y_{12}y_{34}}+k_2\frac{y_{13}y_{24}}{y_{12}y_{34}}\Big)\cdot k_3'-\left\langle\frac{y_{12}}{y_{14}y_{24}}\right\rangle
\Big(k_1\frac{y_{13}y_{24}}{y_{12}y_{34}}+k_2\frac{y_{14}y_{23}}{y_{12}y_{34}}\Big)\cdot k_4'\right]\\
+\left[\Big|\frac{y_{13}y_{24}}{y_{12}y_{34}}\right.&\frac{y_{14}y_{23}}{y_{12}y_{34}}\Big|(k_1'\cdot k_2'+k_3'\cdot k_4')+\Big|\frac{y_{14}y_{23}}{y_{12}y_{34}}\Big|(k_1'\cdot k_3'\\+&\left.k_2'\cdot k_4')+\Big|\frac{y_{13}y_{24}}{y_{12}y_{34}}\Big|(k_1'\cdot k_4'+k_2'\cdot k_3')\right]\Bigg\}\Big|\frac{y_{13}y_{24}}{y_{12}y_{34}}\Big|^{-\a't-2}\Big|\frac{y_{14}y_{23}}{y_{12}y_{34}}\Big|^{-\a'u-2}\ ,
\end{split}
\end{gather}
\end{small}and with the vertex operators
\begin{gather}
\hat{\cV}_s\,=\,H_{\m_1\cdots \m_s}\,\frac{\partial}{\partial k_{\,\m_1}'}\cdots \frac{\partial}{\partial k_{\,\m_s}'}\ .
\end{gather}
As we stressed, given the M\"obius invariance it is possible to fix three points letting $y_1\,=\,0$, $y_2\,=\,1$ and $y_3\,=\,\infty$, while $y_4\,=\,\l$ is to be integrated on the real axis. An independent contribution then comes from the non M\"obius equivalent configuration in which $y_1$ and $y_2$ are interchanged. One is than led to
\begin{small}
\begin{gather}
\begin{split}
Z&\,=\,i\,\frac{g_o^2}{\a'}\,(2\pi)^{d}\,\delta^{\,(d)}(k_{\,1}+k_{\,2}+k_{\,3}+k_{\,4})\int_{-\infty}^{\infty} d\l\ |1-\l|^{-\a't-2}|\l|^{-\a'u-2}\\
\times&\exp\Big\{\sqrt{2\a'}\ \Big[\bra\l\ket(-k_{\,3}\l+k_{\,4}(1-\l))\cdot k_{\,1}'+\bra1-\l\ket(k_{\,3}(1-\l)-k_{\,4}\l)\cdot k_{\,2}'\\+&(k_{\,1}\l-k_{\,2}(1-\l))\cdot k_{\,3}'+\bra\l(1-\l)\ket(-k_{\,1}(1-\l)+k_{\,2}\l)\cdot k_{\,4}'\Big]\\
+&\Big[\l(1-\l)(k_{\,1}'\cdot k_{\,2}'+k_{\,3}'\cdot k_{\,4}')+\l(k_{\,1}'\cdot k_{\,3}'+k_{\,2}'\cdot k_{\,4}')+(1-\l)(k_{\,1}'\cdot k_{\,4}'+k_{\,2}'\cdot k_{\,3}')\Big]\Big\}\\\times&\vphantom{\exp\Big\{\Big\}}+(1\lra 2)\ ,
\end{split}
\end{gather}
\end{small}but in order to consider arbitrary Chan-Paton factors, one must split the complete integration region into three pieces, $[-\infty,0]$, $[0,1]$ and $[1,\infty]$, associating to each of them the corresponding Chan-Paton factor.
In total, one then obtains six contributions and, after the change of variables
\begin{gather}
\begin{split}
(1,\infty)\ra (0,1)&:\ \ \l\ra\frac{1}{1-\l},\ 1-\l\ra-\frac{\l}{1-\l},\ d\l\ra\frac{d\l}{(1-\l)^2}\ ,\\
(-\infty,0)\ra(0,1)&:\ \ \l\ra-1+\frac{1}{\l},\ 1-\l\ra\frac{1}{\l},\ d\l\ra\frac{d\l}{\l^2}\ ,
\end{split}
\end{gather}
and a further redefinition of $k_{\,i}'$, one arrives at the final result described by the six amplitudes
\begin{small}
\begin{gather}
\begin{split}
Z&^{(1)}\,=\,i\,\frac{g_o}{\a'}\,(2\pi)^{d}\,\delta\left(\sum_i k_{\,i}\right)\int_0^1 d\l\ (1-\l)^{-\a't-2}\l^{-\a'u-2}\\
\times&\exp\Big\{\sqrt{2\a'}\ \Big[(-k_{\,3}\l+k_{\,4}(1-\l))\cdot k_{\,1}'+(k_{\,3}(1-\l)-k_{\,4}\l)\cdot k_{\,2}'\\+&(k_{\,1}\l-k_{\,2}(1-\l))\cdot k_{\,3}'+(-k_{\,1}(1-\l)+k_{\,2}\l)\cdot k_{\,4}'\Big]\\
+&\Big[\l(1-\l)(k_{\,1}'\cdot k_{\,2}'+k_{\,3}'\cdot k_{\,4}')+\l(k_{\,1}'\cdot k_{\,3}'+k_{\,2}'\cdot k_{\,4}')+(1-\l)(k_{\,1}'\cdot k_{\,4}'+k_{\,2}'\cdot k_{\,3}')\Big]\Big\}\\\times&Tr[\L^{a_1}\L^{a_4}\L^{a_2}\L^{a_3}]\vphantom{\exp\Big\{\Big\}}\ ,
\end{split}
\end{gather}
\end{small}
\begin{small}
\begin{gather}
\begin{split}
Z&^{(2)}\,=\,i\,\frac{g_o}{\a'}\,(2\pi)^{d}\,\delta\left(\sum_i k_{\,i}\right)\int_0^1 d\l\ (1-\l)^{-\a's-2}\l^{-\a't-2}\\
\times&\exp\Big\{\sqrt{2\a'}\ \Big[-(k_{\,3}+k_{\,4}\l))\cdot k_{\,1}'+(k_{\,3}\l+k_{\,4})\cdot k_{\,2}'\\+&(k_{\,1}+k_{\,2}\l)\cdot k_{\,3}'-(k_{\,1}\l+k_{\,2})\cdot k_{\,4}'\Big]\\
+&\Big[\l(k_{\,1}'\cdot k_{\,2}'+k_{\,3}'\cdot k_{\,4}')+(1-\l)(k_{\,1}'\cdot k_{\,3}'+k_{\,2}'\cdot k_{\,4}')+\l(1-\l)(k_{\,1}'\cdot k_{\,4}'+k_{\,2}'\cdot k_{\,3}')\Big]\Big\}\\\times&Tr[\L^{a_1}\L^{a_2}\L^{a_4}\L^{a_3}]\vphantom{\exp\Big\{\Big\}}\ ,
\end{split}
\end{gather}
\end{small}
\begin{small}
\begin{gather}
\begin{split}
Z&^{(3)}\,=\,i\,\frac{g_o}{\a'}\,(2\pi)^{d}\,\delta\left(\sum_i k_{\,i}\right)\int_0^1 d\l\ (1-\l)^{-\a'u-2}\l^{-\a's-2}\\
\times&\exp\Big\{\sqrt{2\a'}\ \Big[-(k_{\,3}(1-\l)+k_{\,4})\cdot k_{\,1}'+(k_{\,3}+k_{\,4}(1-\l))\cdot k_{\,2}'\\-&(k_{\,1}(1-\l)-k_{\,2})\cdot k_{\,3}'+(k_{\,1}+k_{\,2}(1-\l))\cdot k_{\,4}'\Big]\\
+&\Big[(1-\l)(k_{\,1}'\cdot k_{\,2}'+k_{\,3}'\cdot k_{\,4}')+\l(1-\l)(k_{\,1}'\cdot k_{\,3}'+k_{\,2}'\cdot k_{\,4}')+\l(k_{\,1}'\cdot k_{\,4}'+k_{\,2}'\cdot k_{\,3}')\Big]\Big\}\\\times&Tr[\L^{a_4}\L^{a_1}\L^{a_2}\L^{a_3}]\vphantom{\exp\Big\{\Big\}}\ ,
\end{split}
\end{gather}
\end{small}
\begin{small}
\begin{gather}
\begin{split}
Z&^{(1)'}\,=\,i\,\frac{g_o}{\a'}\,(2\pi)^{d}\,\delta\left(\sum_i k_{\,i}\right)\int_0^1 d\l\ (1-\l)^{-\a't-2}\l^{-\a'u-2}\\
\times&\exp\Big\{-\sqrt{2\a'}\ \Big[(-k_{\,3}\l+k_{\,4}(1-\l))\cdot k_{\,1}'+(k_{\,3}(1-\l)-k_{\,4}\l)\cdot k_{\,2}'\\+&(k_{\,1}\l-k_{\,2}(1-\l))\cdot k_{\,3}'+(-k_{\,1}(1-\l)+k_{\,2}\l)\cdot k_{\,4}'\Big]\\
+&\Big[\l(1-\l)(k_{\,1}'\cdot k_{\,2}'+k_{\,3}'\cdot k_{\,4}')+\l(k_{\,1}'\cdot k_{\,3}'+k_{\,2}'\cdot k_{\,4}')+(1-\l)(k_{\,1}'\cdot k_{\,4}'+k_{\,2}'\cdot k_{\,3}')\Big]\Big\}\\\times&Tr[\L^{a_2}\L^{a_4}\L^{a_1}\L^{a_3}]\vphantom{\exp\Big\{\Big\}}\ ,
\end{split}
\end{gather}
\end{small}
\begin{small}
\begin{gather}
\begin{split}
Z&^{(2)'}\,=\,i\,\frac{g_o}{\a'}\,(2\pi)^{d}\,\delta\left(\sum_i k_{\,i}\right)\int_0^1 d\l\ (1-\l)^{-\a's-2}\l^{-\a'u-2}\\
\times&\exp\Big\{-\sqrt{2\a'}\ \Big[-(k_{\,3}(1-\l)+k_{\,4})\cdot k_{\,1}'+(k_{\,3}+k_{\,4}(1-\l))\cdot k_{\,2}'\\-&(k_{\,1}(1-\l)-k_{\,2})\cdot k_{\,3}'+(k_{\,1}+k_{\,2}(1-\l))\cdot k_{\,4}'\Big]\\
+&\Big[(1-\l)(k_{\,1}'\cdot k_{\,2}'+k_{\,3}'\cdot k_{\,4}')+\l(1-\l)(k_{\,1}'\cdot k_{\,3}'+k_{\,2}'\cdot k_{\,4}')+\l(k_{\,1}'\cdot k_{\,4}'+k_{\,2}'\cdot k_{\,3}')\Big]\Big\}\\\times&Tr[\L^{a_2}\L^{a_1}\L^{a_4}\L^{a_3}]\vphantom{\exp\Big\{\Big\}}\ ,
\end{split}
\end{gather}
\end{small}
\begin{small}
\begin{gather}
\begin{split}
Z&^{(3)'}\,=\,i\,\frac{g_o}{\a'}\,(2\pi)^{d}\,\delta\left(\sum_i k_{\,i}\right)\int_0^1 d\l\ (1-\l)^{-\a's-2}\l^{-\a't-2}\\
\times&\exp\Big\{-\sqrt{2\a'}\ \Big[-(k_{\,3}+k_{\,4}\l))\cdot k_{\,1}'+(k_{\,3}\l+k_{\,4})\cdot k_{\,2}'\\+&(k_{\,1}+k_{\,2}\l)\cdot k_{\,3}'-(k_{\,1}\l+k_{\,2})\cdot k_{\,4}'\Big]\\
+&\Big[\l(k_{\,1}'\cdot k_{\,2}'+k_{\,3}'\cdot k_{\,4}')+(1-\l)(k_{\,1}'\cdot k_{\,3}'+k_{\,2}'\cdot k_{\,4}')+\l(1-\l)(k_{\,1}'\cdot k_{\,4}'+k_{\,2}'\cdot k_{\,3}')\Big]\Big\}\\\times&Tr[\L^{a_4}\L^{a_2}\L^{a_1}\L^{a_3}]\vphantom{\exp\Big\{\Big\}}\ ,
\end{split}
\end{gather}
\end{small}In the last three expressions, we have also made the further change of variable $\l\ra(1-\l)$.

In all these case in order to construct the amplitudes it is necessary to apply four vertex operators to the six contributions to $Z$ and to add the results. In compact notation, we have thus
\begin{gather}
\cA^{(4)}\,=\,\hat{\cV}^{\,(1)}_m\hat{\cV}^{\,(2)}_n\hat{\cV}^{\,(3)}_p\hat{\cV}^{\,(4)}_q\ Z\ .
\end{gather}

\chapter{Higher-Spin String Amplitudes and Currents}
In this chapter we shall exploit generating function techniques to obtain in a compact form all tree-level three-point and four-point amplitudes for open bosonic-string external states of the first Regge trajectory. With the same techniques we shall also identify the currents that determine, to this order, the open string couplings.


\scs{Three-point scattering amplitudes}

In the previous chapter we showed that the generating function for three-point scattering amplitudes can be cast, in terms of the auxiliary coordinates $k_{\,i}'$, in the form
\begin{small}
\begin{gather*}
\begin{split}
Z&\,=\,i\,g_o\,\frac{(2\pi)^{d}}{\a'}\,\delta^{\,(d)}\left(\sum_i k_{\,i}\right)\Bigg[\\&\exp\left\{+\sqrt{\frac{\a'}{2}}\ \Big(k_{\,1}'\cdot k_{\,23}+k_{\,2}'\cdot k_{\,31}+k_{\,3}'\cdot k_{\,12}\Big)+(k_{\,1}'\cdot k_{\,2}'+k_{\,1}'\cdot k_{\,3}'+k_{\,2}'\cdot k_{\,3}')\right\}\\
\times\vphantom{\sqrt{\frac{\a'}{2}}}&Tr[\L^{a_1}\L^{a_2}\L^{a_3}]
\end{split}
\end{gather*}
\begin{gather}
\begin{split}
+&\exp\left\{-\sqrt{\frac{\a'}{2}}\ \Big(k_{\,1}'\cdot k_{\,23}+k_{\,2}'\cdot k_{\,31}+k_{\,3}'\cdot k_{\,12}\Big)+(k_{\,1}'\cdot k_{\,2}'+k_{\,1}'\cdot k_{\,3}'+k_{\,2}'\cdot k_{\,3}')\right\}\\\times&Tr[\L^{a_2}\L^{a_1}\L^{a_3}]\vphantom{\sqrt{\frac{\a'}{2}}}\Bigg]\ ,
\end{split}
\end{gather}
\end{small}
with the vertex operators
\begin{gather}
\cV_s\,=\,H_{\m_1\cdots \m_s}\frac{\partial}{\partial k_{\,\m_1}'}\cdots \frac{\partial}{\partial k_{\,\m_s}'}\ .
\end{gather}
The resulting machinery, however, is rather cumbersome, because of the increasing number of derivatives associated to higher-spin states. Still, it can be encoded in a nice way grouping all totally symmetric polarization tensors into a generating function.

In fact, generating functions give the possibility to deal with contractions among families of tensors in a nice way. Given a pair of generating functions
\begin{gather}
\begin{split}
A\,=\,\sum_{n\,=\,0}^\infty\frac{1}{n!}\ A_{\m_1\cdots\m_n}\,p^{\,\m_1}\cdots p^{\,\m_n}\ ,\\
B\,=\,\sum_{n\,=\,0}^\infty\frac{1}{n!}\ B_{\m_1\cdots\m_n}\,p^{\,\m_1}\cdots p^{\,\m_n}\ ,
\end{split}
\end{gather}
let us define the contraction as
\begin{gather}
A\cdot B\,=\,\sum_{n\,=\,0}\frac{1}{n!}\ A_{\m_1\cdots \m_n} B^{\m_1\cdots \m_n}\ .
\end{gather}
This expression can be turned into the convenient form
\begin{gather}
A\cdot B\,=\,\exp\Big(\frac{\partial}{\partial p}\cdot \frac{\partial}{\partial q}\Big) \,A(p) \, B(q)\Big|_{p\,=\,q\,=\,0}\ .\label{contra}
\end{gather}
One can then consider the Fourier Transform
\begin{gather}
\tilde{A}(p)\,=\,\int \frac{d^d k}{(2\pi)^{d/2}}\ e^{- i p\cdot k}A(k)\ ,\label{F}
\end{gather}
and its inverse
\begin{gather}
A(k)\,=\,\int \frac{d^dp}{(2\pi)^{d/2}}\ e^{ip\cdot k}\tilde{A}(p)\ ,\label{inversF}
\end{gather}
and substituting $(\ref{inversF})$ in $(\ref{contra})$ yields a useful integral form for the contraction
\begin{gather}
\begin{split}
A\cdot B&\,=\,\int \frac{d^dp}{(2\pi)^{d/2}}\ \exp\Big(\frac{\partial}{\partial k_{\,\m}}\frac{\partial}{\partial v^\m}\Big)\ e^{ik\cdot p}\,\tilde{A}(p)\,B(v)\Big|_{v\,=\,k\,=\,0}\\
&\,=\,\int \frac{d^dp}{(2\pi)^{d/2}}\ \exp\Big(ip\cdot \frac{\partial}{\partial v}\Big)\ e^{ik\cdot p}\,\tilde{A}(p)\,B(v)\Big|_{v\,=\,k\,=\,0}\\
&\,=\,\int \frac{d^dp}{(2\pi)^{d/2}}\ \tilde{A}(p)\,B(ip)\ .\label{ContrFormula}
\end{split}
\end{gather}
With this notation, defining the generating functions of polarization tensors with ingoing momenta $k_{\,i}$ as
\begin{gather}
\tilde{H}_i(k_{\,i},p_{\,i})\,=\,\sum_{n\,=\,0}^{\infty}\frac{1}{n!}\ \tilde{H}^{(i)}_{\m_1\cdots\m_n}(k_{\,1})\,p_{\,i}^{\,\m_1}\cdots p_{\,i}^{\,\m_n}\ ,
\end{gather}
the complete amplitude\footnote{This is the formal series of all amplitudes.} turns into the triple contraction
\begin{gather}
\cA^{(3)}\,=\,{\tilde{H}}^{(1)}_m {\tilde{H}}^{(2)}_n {\tilde{H}}^{(3)}_l Z\ ,
\end{gather}
that, making use of $(\ref{ContrFormula})$, takes the suggestive form
\begin{gather}
\begin{split}
\cA&\,=\,i\frac{g_o}{\a'}(2\pi)^{d}\delta^{(d)}(k_{\,1}+k_{\,2}+k_{\,3})\Bigg\{\\
&Tr[\L^{a_1}\L^{a_2}\L^{a_3}]\int \prod_{i\,=\,1}^3\frac{d^dp_{\,i}}{(2\pi)^{d/2}}\ \tilde{Z}_+(k_{\,1},k_{\,2},k_{\,3};p_{\,1},p_{\,2},p_{\,3})\\\times&\tilde{H}_1(k_{\,1},ip_{\,1})\,
\tilde{H}_2(k_{\,2},ip_{\,2})\,\tilde{H}_3(k_{\,3},ip_{\,3})\vphantom{\frac{d^dp_{\,i}}{(2\pi)^{d/2}}}\\
+&Tr[\L^{a_2}\L^{a_1}\L^{a_3}]\int \prod_{i\,=\,1}^3\frac{d^dp_{\,i}}{(2\pi)^{d/2}}\ \tilde{Z}_-(k_{\,1},k_{\,2},k_{\,3};p_{\,1},p_{\,2},p_{\,3})\\\times&\tilde{H}_1(k_{\,1},ip_{\,1})\,
\tilde{H}_2(k_{\,2},ip_{\,2})\,\tilde{H}_3(k_{\,3},ip_{\,3})\Bigg\}\\\,=\,
\vphantom{\frac{d^dp_{\,i}}{(2\pi)^{d/2}}}&i\frac{g_o}{\a'}(2\pi)^{d}
\delta^{(d)}(k_{\,1}+k_{\,2}+k_{\,3})\Big\{\cA_+(k_{\,1},k_{\,2},k_{\,3})Tr[\L^{a_1}\L^{a_2}\L^{a_3}]\\
\vphantom{\frac{d^dp_{\,i}}{(2\pi)^{d/2}}}+&\cA_-(k_{\,1},k_{\,2},k_{\,3})Tr[\L^{a_2}\L^{a_1}\L^{a_3}]\Big\}\ ,\label{ampl}
\end{split}
\end{gather}
that is somehow a triple trace in the auxiliary variables. Here, in order to simplify the discussion, we have expressed the amplitude in terms of the auxiliary amplitudes $\cA_{\pm}$, and we have let
\begin{multline}
Z_{\pm}\,=\,\exp\left\{\pm\sqrt{\frac{\a'}{2}}\,\Big(k_{\,1}'\cdot k_{\,23}+ k_{\,2}'\cdot k_{\,31}+k_{\,3}'\cdot k_{\,12}\Big)\right.\\\left.+(k_{\,1}'\cdot k_{\,2}'+k_{\,1}'\cdot k_{\,3}'+k_{\,2}'\cdot k_{\,3}')\vphantom{\sqrt{\frac{\a'}{2}}}\right\}\ .
\end{multline}
The crucial remark is that $Z_{\pm}$ has a simple exponential form, so that its Fourier transform can be easily computed as a distribution, with the end result
\begin{gather}
\begin{split}
\tilde{Z}_{\pm}&(p_{\,1},p_{\,2},p_{\,3})\,=\,\\\,=\,&\int \prod_{i\,=\,1}^3\frac{dk_{\,i}'}{(2\pi)^{d/2}}\ \exp\left\{\pm\sqrt{\frac{\a'}{2}}\ (k_{\,1}'\cdot k_{\,23}+ k_{\,2}'\cdot k_{\,31}+k_{\,3}'\cdot k_{\,12})\right.\\+&\left.(k_{\,1}'\cdot k_{\,2}'+k_{\,1}'\cdot k_{\,3}'+k_{\,2}'\cdot k_{\,3}')\vphantom{\sqrt{\frac{\a'}{2}}}\right\}\,e^{-ip_{\,1}\cdot k_{\,1}'}\,e^{-ip_{\,2}\cdot k_{\,2}'}\,e^{-ip_{\,3}\cdot k_{\,3}'}\\
\,=\,&\int \prod_{i\,=\,1}^3\frac{dk_{\,i}'}{(2\pi)^{d/2}}\ \exp\left\{\pm i\sqrt{\frac{\a'}{2}}\ (\partial_{p_{\,1}}\cdot k_{\,23}+ \partial_{p_{\,2}}\cdot k_{\,31}+\partial_{p_{\,3}}\cdot k_{\,12})\right.\\-&\left.(\partial_{p_{\,1}}\cdot \partial_{p_{\,2}}+\partial_{p_{\,1}}\cdot \partial_{p_{\,3}}+\partial_{p_{\,2}}\cdot \partial_{p_{\,3}})\sqrt{\frac{\a'}{2}}\right\}\,e^{-ip_{\,1}\cdot k_{\,1}'}\,e^{-ip_{\,2}\cdot k_{\,2}'}\,e^{-ip_{\,3}\cdot k_{\,3}'}\\
\,=\,&(2\pi)^{3d/2}\exp\left\{\pm i\sqrt{\frac{\a'}{2}}\ (\partial_{p_{\,1}}\cdot k_{\,23}+ \partial_{p_{\,2}}\cdot k_{\,31}+\partial_{p_{\,3}}\cdot k_{\,12})\right.\\-&\left.(\partial_{p_{\,1}}\cdot \partial_{p_{\,2}}+\partial_{p_{\,1}}\cdot \partial_{p_{\,3}}+\partial_{p_{\,2}}\cdot \partial_{p_{\,3}})\sqrt{\frac{\a'}{2}}\right\}
\,\delta^{\,(d)}(p_{\,1})\,\delta^{\,(d)}(p_{\,2})\,\delta^{\,(d)}(p_{\,3})\ .\label{Ztransform}
\end{split}
\end{gather}
This expression allows one to compute explicitly the amplitude $(\ref{ampl})$ recalling that the differential operator
\begin{gather}
\exp\left(a\cdot\frac{\partial}{\partial k}\right)\ ,
\end{gather}
acts as the translation operator $k\rightarrow k+a$ while
\begin{gather}
\exp\left(\frac{\partial}{\partial p}\cdot \frac{\partial}{\partial q}\right)\ ,
\end{gather}
acts as a contraction operator\footnote{This operator can be also regarded as the translation operator $q\ra q+\partial_p$ or $p\ra p+\partial_q$, whenever this interpretation is not ambiguous.}.
Keeping in mind these facts one can compute explicitly the auxiliary amplitudes $\cA_{\pm}$, obtaining
\begin{gather}
\begin{split}
\cA_{\pm}&\,=\,\int \prod_{i\,=\,1}^3\frac{d^dp_{\,i}}{(2\pi)^{d/2}}\ \tilde{Z}_+(k_{\,1},k_{\,2},k_{\,3};p_{\,1},p_{\,2},p_{\,3})\\\vphantom{\frac{d^dp_{\,i}}{(2\pi)^{d/2}}}\times
&\tilde{H}_1(k_{\,1},ip_{\,1})\,\tilde{H}_2(k_{\,2},ip_{\,2})\,\tilde{H}_3(k_{\,3},ip_{\,3})\\
\,=\,&\vphantom{\tilde{H}_3\left(k_{\,3},ip_{\,3}\pm\sqrt{\frac{\a'}{2}}\ k_{\,12}\right)}\exp\left\{-(\partial_{p_{\,1}}\cdot \partial_{p_{\,2}}+\partial_{p_{\,1}}\cdot \partial_{p_{\,3}}+\partial_{p_{\,2}}\cdot \partial_{p_{\,3}})\right\}\,\tilde{H}_1\left(k_{\,1},ip_{\,1}\pm\sqrt{\frac{\a'}{2}}\ k_{\,23}\right)\\
\times&\,\tilde{H}_2\left(k_{\,2},ip_{\,2}\pm\sqrt{\frac{\a'}{2}}\ k_{\,31}\right)\,
\tilde{H}_3\left(k_{\,3},ip_{\,3}\pm\sqrt{\frac{\a'}{2}}\ k_{\,12}\right)\Bigg|_{p_{\,i}\,=\,0}\\
\,=\,&\vphantom{\tilde{H}_3\left(k_{\,3},ip_{\,3}\pm\sqrt{\frac{\a'}{2}}\ k_{\,12}\right)}\exp\left\{(\partial_{p_{\,1}}\cdot \partial_{p_{\,2}}+\partial_{p_{\,1}}\cdot \partial_{p_{\,3}}+\partial_{p_{\,2}}\cdot \partial_{p_{\,3}})\right\}\\\times&\tilde{H}_1\left(k_{\,1},p_{\,1}\pm\sqrt{\frac{\a'}{2}}\ k_{\,23}\right)\,
\tilde{H}_2\left(k_{\,2},p_{\,2}\pm\sqrt{\frac{\a'}{2}}\ k_{\,31}\right)\\
\times&\,\tilde{H}_3\left(k_{\,3},p_{\,3}\pm\sqrt{\frac{\a'}{2}}\ k_{\,12}\right)\Bigg|_{p_{\,i}\,=\,0}\ .\label{A1}
\end{split}
\end{gather}
Moreover, regarding the contraction operators as particular translation operators, one can rewrite $(\ref{A1})$ without any ambiguity in the form
\begin{gather*}
\begin{split}
\cA_{\pm}&\,=\,\exp\left(\frac{\partial}{\partial p_{\,1}}\cdot \frac{\partial}{\partial p_{\,2}}+\frac{\partial}{\partial p_{\,1}}\cdot \frac{\partial}{\partial p_{\,3}}+\frac{\partial}{\partial p_{\,2}}\cdot \frac{\partial}{\partial p_{\,3}}\right)\\
\times&\tilde{H}_1\left(k_{\,1},p_{\,1}\pm\sqrt{\frac{\a'}{2}}\ k_{\,23}\right)\,
\tilde{H}_2\left(k_{\,2},p_{\,2}\pm\sqrt{\frac{\a'}{2}}\ k_{\,31}\right)\\
\times&\,\tilde{H}_3\left(k_{\,3},p_{\,3}\pm\sqrt{\frac{\a'}{2}}\ k_{\,12}\right)\Bigg|_{p_{\,i}\,=\,0}
\end{split}
\end{gather*}
\begin{gather}
\begin{split}
\,=\,&\tilde{H}_1\left(k_{\,1},p_{\,1}+\frac{\partial}{\partial p_{\,2}}+\frac{\partial}{\partial p_{\,3}}\pm\sqrt{\frac{\a'}{2}}\ k_{\,23}\right)\,\tilde{H}_2\left(k_{\,2},p_{\,2}+\frac{\partial}{\partial p_{\,3}}\pm\sqrt{\frac{\a'}{2}}\ k_{\,31}\right)\\
\times&\,\tilde{H}_3\left(k_{\,3},p_{\,3}\pm\sqrt{\frac{\a'}{2}}\ k_{\,12}\right)\Bigg|_{p_{\,i}\,=\,0}\ ,
\end{split}
\end{gather}
where derivatives always act to the right. One can now set $p_{\,1}\,=\,0$, since there is no derivative respect to $p_{\,1}$, while $p_{\,2}$ can be interchanged with $\frac{\partial}{\partial p_{\,2}}$, interchanging at the same time $H_1$ and $H_2$, since they produce Kronecker $\d$'s that, as such, are symmetric. One can thus arrive at another form for $\cA_\pm$:
\begin{gather}
\begin{split}
\cA_{\pm}\,=\,&\tilde{H}_2\left(k_{\,2},\frac{\partial}{\partial p_{\,2}}+\frac{\partial}{\partial p_{\,3}}\pm\sqrt{\frac{\a'}{2}}\ k_{\,31}\right)\,\tilde{H}_1\left(k_{\,1},p_{\,2}+\frac{\partial}{\partial p_{\,3}}\pm\sqrt{\frac{\a'}{2}}\ k_{\,23}\right)\\
\times&\,\tilde{H}_3\left(k_{\,3},p_{\,3}\pm\sqrt{\frac{\a'}{2}}\ k_{\,12}\right)\Bigg|_{p_{\,2}\,=\,p_{\,3}\,=\,0}\ .
\end{split}
\end{gather}
Moreover, the two derivatives with respect to $p_{\,2}$ and $p_{\,3}$ act on two different arguments and, on account of Leibniz's rule, can both be replaced with $\frac{\partial}{\partial p}$ if one lets $p_{\,2}\,=\,p_{\,3}\,=\,p$. This yields an equivalent but more compact expression for $\cA_{\pm}$:
\begin{multline}
\cA_{\pm}\,=\,\tilde{H}_1\left(k_{\,1},\frac{\partial}{\partial p}\pm\sqrt{\frac{\a'}{2}}\ k_{\,23}\right)\\\times\tilde{H}_2\left(k_{\,2},p+\frac{\partial}{\partial p}\pm\sqrt{\frac{\a'}{2}}\ k_{\,31}\right)\,\tilde{H}_3\left(k_{\,3},p\pm\sqrt{\frac{\a'}{2}}\ k_{\,12}\right)\Bigg|_{p\,=\,0}\ .\label{A2}
\end{multline}


\scs{Currents}

With the same technique used in the previous paragraph it is also possible to turn the amplitude $(\ref{ampl})$ into a current generating function, whose Fourier transform gives the open string couplings in the form $H\cdot J$. To this end, let us define the current generating function in momentum space by the double contraction
\begin{gather}
\begin{split}
&\tilde{\cJ}\,=\,i\,\frac{g_o}{\a'}\,(2\pi)^{d}\,\delta^{\,(d)}(k_{\,1}+k_{\,2}+k_{\,3})\Bigg\{\\
&Tr[\ \cdot\ \L^{a_2}\L^{a_3}]\int \frac{d^dp_{\,2}}{(2\pi)^{d/2}}\frac{d^dp_{\,3}}{(2\pi)^{d/2}}\ \tilde{Z}_+(k_{\,1},k_{\,2},k_{\,3};k_{\,1}',p_{\,2},p_{\,3})\\
\times&\tilde{H}_2(k_{\,2},ip_{\,2})\tilde{H}_3(k_{\,3},ip_{\,3})\vphantom{\frac{d^dp_{\,2}}{(2\pi)^{d/2}}}\\
+&Tr[\ \cdot\ \L^{a_3}\L^{a_2}]\int \frac{d^dp_{\,2}}{(2\pi)^{d/2}}\frac{d^dp_{\,3}}{(2\pi)^{d/2}}\ \tilde{Z}_-(k_{\,1},k_{\,2},k_{\,3};k_{\,1}',p_{\,2},p_{\,3})\\
\times&\tilde{H}_2(k_{\,2},ip_{\,2})\tilde{H}_3(k_{\,3},ip_{\,3})\Bigg\}\\
\,=\,&i\frac{g_o}{\a'}(2\pi)^{d}\delta^{(d)}(k_{\,1}+k_{\,2}+k_{\,3})\Big\{Tr[\ \cdot\ \L^{a_2}\L^{a_3}]\tilde{J}_+(k_{\,1}')+Tr[\ \cdot\ \L^{a_3}\L^{a_2}]\tilde{J}_-(k_{\,1}')\Big\}\ ,\label{Curr}
\end{split}
\end{gather}
where we have expressed the result in terms of auxiliary currents $\tilde{J}_{\pm}$ in order to deal with more compact expressions.
The Fourier transform of $Z_{\pm}$ with respect to $k_{\,2}'$ and $k_{\,3}'$ but not with respect to $k_{\,1}'$ is given by
\begin{gather}
\begin{split}
&\tilde{Z}_{\pm}(k_{\,1}',p_{\,2},p_{\,3})\,=\,\int \prod_{i\,=\,2}^3\frac{dk_{\,i}'}{(2\pi)^{d/2}}\ \exp\left\{\pm\sqrt{\frac{\a'}{2}}(k_{\,1}'\cdot k_{\,23}+ k_{\,2}'\cdot k_{\,31}+k_{\,3}'\cdot k_{\,12})\right.\\+&\left.\ \ (k_{\,1}'\cdot k_{\,2}'+k_{\,1}'\cdot k_{\,3}'+k_{\,2}'\cdot k_{\,3}')\vphantom{\sqrt{\frac{\a'}{2}}}\right\}e^{-ip_{\,2}\cdot k_{\,2}'}e^{-ip_{\,3}\cdot k_{\,3}'}\vphantom{\sqrt{\frac{\a'}{2}}}\\
\vphantom{\sqrt{\frac{\a'}{2}}}\,=\,&\int \prod_{i\,=\,2}^3\frac{dk_{\,i}'}{(2\pi)^{d/2}}\ \exp\left\{\pm i\sqrt{\frac{\a'}{2}}(-ik_{\,1}'\cdot k_{\,23}+ \partial_{p_{\,2}}\cdot k_{\,31}+\partial_{p_{\,3}}\cdot k_{\,12})\right.\\\vphantom{\sqrt{\frac{\a'}{2}}}-&\left.(-ik_{\,1}'\cdot \partial_{p_{\,2}}-ik_{\,1}'\cdot \partial_{p_{\,3}}+\partial_{p_{\,2}}\cdot \partial_{p_{\,3}})\vphantom{\sqrt{\frac{\a'}{2}}}\right\}e^{-ip_{\,2}\cdot k_{\,2}'}e^{-ip_{\,3}\cdot k_{\,3}'}\\
\vphantom{\sqrt{\frac{\a'}{2}}}\,=\,&(2\pi)^d\exp\left\{\pm i\sqrt{\frac{\a'}{2}}(-ik_{\,1}'\cdot k_{\,23}+ \partial_{p_{\,2}}\cdot k_{\,31}+\partial_{p_{\,3}}\cdot k_{\,12})\right.\\-&\left.(-ik_{\,1}'\cdot \partial_{p_{\,2}}-ik_{\,1}'\cdot \partial_{p_{\,3}}+\partial_{p_{\,2}}\cdot \partial_{p_{\,3}})\vphantom{\sqrt{\frac{\a'}{2}}}\right\}\delta^{(d)}(p_{\,2})\delta^{(d)}(p_{\,3})\ ,
\end{split}
\end{gather}
and the $dp_{\,2}$ and $dp_{\,3}$ integrals in $(\ref{Curr})$ yield
\begin{multline}
\tilde{J}_\pm\,=\,\exp\left\{\pm \sqrt{\frac{\a'}{2}}\ k_{\,1}'\cdot k_{\,23}\right\}\exp\left(\partial_{p_{\,2}}\cdot \partial_{p_{\,3}}\right)\\\times\tilde{H}_2\left(k_{\,2},p_{\,2}+k_{\,1}'\pm\sqrt{\frac{\a'}{2}}\ k_{\,31}\right)\,
\tilde{H}_3\left(k_{\,3},p_{\,3}+k_{\,1}'\pm\sqrt{\frac{\a'}{2}}\ k_{\,12}\right)\Bigg|_{p_{\,2}\,=\,p_{\,3}\,=\,0}\ ,
\end{multline}
where we have performed the further rescaling
\begin{gather}
p_{\,1}\ra -ip_{\,1}\ ,\\
p_{\,2}\ra -ip_{\,2}\ .
\end{gather}
Here, interpreting the contraction operators as a translation operators, the generating function $\tilde{J}_\pm$ can be turned into the equivalent expression
\begin{multline}
\tilde{J}_\pm\,=\,\exp\left\{\pm \sqrt{\frac{\a'}{2}}\ k_{\,1}'\cdot k_{\,23}\right\}\\\times\tilde{H}_2\left(k_{\,2},\partial_p+k_{\,1}'\pm\sqrt{\frac{\a'}{2}}\ k_{\,31}\right)
\tilde{H}_3\left(k_{\,3},p+k_{\,1}'\pm\sqrt{\frac{\a'}{2}}\ k_{\,12}\right)\Bigg|_{p\,=\,0}\ .
\end{multline}

In coordinate space the current is given by the Fourier transform
\begin{gather}
\cJ(x,k_{\,1}')\,=\,i\,\frac{g_o}{\a'}\int \frac{dk_{\,2}}{(2\pi)^{d}} \frac{dk_{\,3}}{(2\pi)^d}\ e^{ix\cdot (k_{\,2}+k_{\,3})}\tilde{\cJ}(k_{\,1},k_{\,2},k_{\,3};k_{\,1}')\ ,
\end{gather}
where the integrations in $dk_{\,2}$ and $dk_{\,3}$ give
\begin{multline}
J_{\pm}(x,k_{\,1}')\,=\,\int \frac{d^dk_{\,2}}{(2\pi)^{d}} \frac{d^dk_{\,3}}{(2\pi)^d}\ e^{ix\cdot (k_{\,2}+k_{\,3})}\tilde{J}_{\pm}(k_{\,1},k_{\,2},k_{\,3};k_{\,1}')\\\,=\,\int \frac{d^d k_{\,2}}{(2\pi)^{d}} \frac{d^d k_{\,3}}{(2\pi)^d}\ \exp\left[i\left(x\ \mp\ i\sqrt{\frac{\a'}{2}}\ k_{\,1}'\right)\cdot k_{\,2}\right]\exp\left[i\left(x\ \pm\ i\sqrt{\frac{\a'}{2}}k_{\,1}'\right)\cdot k_{\,3}\right]\\\times\tilde{H}_2\left(k_{\,2},\frac{\partial}{\partial p}+k_{\,1}'\pm\sqrt{2\a'}\ k_{\,3}\right)\,\tilde{H}_3\left(k_{\,3},p+k_{\,1}'\mp\sqrt{2\a'}\ k_{\,2}\vphantom{\frac{\partial}{\partial p}}\right)\Bigg|_{p\,=\,0}\ .
\end{multline}
Finally, using the identities
\begin{gather}
k_{\,2}\,=\, -i\,\partial\, e^{ix\cdot k_{\,2}}\ ,\\
k_{\,3}\,=\, -i\,\partial\, e^{ix\cdot k_{\,3}}\ ,
\end{gather}
inside the second argument of the generating function and integrating, one obtains\footnote{The subscripts on the generating function and on the derivatives with respect to $x$ indicate on which generating function the derivative acts.}
\begin{multline}
J_{\pm}(x,k_{\,1}')\,=\,{H}_2\left(x\ \mp\ i\sqrt{\frac{\a'}{2}}\ k_{\,1}',\ \frac{\partial}{\partial p}+k_{\,1}'\mp i\sqrt{2\a'}\ \partial_{\,3}\right)\\\times{H}_3\left(x\ \pm\ i\sqrt{\frac{\a'}{2}}\ k_{\,1}',\ p+k_{\,1}'\pm i\sqrt{2\a'}\ \partial_{\,2}\right)\Bigg|_{p\,=\,0}\ ,\label{offshell}
\end{multline}
where $H_i$ denotes the inverse Fourier transform of the generating function of polarizations, namely the generating function of the fields. Notice that we have not considered the integration in $d^dk_{\,1}$. In this way the generating function that we just obtained can be contracted with a generating function of fields, rather than of polarizations, to build the coupling.

The complete result for the current is a combination of $J_+$ and $J_-$,
\begin{gather}
\cJ(x,k_{\,1}')\,=\,i\,\frac{g_o}{\a'}\Big\{J_+(x,k_{\,1}')\,Tr[\ \cdot\ \L^{a_2}\L^{a_3}]\,+\,J_-(x,k_{\,1}')\,Tr[\ \cdot\ \L^{a_3}\L^{a_2}]\Big\}\ ,
\end{gather}
where Chan-Paton factors are also included.

This current codifies string couplings and it is generically not conserved.
In order to study current conservation for the generating function, one should verify that
\begin{gather}
\left(\eta^{\m\n}\frac{\partial}{\partial x^\m}\frac{\partial}{\partial k_{\,1}^{'\n}}\right)\,\cJ(x,k_{\,1}')\simeq0\ ,\label{Cons}
\end{gather}
where the symbol $\simeq$ means ``on-shell''.
In order to perform this computation in a simple way, let us define $a_{\pm}$, $b_{\pm}$ and $c_{\pm}$ in such a way that
\begin{gather}
J_{\pm}(x,k_{\,1}')\,=\,{H}_2(a_{\mp},b_\mp){H}_3(a_{\pm},c_{\pm})\ .
\end{gather}
As a result
\begin{gather}
\begin{split}
\partial_x J_{\pm}(x,k_{\,1}')\,=\,\partial_{a_\mp}{H}_2(a_{\mp},b_\mp){H}_3(a_{\pm},c_{\pm})+
{H}_2(a_{\mp},b_\mp)\partial_{a_{\pm}}{H}_3(a_{\pm},c_{\pm})\ ,
\end{split}
\end{gather}
and
\begin{small}
\begin{gather}
\begin{split}
&\partial_{k_{\,1}'}\cdot\partial_x\ J_{\pm}(x,k_{\,1}')\,=\,\\&\mp i\sqrt{\frac{\a'}{2}}\Big[\partial_{a_\mp}^2H_2(a_{\mp},b_\mp)\Big]{H}_3(a_{\pm},c_{\pm})+
\Big[\partial_{a_\mp}\cdot \partial_{b_\mp}{H}_2(a_{\mp},b_\mp)\Big]{H}_3(a_{\pm},c_{\pm})\\
&\pm i\sqrt{\frac{\a'}{2}}\Big[\partial_{a_\mp}H_2(a_{\mp},b_\mp)\Big]
\Big[\partial_{a_\pm}{H}_3(a_{\pm},c_{\pm})\Big]+
\Big[\partial_{a_\mp}H_2(a_{\mp},b_\mp)\Big]\Big[\partial_{c_\pm}{H}_3(a_{\pm},c_{\pm})\Big]\\
&\mp i\sqrt{\frac{\a'}{2}}\Big[\partial_{a_\mp}H_2(a_{\mp},b_\mp)\Big]
\Big[\partial_{a_\pm}{H}_3(a_{\pm},c_{\pm})\Big]+
\Big[\partial_{b_\mp}{H}_2(a_{\mp},b_\mp)\Big]\Big[\partial_{a_\pm}{H}_3(a_{\pm},c_{\pm})\Big]\\
&\pm i\sqrt{\frac{\a'}{2}}H_2(a_\mp,b_\mp)\Big[\partial_{a_\pm}^2H_3(a_\pm,c_\pm)\Big]+
H_2(a_\mp,b_\mp)\Big[\partial_{a_\pm}\cdot\partial_{c_\pm}H_3(a_\pm,c_\pm)\Big]\ ,\label{var}
\end{split}
\end{gather}
\end{small}where the derivatives are always contracted together, even when this is not explicitly indicated.
On shell, at least for massive fields, one can use
\begin{gather}
\begin{split}
(\partial_a^2+m^2)H(a,b)&\simeq 0\ ,\\
\partial_a\cdot \partial_b H(a,b)&\simeq 0\ ,
\end{split}
\end{gather}
that are the transcriptions of the Klein-Gordon equation and of the transversality condition in this language. In this way $(\ref{var})$ finally turns into
\begin{multline}
\partial_{k_{\,1}'}\cdot\partial_x\ J_{\pm}(x,k_{\,1}')\simeq\\\simeq\Big[\partial_{a_\mp}H_2(a_{\mp},b_\mp)\Big]
\Big[\partial_{c_\pm}{H}_3(a_{\pm},c_{\pm})\Big]+\Big[\partial_{b_\mp}{H}_2(a_{\mp},b_\mp)\Big]
\Big[\partial_{a_\pm}{H}_3(a_{\pm},c_{\pm})\Big]\ .
\end{multline}
This equation implies that, in general, there are some non conserved portions in the currents that we have identified. In the next chapter we shall see how, in some particular cases, terms of this type are generated by the massive Klein-Gordon equation for the massive external states. We expect that this will be the general structure of all string couplings.

This on-shell result implies that the truly conserved part of the current is simply
\begin{multline}
J_{\pm}(x,k_{\,1}')\,=\,{H}_2\left(x\ \mp\ i\sqrt{\frac{\a'}{2}}\ k_{\,1}',\ \frac{\partial}{\partial p}\mp i\sqrt{2\a'}\ \partial_{\,3}\right)\\\times{H}_3\left(x\ \pm\ i\sqrt{\frac{\a'}{2}}\ k_{\,1}',\ p\pm i\sqrt{2\a'}\ \partial_{\,2}\right)\Bigg|_{p\,=\,0}\ ,\label{off-shell}
\end{multline}
while non conserved terms can arise every time $k_{\,1}'$ appears in the second argument of $H_i$.
Moreover, in the high energy limit $\a'\ra\infty$, one is led to perform the redefinition $k_{\,i}'\rightarrow \sqrt{\frac{1}{2\a'}}\ k_{\,i}'$, turning $(\ref{off-shell})$ into
\begin{multline}
J_{\pm}(x,k_{\,1}')\,=\,{H}_2\left(x\ \mp\ i\frac{k_{\,1}'}{2},\ \frac{\partial}{\partial p}+\frac{k_{\,1}'}{\sqrt{2\a'}}\mp i\sqrt{2\a'}\ \partial_{3}\right)\\\times{H}_3\left(x\ \pm\ i\frac{k_{\,1}'}{2},\ p+\frac{k_{\,1}'}{\sqrt{2\a'}}\pm i\sqrt{2\a'}\ \partial_{2}\right)\Bigg|_{p\,=\,0}\ ,
\end{multline}
so that the relative weight of the various contributions becomes manifest.


\scs{Four-point scattering amplitudes}
The same machinery used in the previous paragraph can be extended in a straightforward way to amplitudes with four external particles. The starting point is the generating function of four-particle scattering amplitudes computed in the previous chapter and resulting from the six contributions
\begin{small}
\begin{gather}
\begin{split}
Z&^{(1)}\,=\,i\,\frac{g_o}{\a'}\,(2\pi)^{d}\,\delta\left(\sum_i k_{\,i}\right)\int_0^1 d\l\ (1-\l)^{-\a't-2}\l^{-\a'u-2}\\
\times&\vphantom{\int_0^1}\exp\Big\{\sqrt{2\a'}\Big[(-k_{\,3}\l+k_{\,4}(1-\l))\cdot k_{\,1}'+(k_{\,3}(1-\l)-k_{\,4}\l)\cdot k_{\,2}'\\+\vphantom{\int_0^1}&(k_{\,1}\l-k_{\,2}(1-\l))\cdot k_{\,3}'+(-k_{\,1}(1-\l)+k_{\,2}\l)\cdot k_{\,4}'\Big]\\
+&\vphantom{\int_0^1}\Big[\l(1-\l)(k_{\,1}'\cdot k_{\,2}'+k_{\,3}'\cdot k_{\,4}')+\l(k_{\,1}'\cdot k_{\,3}'+k_{\,2}'\cdot k_{\,4}')+(1-\l)(k_{\,1}'\cdot k_{\,4}'+k_{\,2}'\cdot k_{\,3}')\Big]\Big\}\\\vphantom{\int_0^1}\times&Tr[\L^{a_1}\L^{a_4}\L^{a_2}\L^{a_3}]\ ,
\end{split}
\end{gather}
\end{small}
\begin{small}
\begin{gather}
\begin{split}
Z&^{(2)}\,=\,i\,\frac{g_o}{\a'}\,(2\pi)^{d}\,\delta\left(\sum_i k_{\,i}\right)\int_0^1 d\l\ (1-\l)^{-\a's-2}\l^{-\a't-2}\\
\vphantom{\int_0^1}\times&\exp\Big\{\sqrt{2\a'}\Big[-(k_{\,3}+k_{\,4}\l))\cdot k_{\,1}'+(k_{\,3}\l+k_{\,4})\cdot k_{\,2}'\\\vphantom{\int_0^1}+&(k_{\,1}+k_{\,2}\l)\cdot k_{\,3}'-(k_{\,1}\l+k_{\,2})\cdot k_{\,4}'\Big]\\
\vphantom{\int_0^1}+&\Big[\l(k_{\,1}'\cdot k_{\,2}'+k_{\,3}'\cdot k_{\,4}')+(1-\l)(k_{\,1}'\cdot k_{\,3}'+k_{\,2}'\cdot k_{\,4}')+\l(1-\l)(k_{\,1}'\cdot k_{\,4}'+k_{\,2}'\cdot k_{\,3}')\Big]\Big\}\\\vphantom{\int_0^1}\times&Tr[\L^{a_1}\L^{a_2}\L^{a_4}\L^{a_3}]\ ,
\end{split}
\end{gather}
\end{small}
\begin{small}
\begin{gather}
\begin{split}
Z&^{(3)}\,=\,i\,\frac{g_o}{\a'}\,(2\pi)^{d}\,\delta\left(\sum_i k_{\,i}\right)\int_0^1 d\l\ (1-\l)^{-\a'u-2}\l^{-\a's-2}\\
\vphantom{\int_0^1}\times&\exp\Big\{\sqrt{2\a'}\Big[-(k_{\,3}(1-\l)+k_{\,4})\cdot k_{\,1}'+(k_{\,3}+k_{\,4}(1-\l))\cdot k_{\,2}'\\\vphantom{\int_0^1}-&(k_{\,1}(1-\l)-k_{\,2})\cdot k_{\,3}'+(k_{\,1}+k_{\,2}(1-\l))\cdot k_{\,4}'\Big]\\
\vphantom{\int_0^1}+&\Big[(1-\l)(k_{\,1}'\cdot k_{\,2}'+k_{\,3}'\cdot k_{\,4}')+\l(1-\l)(k_{\,1}'\cdot k_{\,3}'+k_{\,2}'\cdot k_{\,4}')+\l(k_{\,1}'\cdot k_{\,4}'+k_{\,2}'\cdot k_{\,3}')\Big]\Big\}\\ \times&\vphantom{\int_0^1}Tr[\L^{a_4}\L^{a_1}\L^{a_2}\L^{a_3}]\ ,
\end{split}
\end{gather}
\end{small}
\begin{small}
\begin{gather}
\begin{split}
Z&^{(1)'}\,=\,i\,\frac{g_o}{\a'}\,(2\pi)^{d}\,\delta\left(\sum_i k_{\,i}\right)\int_0^1 d\l\ (1-\l)^{-\a't-2}\l^{-\a'u-2}\\
\vphantom{\int_0^1}\times&\exp\Big\{-\sqrt{2\a'}\Big[(-k_{\,3}\l+k_{\,4}(1-\l))\cdot k_{\,1}'+(k_{\,3}(1-\l)-k_{\,4}\l)\cdot k_{\,2}'\\\vphantom{\int_0^1}+&(k_{\,1}\l-k_{\,2}(1-\l))\cdot k_{\,3}'+(-k_{\,1}(1-\l)+k_{\,2}\l)\cdot k_{\,4}'\Big]\\
\vphantom{\int_0^1}+&\Big[\l(1-\l)(k_{\,1}'\cdot k_{\,2}'+k_{\,3}'\cdot k_{\,4}')+\l(k_{\,1}'\cdot k_{\,3}'+k_{\,2}'\cdot k_{\,4}')+(1-\l)(k_{\,1}'\cdot k_{\,4}'+k_{\,2}'\cdot k_{\,3}')\Big]\Big\}\\ \times&\vphantom{\int_0^1}Tr[\L^{a_2}\L^{a_4}\L^{a_1}\L^{a_3}]\ ,
\end{split}
\end{gather}
\end{small}
\begin{small}
\begin{gather}
\begin{split}
Z&^{(2)'}\,=\,i\,\frac{g_o}{\a'}\,(2\pi)^{d}\,\delta\left(\sum_i k_{\,i}\right)\int_0^1 d\l\ (1-\l)^{-\a's-2}\l^{-\a'u-2}\\
\vphantom{\int_0^1}\times&\exp\Big\{-\sqrt{2\a'}\Big[-(k_{\,3}(1-\l)+k_{\,4})\cdot k_{\,1}'+(k_{\,3}+k_{\,4}(1-\l))\cdot k_{\,2}'\\\vphantom{\int_0^1}-&(k_{\,1}(1-\l)-k_{\,2})\cdot k_{\,3}'+(k_{\,1}+k_{\,2}(1-\l))\cdot k_{\,4}'\Big]\\
\vphantom{\int_0^1}+&\Big[(1-\l)(k_{\,1}'\cdot k_{\,2}'+k_{\,3}'\cdot k_{\,4}')+\l(1-\l)(k_{\,1}'\cdot k_{\,3}'+k_{\,2}'\cdot k_{\,4}')+\l(k_{\,1}'\cdot k_{\,4}'+k_{\,2}'\cdot k_{\,3}')\Big]\Big\}\\ \times&\vphantom{\int_0^1}Tr[\L^{a_2}\L^{a_1}\L^{a_4}\L^{a_3}]\ ,
\end{split}
\end{gather}
\end{small}
\begin{small}
\begin{gather}
\begin{split}
Z&^{(3)'}\,=\,i\,\frac{g_o}{\a'}\,(2\pi)^{d}\,\delta\left(\sum_i k_{\,i}\right)\int_0^1 d\l\ (1-\l)^{-\a's-2}\l^{-\a't-2}\\
\vphantom{\int_0^1}\times&\exp\Big\{-\sqrt{2\a'}\Big[-(k_{\,3}+k_{\,4}\l))\cdot k_{\,1}'+(k_{\,3}\l+k_{\,4})\cdot k_{\,2}'\\\vphantom{\int_0^1}+&(k_{\,1}+k_{\,2}\l)\cdot k_{\,3}'-(k_{\,1}\l+k_{\,2})\cdot k_{\,4}'\Big]\\
\vphantom{\int_0^1}+&\Big[\l(k_{\,1}'\cdot k_{\,2}'+k_{\,3}'\cdot k_{\,4}')+(1-\l)(k_{\,1}'\cdot k_{\,3}'+k_{\,2}'\cdot k_{\,4}')+\l(1-\l)(k_{\,1}'\cdot k_{\,4}'+k_{\,2}'\cdot k_{\,3}')\Big]\Big\}\\\times&\vphantom{\int_0^1}Tr[\L^{a_4}\L^{a_2}\L^{a_1}\L^{a_3}]\ ,
\end{split}
\end{gather}
\end{small}or, in a different but equivalent form, by
\begin{small}
\begin{gather}
\begin{split}
Z&\,=\,\vphantom{\exp\Big\{\Big\}}i\,\frac{g_o^2}{\a'}\,(2\pi)^{d}\,\delta^{(d)}(k_{\,1}+k_{\,2}+k_{\,3}+k_{\,4})
\int_{-\infty}^{\infty} d\l\ (1-\l)^{-\a't-2}\l^{-\a'u-2}\\
\vphantom{\exp\Big\{\Big\}}\times&\exp\Big\{\sqrt{2\a'}\Big[\bra\l\ket(-k_{\,3}\l+k_{\,4}(1-\l))\cdot k_{\,1}'+\bra1-\l\ket(k_{\,3}(1-\l)-k_{\,4}\l)\cdot k_{\,2}'\\\vphantom{\exp\Big\{\Big\}}+&(k_{\,1}\l-k_{\,2}(1-\l))\cdot k_{\,3}'+\bra\l(1-\l)\ket(-k_{\,1}(1-\l)+k_{\,2}\l)\cdot k_{\,4}'\Big]\\
\vphantom{\exp\Big\{\Big\}}+&\Big[\l(1-\l)(k_{\,1}'\cdot k_{\,2}'+k_{\,3}'\cdot k_{\,4}')+\l(k_{\,1}'\cdot k_{\,3}'+k_{\,2}'\cdot k_{\,4}')+(1-\l)(k_{\,1}'\cdot k_{\,4}'+k_{\,2}'\cdot k_{\,3}')\Big]\Big\}\\\times &\vphantom{\exp\Big\{\Big\}}Tr[\L^{a_1}\L^{a_4}\L^{a_2}\L^{a_3}]+(1\lra 2)\ ,\label{Four}
\end{split}
\end{gather}
\end{small}where, by convention,
\begin{gather}
\bra a\ket\,=\,sign(a)\ .
\end{gather}
Grouping all polarization tensors into a generating function as before, the four-point amplitude turns into the quadruple contraction
\begin{gather}
\cA^{(4)}\,=\,{\tilde{H}}^{(1)}_m {\tilde{H}}^{(2)}_n{\tilde{H}}^{(3)}_p{\tilde{H}}^{(4)}_q\cdot Z\ .\label{four-point}
\end{gather}
Here, in order to do the computation in general for all the various contributions, one can choose as a starting point the parametric expression for $Z$
\begin{multline}
Z^{(i)}\,=\,\exp\Big\{\sqrt{2\a'}\Big[k_{\,34}^{(i)}(\l)\cdot k_{\,1}'+k_{\,43}^{(i)}(\l)\cdot k_{\,2}'+k_{\,12}^{(i)}(\l)\cdot k_{\,3}'+k_{\,21}^{(i)}(\l)\cdot k_{\,4}'\Big]\\+\Big[A_\l^{(i)}(k_{\,1}'\cdot k_{\,2}'+k_{\,3}'\cdot k_{\,4}')+B_\l^{(i)}(k_{\,1}'\cdot k_{\,3}'+k_{\,2}'\cdot k_{\,4}')+C_\l^{(i)}(k_{\,1}'\cdot k_{\,4}'+k_{\,2}'\cdot k_{\,3}')\Big]\Big\}\ ,
\end{multline}
where the parameters $k_{\,ij}(\l)$, $A_\l$, $B_\l$ and $C_\l$ are defined case by case by comparison with the previous formulas for $Z$.
Using $(\ref{ContrFormula})$ the contraction $(\ref{four-point})$ turns into
\begin{multline}
\cA^{(i)}\,=\,\int \prod_{i\,=\,1}^{4}\frac{d^d p_{\,i}}{(2\pi)^{d/2}}\ \tilde{Z}^{(i)}(k_{\,1},\cdots,k_{\,4};p_{\,1},\cdots,p_{\,4})\\
\times\tilde{H}_1(k_{\,1},ip_{\,1})\tilde{H}_2(k_{\,2},ip_{\,2})\tilde{H}_3(k_{\,3},ip_{\,3})\tilde{H}_4(k_{\,4},ip_{\,4})\ ,
\end{multline}
while the Fourier transform $\tilde{Z}^{(i)}$ is given by the distribution
\begin{gather}
\begin{split}
\tilde{Z}^{(i)}\,=\,&(2\pi)^{2d}\exp\Big\{i\sqrt{2\a'}\Big[k_{\,34}^{(i)}(\l)\cdot \partial_{p_{\,1}}+k_{\,43}^{(i)}(\l)\cdot \partial_{p_{\,2}}+k_{\,12}^{(i)}(\l)\cdot \partial_{p_{\,3}}\\&+k_{\,21}^{(i)}(\l)\cdot \partial_{p_{\,4}}\Big]\\
-&\Big[A_\l^{(i)}(\partial_{p_{\,1}}\cdot \partial_{p_{\,2}}+\partial_{p_{\,3}}\cdot \partial_{p_{\,4}})+B_\l^{(i)}(\partial_{p_{\,1}}\cdot \partial_{p_{\,3}}+\partial_{p_{\,2}}\cdot \partial_{p_{\,4}})\\+&C_\l^{(i)}(\partial_{p_{\,1}}\cdot \partial_{p_{\,4}}+\partial_{p_{\,2}}\cdot \partial_{p_{\,3}})\Big]\Big\}
\delta^{(d)}(p_{\,1})\delta^{(d)}(p_{\,2})\delta^{(d)}(p_{\,3})\delta^{(d)}(p_{\,4})\ .
\end{split}
\end{gather}
Finally, noticing that $\tilde{Z}^{(i)}$ can be interpreted as a product of contraction and translation operators, one is led to
\begin{gather}
\begin{split}
\cA&^{(i)}\,=\,\exp\Big\{\Big[A_\l^{(i)}(\partial_{p_{\,1}}\cdot \partial_{p_{\,2}}+\partial_{p_{\,3}}\cdot \partial_{p_{\,4}})+B_\l^{(i)}(\partial_{p_{\,1}}\cdot \partial_{p_{\,3}}+\partial_{p_{\,2}}\cdot \partial_{p_{\,4}})\\+&C_\l^{(i)}(\partial_{p_{\,1}}\cdot \partial_{p_{\,4}}+\partial_{p_{\,2}}\cdot \partial_{p_{\,3}})\Big]\Big\}\tilde{H}_1(k_{\,1},p_{\,1}+\sqrt{2\a'}k_{\,34}^{(i)}(\l))\\\times
&\tilde{H}_2(k_{\,2},p_{\,2}+\sqrt{2\a'}k_{\,43}^{(i)}(\l))\tilde{H}_3(k_{\,3},p_{\,3}+\sqrt{2\a'}k_{\,12}^{(i)}(\l))\\
\times&\tilde{H}_4(k_{\,4},p_{\,4}+\sqrt{2\a'}k_{\,21}^{(i)}(\l))\Big|_{p_{\,i}\,=\,0}\ ,\label{A}
\end{split}
\end{gather}
where we have redefined $ip_{\,i}\ra p_{\,i}$. 
The complete amplitude is given by the integral of $(\ref{A})$ with the corresponding measure.

For example, starting from $(\ref{Four})$, one obtains the integral
\begin{gather}
\begin{split}
\cA&\,=\,i\frac{g_o^2}{\a'}(2\pi)^{d}\delta^{(d)}(k_{\,1}+k_{\,2}+k_{\,3}+k_{\,4})\int_{-\infty}^{\infty} d\l\ |1-\l|^{-\a't-2}|\l|^{-\a'u-2}\\\times& \exp\Big\{\Big[A_{\l}^{(i)}(\partial_{p_{\,1}}\cdot \partial_{p_{\,2}}+\partial_{p_{\,3}}\cdot \partial_{p_{\,4}})+B_{\l}^{(i)}(\partial_{p_{\,1}}\cdot \partial_{p_{\,3}}+\partial_{p_{\,2}}\cdot \partial_{p_{\,4}})\\+&C_{\l}^{(i)}(\partial_{p_{\,1}}\cdot \partial_{p_{\,4}}+\partial_{p_{\,2}}\cdot \partial_{p_{\,3}})\Big]\Big\}\tilde{H}_1(k_{\,1},p_{\,1}+\sqrt{2\a'}k_{\,34}^{(i)}(\l))\\\times
&\tilde{H}_2(k_{\,2},p_{\,2}+\sqrt{2\a'}k_{\,43}^{(i)}(\l))\,\tilde{H}_3(k_{\,3},p_{\,3}+\sqrt{2\a'}
k_{\,12}^{(i)}(\l))\\\times&\tilde{H}_4(k_{\,4},p_{\,4}+\sqrt{2\a'}k_{\,21}^{(i)}(\l))\Big|_{p_{\,i}\,=\,0}\ .\label{HighEnergy}
\end{split}
\end{gather}


\scss{High energy limit of four-particle amplitudes}

The last expression for the amplitudes is very useful to study their high-energy behavior. A similar analysis was performed by Gross and Mende \cite{Gross:1987kza} for tachyons, and more recently by Moeller and West \cite{Moeller:2005ez}, who extended it to all string excitations. A similar analysis was also done by Amati, Ciafaloni and Veneziano, mostly in the Regge limit, in \cite{Amati:1987wq}.

In the limit $s\ra \infty$ at fixed angle all Mandelstam variables tend to $\pm\infty$, so that one has the approximate relation $s+t+u\sim 0$, while $t/s$ is fixed and the integral $(\ref{HighEnergy})$ can be approximated by a saddle point technique. The dominant term is given by
\begin{gather}
\int_{-\infty}^{\infty} d\l\ |1-\l|^{-\a't}|\l|^{-\a'u}\,=\,\int_{-\infty}^{\infty}d\l\ \exp\left[-\a's\left(\frac{t}{s}\ln|1-\l|+\frac{u}{s}\ln|\l|\right)\right]\ ,\label{exp}
\end{gather}
while the other $\l$-dependent terms can be computed at the saddle, since they are bounded when $s\ra\infty$.
Moreover, the contraction operator is manifestly sub-dominant with respect to the momenta, so that the dominant contribution will be given by the term containing the maximum number of momenta.
The saddle point can be computed extremizing the exponent in $(\ref{exp})$, with the result
\begin{gather}
\l_0\,=\,-\frac{u}{s}\ ,\\
1-\l_0\,=\,-\frac{t}{s}\ .
\end{gather}
The dominant behavior is given by
\begin{gather}
\begin{split}
\cA&\ \sim\ i\,\frac{g_o^2}{\a'}\,(2\pi)^{d}\,\delta^{\,(d)}(k_{\,1}+k_{\,2}+k_{\,3}+k_{\,4})e^{-\a's\ln \a's-\a't\ln\a't-\a'u\ln\a'u}\\\times& \exp\Big\{\Big[A_{\l_0}(\partial_{p_{\,1}}\cdot \partial_{p_{\,2}}+\partial_{p_{\,3}}\cdot \partial_{p_{\,4}})+B_{\l_0}(\partial_{p_{\,1}}\cdot \partial_{p_{\,3}}+\partial_{p_{\,2}}\cdot \partial_{p_{\,4}})\\+&C_{\l_0}(\partial_{p_{\,1}}\cdot \partial_{p_{\,4}}+\partial_{p_{\,2}}\cdot \partial_{p_{\,3}})\Big]\Big\}\tilde{H}_1(k_{\,1},p_{\,1}+\sqrt{2\a'}k_{\,34}(\l_0))\\\times&
\tilde{H}_2(k_{\,2},p_{\,2}+\sqrt{2\a'}k_{\,43}(\l_0))\,\tilde{H}_3(k_{\,3},p_{\,3}+\sqrt{2\a'}k_{\,12}(\l_0))\\
\times&\tilde{H}_4(k_{\,4},p_{\,4}+\sqrt{2\a'}k_{\,21}(\l_0))\Big|_{p_{\,i}\,=\,0}\ ,
\end{split}
\end{gather}
or, neglecting the contraction operators, by
\begin{gather}
\begin{split}
\cA\ \sim\ &i\,\frac{g_o^2}{\a'}\,(2\pi)^{d}\,\delta^{\,(d)}(k_{\,1}+k_{\,2}+k_{\,3}+k_{\,4})\ e^{-\a's\ln \a's-\a't\ln\a't-\a'u\ln\a'u}\\\times&\tilde{H}_1(k_{\,1},\sqrt{2\a'}k_{\,34}^{(i)}(\l_0))
\tilde{H}_2(k_{\,2},\sqrt{2\a'}k_{\,43}^{(i)}(\l_0))\\\times &\tilde{H}_3(k_{\,3},\sqrt{2\a'}k_{\,12}^{(i)}(\l_0))\tilde{H}_4(k_{\,4},\sqrt{2\a'}k_{\,21}^{(i)}(\l_0))\ .
\end{split}
\end{gather}
This behavior shows that the amplitude is exponentially depressed in the UV, which can be regarded as evidence for the UV finiteness of String Theory. In fact, the complete result for the sum of all higher-spin contributions is better behaved then any single contribution coming from the exchange of a single spin-$s$ field. This may be regarded as a convincing argument in favor of Higher Spin Field Theory as a proper framework to better understand these properties of String Theory.

\chapter{Applications}
In this chapter we shall analyze in more detail the string couplings and the corresponding currents involving higher-spin states that we have just identified. In particular, we shall compute on the Quantum Field Theory side higher-spin current exchanges and tree-level four-point scattering amplitudes in which infinitely many higher-spin particles are interchanged, with special emphasis on their behavior and on the possibility to construct consistent theories deforming the abelian gauge symmetry. In deriving these results, we shall also clarify the origin of the current exchange amplitudes of \cite{Francia:2007qt} and their extension to the case of mixed symmetry fields. Some explicit cubic couplings will be also analyzed in order to compare the results obtained in the massless limit with the coupling described in \cite{Boulanger:2008tg} and the conserved currents of Berends, Burgers and van Dam \cite{Berends:1985xx}.


\scs{Higher-Spin Current Exchanges}
For later convenience we would like to begin by deriving an explicit compact form for the generating function of the current exchange amplitudes for totally symmetric higher-spin bosonic fields.

Moreover, we shall present, for the first time, all mixed-symmetry propagators and the extension of the result to the case of external currents that are not conserved, that as we have seen can be of interest for massive fields.

By definition, the spin-$m$ propagator takes the general form
\begin{gather}
\cP^{(m)}_{\m_1\cdots\m_m;\n_1\cdots\n_m}\,=\,-\,\frac{1}{k^{\,2}+M^{\,2}}\,P_{\m_1\cdots\m_m;\n_1\cdots\n_m}\ ,
\end{gather}
where we are using the mostly-plus convention for the space-time signature and we have factored out the pole part while also highlighting its tensorial structure with the two groups of indices associated, respectively, to incoming and outgoing currents.

The propagator for a spin-$m$ particle can be uniquely determined imposing that it entails the correct current exchange, namely requiring that its tensorial structure, encoded in $P_{\m_1\cdots\m_m;\n_1\cdots\n_m}$, be symmetric under the interchange of the two groups of indices and that it project an arbitrary incoming or outgoing rank-$m$ current onto the spin-$m$ unitary irreducible representation of the Poincar\'e group.

For massive particles the irreducible representations are associated to particular transverse and traceless Young projections. On the other hand, for massless particles the gauge symmetry naturally imposes transversality constraint on each current so that, in a covariant gauge, one needs only to account for the Young projection and for tracelessness on the $(d-2)$-dimensional transverse space.


For simplicity let us begin by considering the simpler case of transverse totally symmetric currents associated with propagators for totally symmetric spin-$m$ fields introducing auxiliary variables $p_\m$ associated to incoming indices and $q_\m$ associated to outgoing ones. Remarkably, a simple modification of the totally symmetric spin-$m$ propagator will then suffice to recover a generating function for all mixed-symmetry ones. As we shall see, it is also possible to address the general case in which the currents are not conserved, both in the totally-symmetric and in the mixed-symmetry cases.

In the totally symmetric case, when sandwiched between transverse currents, the propagator polynomial is very simple, because it can be built solely out of $p^{\,2}q^{\,2}$ and $p\cdot q$. Moreover, one has only to account for the tracelessness in the transverse subspace, since the symmetrization is automatically realized thanks to the commuting nature of the auxiliary variables $p_\m$ and $q_\m$. The trace operator in this formalism coincides with the Laplace operator $\partial_p\cdot\partial_p$, so that tracelessness in the transverse subspace translates into the differential equation
\begin{gather}
(\partial_p\cdot\partial_p)\,\cP^{(m)}(p,q)\,=\,0\ ,
\end{gather}
where we recall that $\cP^{(m)}$ is a homogeneous polynomial of degree $m$ in both $p$ and $q$ while the auxiliary coordinates are effectively projected onto the transverse subspace by the external currents, so that for instance
\begin{gather}
(\partial_p\cdot\partial_p)\,p^{\,2}\,=\,2(d-2)\ ,
\end{gather}
since $p$ becomes a $(d-2)$-dimensional vector after the projection. These arguments show that one can associate to the massless propagator for totally symmetric spin-$m$ tensors a harmonic homogeneous polynomial of degree $m$ in both $p$ and $q$.

Using spherical coordinates, it is natural to separate the angular dependence from the radial one. From the auxiliary coordinates $p$ and $q$ one can thus construct an angle $\theta$, with
\begin{gather}
y\,=\,\cos\theta\,=\,\frac{p\cdot q}{\sqrt{p^{\,2} q^{\,2}}}\ ,\label{y}
\end{gather}
so that the most natural ansatz for the propagator of a totally symmetric spin-$m$ massless particle is of the form
\begin{gather}
\cP^{(m)}(p,q)\,=\,K\ (p^{\,2}q^{\,2})^{m/2}f\left(\frac{p\cdot q}{\sqrt{p^{\,2} q^{\,2}}}\right)\ ,
\end{gather}
where the exponent $m/2$ is determined by the condition that the polynomial be homogeneous of degree $m$ and $K$ is a normalization factor.

Applying the Laplace operator to $\cP^{\,(m)}$ yields an ordinary differential equation for $f(x)$,
\begin{gather}
(1-y^{\,2})f''(y)\,-\,(2\a+1)\,yf'(y)\,+\,m(m+2\a)f(y)\,=\,0\ ,\label{Laplace}
\end{gather}
with
\begin{gather}
\a\,=\,\frac{d}{2}-2\ .
\end{gather}
For $\a>0$ this equation has indeed the polynomial solutions
\begin{gather}
f^{[\a]}_m(x)\,=\,C_m^{[\a]}(x)\ ,
\end{gather}
where $C_m^{[\a]}(x)$ is a Gegenbauer polynomial, while for $\a\,=\,0$, that corresponds to $d\,=\,4$, it has the polynomial solution
\begin{gather}
f^{[0]}_m(x)\,=\,T_m(x)\ ,
\end{gather}
with $T_m(x)$ a Chebyshev polynomial of the first kind.

One can now determine $K$ demanding that the coefficient of the monomial $(p\cdot q)^m$ in the propagator be $(m!)^{-1}$, which is a convenient normalization. As a result, for $\a>0$
\begin{gather}
K\,=\,\frac{1}{2^m}\frac{\Gamma(\a)}{\Gamma(\a+m)}\ ,
\end{gather}
while for $\a\,=\,0$
\begin{align}
K&\,=\,\frac{2}{m!2^m}\ ,& m&\geq 1\ ,\\
K&\,=\,1\ ,& m&\,=\,0\ .
\end{align}
In conclusion, the massless propagator for totally symmetric spin-$m$ fields can be cast in the form
\begin{gather}
\cP^{(m)}(p,q)\,=\,-\frac{1}{k^{\,2}}\Bigg\{K\ \Big(p^{\,2}\,q^{\,2}\Big)^{m/2}f^{[\a]}_m\left(\frac{p\cdot q}{\sqrt{p^{\,2} q^{\,2}}}\right)\Bigg\}\ ,
\end{gather}
where $k$ is the exchanged momentum and $f_m^{[\a]}(x)$ is the polynomial solution of $(\ref{Laplace})$.

Summarizing, we have seen that each spin-$m$ propagator can be split into a radial part times a spherical harmonic that, for totally symmetric fields in $d$ dimensions, is a Gegenbauer or Chebyshev polynomial.

In the mixed symmetry case the trace conditions become the family of constraints
\begin{gather}
\partial_{p_{\,i}}\cdot\partial_{p_{\,j}}\ \cP(p_{\,l},q_{\,n})\,=\,0\ ,\ \ \forall\ i,j\ ,
\end{gather}
that eliminate all traces within any of the families or across pairs of them.
Most notably, simple polynomial solutions of these equations are then given by
\begin{multline}
P_m(p_{\,i},q_{\,j})\,=\,\left[(p_{\,1}+\dots+p_{\,N})^2(q_{\,1}+\dots +q_{\,N})^{\,2}\right]^{m/2}\\\times f^{[\a]}_m\left(\frac{(p_{\,1}+\dots+p_{\,N})\cdot (q_{\,1}+\dots +q_{\,N})}{\sqrt{(p_{\,1}+\dots+p_{\,N})^2 (q_{\,1}+\dots +q_{\,N})^2}}\right)\ ,\label{Mixed}
\end{multline}
so that, extracting from $(\ref{Mixed})$ the monomials of the form \begin{small}$p_{\,1}^{n_1}\cdots p_N^{n_N}q_1^{n_1}\cdots q_N^{n_N}$\end{small} where $m\,=\,n_1+\cdots +n_N$, one obtains, up to a normalization factor, the projector onto the traceless part of a generic transverse tensor bearing $k$ index families of length $\{n_i\}$. In this fashion, $(\ref{Mixed})$ can be regarded as a set of generating functions for mixed symmetry propagators. For instance, for two-family tensors of type $(n_1,n_2)$ the propagator polynomial that computes the current exchange for the corresponding conserved currents is given by
\begin{small}
\begin{multline}
\!\!\!\!\!\cP^{(n_1,n_2)}(p,q)\,=\,-\,\frac{K^{\a}_{n_1,n_2}}{k^{\,2}}\,(\partial_{\l_1}\partial_{\eta_1})^{n_1}
(\partial_{\l_2}\partial_{\eta_2})^{n_2}\Bigg\{\Big[(\l_1p_{\,1}+\l_2p_{\,2})^2(\eta_1q_1+\eta_2 q_2)^2\Big]^{\frac{n_1+n_2}{2}}\\\times f^{[\a]}_{n_1+n_2}\left(\frac{(\l_1p_{\,1}+\l_2 p_{\,2})\cdot (\eta_1 q_1+\eta_2 q_2)}{\sqrt{(\l_1p_{\,1}+\l_2 p_{\,2})^2 (\eta_1q_1+\eta_2 q_2)^2}}\right)\Bigg\}\ ,
\end{multline}
\end{small}where it is not necessary to let $\l_i\,=\,\n_i\,=\,0$ since the result is automatically independent of $\l$ and $\eta$. An analogous relation holds for other propagators for fields bearing arbitrary numbers of index families.

The results obtained so far are only valid for massless particles, but they can be extended straightforwardly to the massive case provided the currents are still transverse. This can be done with the formal substitution $d\ra d+1$ or $\a\ra\a+\frac{1}{2}$, that amounts to considering the massive little group $SO(d-1)$ rather than $SO(d-2)$. In this fashion one can obtain the massive totally symmetric higher-spin propagators
\begin{gather}
\cP^{(m)}\,=\,-\,\frac{1}{k^{\,2}+M_m^{\,2}}\Bigg\{K\ \Big(p^{\,2}\,q^{\,2}\Big)^{m/2}f^{[\a+\frac{1}{2}]}_m\left(\frac{p\cdot q}{\sqrt{p^{\,2} q^{\,2}}}\right)\Bigg\}\ ,
\end{gather}
and corresponding expressions in the mixed symmetry case.

In order to guarantee the correct current exchange also in the presence of non-transverse currents, however, the propagator must project them onto their transverse parts. Remarkably, this issue can also be conveniently addressed in our formalism. In fact, in order to project the current onto the transverse subspace, it is sufficient to project the auxiliary coordinates $p_{\,i}$ and $q_{\,i}$ according to
\begin{gather}
\begin{split}
p_{\,i}\ra \hat{p_{\,i}}\,=\,p_{\,i}\,+\,k\,\frac{p_{\,i}\cdot k}{M^{\,2}}\ ,\\
q_{\,i}\ra \hat{q_{\,i}}\,=\,q_{\,i}\,+\,k\,\frac{q_{\,i}\cdot k}{M^{\,2}}\ .
\end{split}
\end{gather}
Therefore, making the substitutions $q_{\,i}\ra \hat{q}_i$ and $p_{\,i}\ra \hat{p}_i$ in each of the expressions previously obtained, one recovers the complete propagators for massive higher-spin particles. For totally symmetric fields the general result is thus
\begin{gather}
\cP^{(m)}\,=\,-\,\frac{1}{k^{\,2}+M_m^{\,2}}\Bigg\{K\Big(\hat{p}^{\,2}\,\hat{q}^{\,2}\Big)^{m/2} f^{[\a+\frac{1}{2}]}_m\left(\frac{\hat{p}\cdot \hat{q}}{\sqrt{\hat{p}^{\,2} \, \hat{q}^{\,2}}}\right)\Bigg\}\ ,
\end{gather}
while similar relations, with $q_{\,i}\ra \hat{q}_i$ and $p_{\,i}\ra \hat{p}_i$, hold in the mixed-symmetry case.

Finally, one can consider the generating function of all totally symmetric higher-spin propagators
\begin{gather}
\hat{\cP}\,=\,\sum_{m\,=\,0}^{\infty}-\,\frac{1}{k^{\,2}+M_m^{\,2}} \Bigg\{K^{\a(m)}_{m}\left(\hat{p}^{\,2}\,\hat{q}^{\,2}\right)^{m/2}
f^{[\a(m)]}_m\left(\frac{\hat{p}\cdot \hat{q}}{\sqrt{\hat{p}^{\,2} \, \hat{q}^{\,2}}}\right)\Bigg\}\ ,\label{propSum1}
\end{gather}
where
\begin{gather}
\a(m)\,=\,\frac{d}{2}\,-\,2\ ,
\end{gather}
if $M^2_m\,=\,0$ and
\begin{gather}
\a(m)\,=\,\frac{d+1}{2}\,-\,2\ ,
\end{gather}
if $M^{\,2}_m\,\neq\, 0$ while $f^{[\a]}_m(x)$ and $K_m^\a$ are defined as before.
\\
For our application to the string currents computed in the previous chapter, or to currents containing in their definition the coupling constants, we can limit ourselves to eq.$\ (\ref{propSum1})$. In general, however, it is possible to consider arbitrary spin-dependent coupling constants $a_m>0$, that can be encoded in a generating function
\begin{gather}
a(z)\,=\,\sum_{n\,=\,0}^{\infty}\frac{a_n}{n!}\ z^n\ ,\label{cau}
\end{gather}
so that
\begin{gather}
\hat{\cP}(a)\,=\,\sum_{m\,=\,0}^{\infty}-\,\frac{a_m}{k^{\,2}+M_m^{\,2}} \Bigg\{K^{\a(m)}_{m}\Big(\hat{p}^{\,2}\,\hat{q}^{\,2}\Big)^{m/2}
f^{[\a(m)]}_m\left(\frac{\hat{p}\cdot \hat{q}}{\sqrt{\hat{p}^{\,2} \, \hat{q}^{\,2}}}\right)\Bigg\}\ .
\end{gather}


\scss{Generating Function of Massless Exchanges}

In this section we consider the simplest case, with all massless higher-spin propagating particles, so that the generating function of propagators for totally symmetric fields reduces to
\begin{gather}
\hat{\cP}(a)\,=\,\sum_{m\,=\,0}^{\infty}-\,\frac{a_m}{k^{\,2}}\Bigg\{K^{\a}_{m}\Big(p^{\,2}\,q^{\,2}\Big)^{m/2} f^{[\a]}_m\left(\frac{p\cdot q}{\sqrt{p^{\,2}\, q^{\,2}}}\right)\Bigg\}\ .\label{Propsum}
\end{gather}
We shall first consider the case $d\,=\,4$, in which $(\ref{Propsum})$ simplifies further and becomes
\begin{gather}
\hat{\cP}(a)\,=\,\sum_{m\,=\,0}^{\infty}-\,\frac{1}{k^{\,2}}\Bigg\{\frac{2\, a_m}{m!}\ \left(\frac{p^{\,2}\,q^{\,2}}{4}\right)^{m/2}\,T_m\left(\frac{p\cdot q}{\sqrt{p^{\,2}\, q^{\,2}}}\right)\,-\,a_0\Bigg\}\ ,\label{D4}
\end{gather}
and then we shall briefly turn to the generalization of the result to space-time dimensions $d>4$.

In the four-dimensional case one can sum the series via the generating function of Chebyshev polynomials
\begin{gather}
\cT(y,t)\,=\,\sum_{m\,=\,0}^{\infty}T_m(y)\,t^m\,=\,\frac{1-yt}{1-2yt+t^2}\ ,\label{g1}
\end{gather}
with $y$ as in eq.$\ (\ref{y})$, observing that the sum in $(\ref{D4})$, with $a_m\,=\,1$ for all $m$, is exactly of the form
\begin{gather}
f(y,t)\,=\,\sum_{n\,=\,0}^{\infty}\frac{1}{n!}\ T_n(y)t^n\ .\label{g2}
\end{gather}
In going from $(\ref{g1})$ to $(\ref{g2})$, it is convenient to resort to the Hankel contour integral for the Euler $\G$ function
\begin{gather}
\frac{1}{\Gamma(q)}\,=\,\frac{1}{2\pi i}\oint_C e^{\,z} z^{-q}\,dz\ ,\label{Hankel}
\end{gather}
where $C$ is a contour encircling the negative real axis, starting and ending at $-\infty$ and turning around the origin counterclockwise. One then finds
\begin{multline}
\sum_{n\,=\,0}^{\infty}\frac{1}{\Gamma(n+1)}\ T_n(y)\ t^n\,=\,\frac{1}{2\pi i}\sum_{n\,=\,0}^{\infty}\oint_C e^z z^{-n-1}\ T_n(y)\ t^n\\
\,=\,\frac{1}{2\pi i}\oint_C dz\ \sum_{n\,=\,0}^{\infty}e^{z}z^{-1}T_n(y)\Big(\frac{t}{z}\Big)^n\,=\,\frac{1}{2\pi i}\oint_C dz\ e^z\frac{z-yt}{z^2-2yzt+t^2}\ .
\end{multline}
Thanks to the exponentially suppressed behavior of the integrand at $-\infty$, the contour integral is simply given by the sum of the residues at
\begin{gather}
z_{\pm}\,=\,yt\pm t\sqrt{y^2-1}\ ,
\end{gather}
and the final result is
\begin{gather}
f(y,t)\,=\,\sum_{n\,=\,0}^{\infty}\frac{1}{n!}\ T_n(y)\,t^n\,=\,\frac{1}{2}\,\left(e^{\left(y+\sqrt{y^2-1}\right)t}+e^{\left(y-\sqrt{y^2-1}\right)t}\right)\ .\label{eq}
\end{gather}
Moreover, one can also address in this fashion the case of generic $a_n$ couplings, considering, with reference to eq.$\ (\ref{cau})$,
\begin{gather}
g(y,t)\,=\,\sum_{n\,=\,0}^{\infty}\frac{a_n}{n!}\ T_n(y)\ t^n\ .
\end{gather}
Computing the Fourier Transform of $f(y,t)$ with respect to $t$, that is given by
\begin{gather}
\tilde{f}(y,q)\,=\,\frac{1}{2}\left(e^{i\left(y+\sqrt{y^2-1}\right)\partial_q}+
e^{i\left(y-\sqrt{y^2-1}\right)\partial_q}\right)\,\sqrt{2\pi}\,\delta(q)\ ,
\end{gather}
and resorting the identity
\begin{gather}
\sum_n\frac{a_n}{n!}\ T_n(y)\,=\,\int \frac{dz}{\sqrt{2\pi}}\ a(izt)\tilde{f}(y,z)\ ,
\end{gather}
yields
\begin{gather}
g(y,t)\,=\,\frac{1}{2}\left[a\left(ty+t\sqrt{y^2-1}\right)+a\left(ty-t\sqrt{y^2-1}\right)\right]\ ,
\end{gather}
where $a(z)$ is defined in eq. $(\ref{cau})$. Finally, the generating function of massless totally symmetric exchanges $(\ref{D4})$ is
\begin{multline}
\hat{\cP}\,=\,-\frac{1}{k^{\,2}}\left[a\left(\frac{1}{2}\,p\cdot q+\frac{1}{2}\,\sqrt{(p\cdot q)^2-p^{\,2}\,q^{\,2}}\right)\right.\\\left.+\,a\left(\frac{1}{2}\,p\cdot q-\frac{1}{2}\,\sqrt{(p\cdot q)^2-p^{\,2}\,q^{\,2}}\right)\,-\,a_0\right]\ ,
\end{multline}
and for $a(z)\,=\,e^z$ it reduces to $(\ref{eq})$. This result was presented recently by Bekaert, Mourad and Joung in \cite{Bekaert:2009ud} with reference to the scalar currents obtained first in \cite{Berends:1985xx}. Here we have recovered it following a different and more general route.\\

The general case with $d>4$ (or equivalently $\a>0$) can be similarly addressed, but the end result is rather cumbersome, since the generating function of all higher-spin massless propagators is given by
\begin{gather}
\hat{\cP}(a)\,=\,\sum_{m\,=\,0}^{\infty}-\,\frac{a_m}{k^{\,2}}\Bigg\{\frac{\Gamma(\a)}{\Gamma(\a+m)} \left(\frac{p^{\,2}\,q^{\,2}}{4}\right)^{m/2}\ C^{[\a]}_m\left(\frac{p\cdot q}{\sqrt{p^{\,2}\, q^{\,2}}}\right)\Bigg\}\ ,\label{Propsum2}
\end{gather}
where the $C_m^{[\a]}(x)$ are Gegenbauer polynomials. In order to sum this series it is convenient, as before, to start from the generating function, that in this case is
\begin{gather}
\cG^{[\a]}(x,t)\,=\,\sum_{n\,=\,0}^{\infty}\ G_n^{[\a]}(x)\ t^n\,=\,\frac{1}{(1-2xt+t^2)^\a}\ .
\end{gather}
In fact, for $a_m\,=\,1$ for all $m$, the sum over $m$ in $(\ref{Propsum2})$ is exactly of the form
\begin{gather}
k^{[\a]}(x,t)\,=\,\sum_{n\,=\,0}^{\infty}\frac{1}{\Gamma(\a+m)}\ G_n^{[\a]}(x)\ t^n\ ,
\end{gather}
and to compute $k^{[\a]}(x,t)$ one can use, as before, the Hankel contour integral of eq. $(\ref{Hankel})$, with the result
\begin{multline}
k^{[\a]}(x,t)\,=\,\sum_{n\,=\,0}^{\infty}\frac{1}{\Gamma(\a+m)}\ G_n^{[\a]}(x) \, t^n\\\,=\,\frac{1}{2\pi i}\oint_C dz\sum_{n\,=\,0}^{\infty} e^z z^{-\a-m}\ G_n^{[\a]}(x) \, t^n\,=\,\frac{1}{2\pi i}\oint_C dz\ \frac{e^{z} z^{\a}}{(z^2-2xtz+t^2)^\a}\ .
\end{multline}
Actually, in all cases when the space-time dimension $d$ is even the integral has only poles, and one can evaluate it summing residues, with the end result
\begin{gather}
k^{[\a]}(x,t)\,=\,\frac{1}{\Gamma(\a)}\left[\partial_z^{\,\a-1}\left(\frac{e^z z^\a}{(z-z_-)^\a}\right)\Big|_{z\,=\,z_+}+\partial_z^{\,\a-1}\left(\frac{e^z z^\a}{(z-z_+)^\a}\right)\Big|_{z\,=\,z_-}\right]\ ,\label{resul}
\end{gather}
where
\begin{gather}
z_{\pm}\,=\,xt\pm t\sqrt{x^2-1}\ .
\end{gather}
One can also simplify eq. $(\ref{resul})$, resorting to the identity
\begin{gather}
(e^zg(z))^{(n)}\,=\,e^z\sum_{k\,=\,0}^{n}\binom{n}{k}g^{(k)}(z)\ .
\end{gather}
As a result
\begin{multline}
k^{[\a]}(x,t)\,=\,\frac{1}{\Gamma(\a)}\left[e^{xt+ t\sqrt{x^2-1}}\ \sum_{k\,=\,0}^{\a-1}\binom{\a-1}{k}\ g_-^{(k)}(xt+t\sqrt{x^2-1})\right.\\\left.+ e^{xt- t\sqrt{x^2-1}}\ \sum_{k\,=\,0}^{\a-1}\binom{\a-1}{k}\ g_+^{(k)}(xt- t\sqrt{x^2-1})\right]\ ,
\end{multline}
where we have defined
\begin{gather}
g_\pm(z)\,=\,\left(\frac{z}{z-z_\pm}\right)^{\,\a}\ ,
\end{gather}
and we have pulled out the exponential factor that plays a dominant role in the asymptotic behavior of higher-spin amplitudes.
Finally, for $d$ even, the propagator generating function takes the compact form
\begin{small}
\begin{multline}
\hat{\cP}\,=\,-\frac{1}{k^{\,2}}\left[e^{\frac{1}{2}\,p\cdot q+\frac{1}{2}\sqrt{(p\cdot q)^2-p^{\,2}\,q^{\,2}}}\sum_{k\,=\,0}^{\a-1}\binom{\a-1}{k}\ g_-^{(k)}\left(\frac{1}{2}\,p\cdot q+\frac{1}{2}\sqrt{(p\cdot q)^2-p^{\,2}\,q^{\,2}}\right)\right.\\\left.+ e^{\frac{1}{2}\,p\cdot q-\frac{1}{2}\sqrt{(p\cdot q)^2-p^{\,2}\,q^{\,2}}}\sum_{k\,=\,0}^{\a-1}\binom{\a-1}{k}\ g_+^{(k)}\left(\frac{1}{2}\,p\cdot q-\frac{1}{2}\sqrt{(p\cdot q)^2-p^{\,2}\,q^{\,2}}\right)\right]\ .
\end{multline}
\end{small}For $d$ odd, or for massive particles in even dimensions, things are more subtle because the integrand has a cut and one must take into account the contribution of the corresponding discontinuity. Although this can be done relatively easily in some simple cases with special choices of the $a_n$, we do not consider this case here because it is not necessary for the ensuing discussion.

From now on we shall concentrate on the simpler case of massless higher-spin particles in $d\,=\,4$ to study some properties of the scattering amplitudes involving the string currents computed in the previous chapter in the massless limit.


\scs{String Couplings}
In this section we shall study the amplitude $(\ref{A2})$ and the currents $(\ref{Curr})$, trying to extract from them some truly off-shell couplings. The aim is to construct string-inspired cubic interactions of the form
\begin{gather}
\cL_3\,=\,\cJ\cdot H\ ,
\end{gather}
and to study their massless limit.
The starting point is the explicit expansion of the generating function $(\ref{A2})$, computed using the standard binomial and the trinomial formulas
\begin{gather}
\begin{split}
(a+b+c)^n\,=\,&\sum_{i,j\,=\,0}^n\frac{n!}{i!j!k!}\ a^i\ b^j\ c^k\ \d_{i+j+k,n}\ ,\\
(a+b)^n\,=\,&\sum_{k\,=\,0}^n\frac{n!}{k!(n-k)!}\ a^k\ b^{n-k}\ ,
\end{split}
\end{gather}
that lead to the expansion
\begin{multline}
\cA_{\pm}\,=\,\tilde{H}_1\,\tilde{H}_2\,\tilde{H}_3\!\!\!\sum_{i\,,\,j\,,\,k\,\in\,\cI}\,
\left(\pm\sqrt{\frac{\a^{\,\prime}\!}{2}}\right)^{s_1+s_2+s_3-2i-2j-2k}\\
\times\,\left[
\frac{s_1!\,s_2!\,s_3!}{i!\,j!\,k!\,(s_1\,-\,j\,-\,k)!\,(s_2\,-\,k\,-\,i)!\,(s_3\,-\,i\,-\,j)!}\right.\\\left.
p_{\,23}^{\,s_1-j-k}\ p_{\,31}^{\,s_2-k-i}\ p_{\,12}^{\,s_3-i-j}\ \delta_{\,23}^{\,i}\
\delta_{\,31}^{\,j}\ \delta_{\,12}^{\,k}\vphantom{\sqrt{\frac{a}{b}}}\right]\ .\label{expl}
\end{multline}
Here $\tilde{H}_1$, $\tilde{H}_2$ and $\tilde{H}_3$ are totally symmetric polarization tensors of spin $n$, $m$ and $l$ respectively,
\be
\cI\,=\,\left\{i\,,\,j\,,\,k\,\in\,\mathbb{N}\,\Big|\ s_1-j-k\,\geq\,0\,;\,s_2-k-i\,\geq\,0\,;\,s_3-i-j\,\geq\,0\,\right\}\ .
\ee
and we have resorted to a shorthand notation so that $k_{\,ij}$ is always contracted with $\tilde{H}_k$ with $k$ different from $i$ and $j$. Finally, the $\d_{ij}$ compute contractions between $\tilde{H}_i$ and $\tilde{H}_j$.

\scss{$0-0-s$ Couplings}
This case corresponds to the coupling $0-0-s$ and to the string current $(\ref{Curr})$
\begin{gather}
\cJ\,=\,i\frac{g_o}{\a'}\left\{J_+(x,k')\,Tr[\ \cdot\ \L^{a_1}\L^{a_2}]\,+\,J_-(x,k')\,Tr[\ \cdot\ \L^{a_2}\L^{a_1}]\right\}\ ,
\end{gather}
with
\begin{gather}
J_{\pm}(k_{\,1},k_{\,2},k')\,=\,\exp\left(\pm\sqrt{\frac{\a'}{2}}\ k_{\,12}\cdot k'\right)\,\tilde{\phi}(k_{\,1})\,\tilde{\phi}(k_{\,2})\ ,
\end{gather}
in momentum space, or
\begin{gather}
J_{\pm}(x,k')\,=\,\phi\left(x\pm i\sqrt{\frac{\a'}{2}}\ k'\right)\,\phi\left(x\mp i\sqrt{\frac{\a'}{2}}\ k'\right)\ .\label{scalar}
\end{gather}
In coordinate space, these currents are very simple and automatically conserved. They were identified long ago by Berends, Burgers and van Dam \cite{Berends:1985xx} and were recently reconsidered by Bekaert, Joung and Mourad \cite{Bekaert:2009ud}. Here we see them emerging as effective tree-level couplings of the open string. It is important to stress that, in this simple case, the coupling is exactly the highest derivative one, simply because the only way to couple a pair of scalar fields with a spin-$s$ tensor is via $s$ derivatives.


\scss{$1-1-s$ Couplings}

This case corresponds to the $1-1-s$ coupling, but the current generating function $(\ref{Curr})$ does not give directly a result that is manifestly gauge invariant off-shell for the spin-$1$ external legs. To overcome this difficulty one can use the explicit on-shell expression $(\ref{expl})$, that in this particular case simplifies and becomes
\begin{gather}
\begin{split}
\cA^{\pm}&_{s-1-1}\,=\,\left[\pm\sqrt{\frac{\a'}{2}}\right]^s A^2 H\cdot k_{\,12}^s+s(s-1)\left[\pm\sqrt{\frac{\a'}{2}}\right]^{s-2}A_{\m}^{(1)}A_{\n}^{(2)}H^{\m\n\cdots}k_{\,12}^{s-2}\\
+&\left[\pm\sqrt{\frac{\a'}{2}}\right]^ssA^{(1)}\cdot k_{\,23}A_\n^{(2)}H^{\n\cdots}k_{\,12}^{s-1}+
\left[\pm\sqrt{\frac{\a'}{2}}\right]^s sA^{(2)}\cdot k_{\,31} A_\n^{(1)}H^{\n\cdots}k_{\,12}^{s-1}\\
+&\left[\pm\sqrt{\frac{\a'}{2}}\right]^{s+2}A^{(1)}\cdot k_{\,23}A^{(2)}\cdot k_{\,31} H\cdot k_{\,12}^s\ .
\end{split}
\end{gather}
Here gauge invariance with respect to the spin-$1$ external legs is manifest and one can guess that this amplitude must be parameterized by terms involving $F_{\m\n}$ of the form
\begin{gather}
\cA_{s-1-1}\,=\,a\,F^2\, H\cdot k_{\,12}^s\,+\,b\,F_{\m\a}\,{F_\n}^{\a}\,H^{\m\n\cdots}\,k_{\,12}^{s-2}\ .
\end{gather}
A direct comparison with the explicit relations
\begin{gather}
\begin{split}
F^2\,=\,&2\,k_{\,1}\cdot k_{\,2}\,A^2\,+\,\frac{1}{2}\,A^{(1)}\cdot k_{\,23}\, A^{(2)}\cdot k_{\,31}\ ,\\
F_{\m\a}\,{F_\n}^\a\,=\,&-\,\frac{1}{4}\,\left(k_{\,12\m}k_{\,12\m}\,A^2\,+\,A^{(1)}\cdot k_{\,23}\,A^{(2)}_\n k_{\,12\m}\,+\,A^{(2)}\cdot k_{\,31} \,A_{\m}^{(1)} k_{\,12\n}\right)\\
&+\,k_{\,1}\cdot k_{\,2} \,A_{\m}^{(1)}\,A_{\n}^{(2)}\ ,
\end{split}
\end{gather}
determines the coefficients
\begin{align}
a&\,=\,2\left[\pm\sqrt{\frac{\a'}{2}}\right]^{s+2}\ ,& b&\,=\,-4s\left[\pm\sqrt{\frac{\a'}{2}}\right]^s\ .
\end{align}
so that, finally,
\begin{gather}
\cA^{\pm}_{s-1-1}\,=\,2\left[\pm\sqrt{\frac{\a'}{2}}\right]^{s+2} F^2\, H\cdot k_{\,12}^s\,-\,4s\,\left[\pm\sqrt{\frac{\a'}{2}}\right]^s\,F_{\m\a}\,{F_\n}^{\a}\,H^{\m\n\cdots}\,k_{\,12}^{s-2}\ .\label{onshell}
\end{gather}
It is important to stress that, even if we started with an interaction that was not gauge invariant with respect to the spin-$s$ field, the final form is manifestly gauge invariant in the massless limit. This means that, for both massive and massless higher spins, the $1-1-s$ coupling is actually induced by a conserved current, exactly as was the case for the scalar coupling $(\ref{scalar})$. In this respect the crucial observation is that gauge invariance is broken by the on-shell equation
\begin{gather}
k_{\,1}\cdot k_{\,2}\,=\,\frac{1}{2\a'}\,(l-m-n+1)\,\neq\, 0\ ,\label{shell}
\end{gather}
where we have resorted to the mass parametrization used in the previous chapter
\begin{align}
-\a'k_{\,1}^2&\,=\,m-1\ ,& -\a' k_{\,2}^2&\,=\,n-1\ ,& -\a' k_{\,3}^2&\,=\,l-1\ .
\end{align}
If this picture is valid in general, we expect that manifestly gauge invariant results are obtained in the massless limit for all other couplings.

Analyzing the particular form of the interaction, one can recognize a higher derivative term proportional to the squared curvature of the external fields and a term that resembles the energy-momentum tensor of the spin-$1$ field, aside from the term involving the trace of $F^2$ that cannot appear simply because $H$ is traceless on-shell.

The current generating functions giving rise to these two couplings can be cast in the exponential form
\begin{gather}
\cJ^{\pm}(k')\,=\,\a'\exp\left(\pm\sqrt{\frac{\a'}{2}}\ k_{\,12}\cdot k'\right)\,F_{\m\n}\,F^{\m\n}\ ,\label{J10}
\end{gather}
for the first and
\begin{gather}
\cJ^{\pm}(k')\,=\,-4\partial_\l\left(\partial_q\cdot\partial_q\right)
\exp\left(\pm(1+\l)\sqrt{\frac{\a'}{2}}\ [k_{\,12\m}\,+\,F_{\m\a}\,q^\a]\,k^{'\m}\right)\Bigg|_{q\,=\,\l\,=\,0}\ ,\label{J20}
\end{gather}
for the second, where the differential operator $\partial_\l$ accounts for the factor $s$ in $(\ref{onshell})$.
It is important to stress that the coupling involving the energy-momentum tensor just found in $(\ref{onshell})$ is exactly the one conjectured in this case by Berends, Burgers and van Dam \cite{Berends:1985xx}. The emergence of the gauge invariant structure of the couplings resonates with the long-held suspicion that the string masses originate from a symmetry breaking phenomenon.


\scss{$1-s-s$ Couplings}
We now consider the $1-s-s$ couplings that codify the electromagnetic interactions of spin-$s$ currents. In this particular case eq.$\ (\ref{expl})$ can be computed explicitly via the binomial expansion and as a result one obtains the expression
\begin{small}
\begin{gather}
\begin{split}
\cA^{\pm}_{1-s-s}&\,=\,\left[\sum_{k\,=\,0}^{s-1}\frac{2(s!)^2}{k!(s-k)!(s-k-1)!} \, \left(\pm\sqrt{\frac{\a'}{2}}\right)^{2s-2k-1}\right.\\
&\left.\times k_{\,23}^{s-k-1}\cdot H^{(1)}_{\a_{s-k}\cdots\a_{s-1}\m}{H^{(3)\a_{s-k}\cdots\a_{s-1}}}_{\n}\cdot k_{\,12}^{s-k-1}\vphantom{\left(\pm\sqrt{\frac{\a'}{2}}\right)^{2s-2k-1}}\right]\,F^{(2)\m\n}\\
&+\left[\sum_{k\,=\,0}^s\frac{(s!)^2}{k!(s-k)!^2}\,\left(\pm\sqrt{\frac{\a'}{2}}\right)^{2s-2k+1}\right.\\&\left.\times k_{\,23}^{s-k}\cdot H^{(1)}_{\a_{s-k+1}\cdots\a_s}{H^{(3)\a_{s-k+1}\cdots\a_{s}}}\cdot k_{\,12}^{s-k}\,A\cdot k_{\,31}\vphantom{\left(\pm\sqrt{\frac{\a'}{2}}\right)^{2s-2k-1}}\right]\ ,\label{1ss}
\end{split}
\end{gather}
\end{small}where $H^{(1)}$ and $H^{(2)}$ are totally symmetric spin-$s$ irreducible polarization tensors, $A_\m$ is the spin-$1$ polarization vector and
\begin{gather}
F^{(2)\m\n}\,=\,k_{\,2}^\m A^\n-k_{\,2}^\n A^\m\ .
\end{gather}
From $(\ref{1ss})$ one can extract, in particular, the $1-3-3$ coupling, that was recently studied in \cite{Boulanger:2008tg} for massless particles, obtaining in this case
\begin{small}
\begin{gather}
\begin{split}
\cA^{\pm}_{1-3-3}&\,\simeq\,\\\,=\,&\left(\pm\sqrt{\frac{\a'}{2}}\right)\left[H^{(1)}_{\a_1\a_2\m}{H^{(3)\a_1\a_2}}_\n\ F^{(2)\m\n}+\frac{1}{6}\,H^{(1)}\cdot H^{(3)}\  A\cdot k_{\,31}\right]\\
+&\left(\pm\sqrt{\frac{\a'}{2}}\right)^3\left[k_{\,23}\cdot H^{(1)}_{\a_1\m}{H^{(3)\a_1}}_\n\cdot k_{\,12}\ F^{(2)\m\n}\right.\\&\left.+\frac{1}{2}\,k_{\,23}\cdot H^{(1)}_{\a_1\a_2} H^{(3)\a_1\a_2}\cdot k_{\,12}\  A\cdot k_{\,31}\right]\\
+&\left(\pm\sqrt{\frac{\a'}{2}}\right)^5\left[\frac{1}{6}\,k_{\,23}^2\cdot H^{(1)}_{\m}{H^{(3)}}_\n\cdot k_{\,12}^2\ F^{(2)\m\n}\right.\\&\left.+\frac{1}{4}\,k_{\,23}^2\cdot H^{(1)}_{\a_1} H^{(3)\a_1}\cdot k_{\,12}^2\  A\cdot k_{\,31}\right]\\
+&\left(\pm\sqrt{\frac{\a'}{2}}\right)^7\left[\frac{1}{36}\,k_{\,23}^3\cdot H^{(1)} H^{(3)}\cdot k_{\,12}^3\  A\cdot k_{\,31}\right]\ ,\label{133}
\end{split}
\end{gather}
\end{small}where we have grouped together terms with the same number of derivatives. From $(\ref{133})$ and $(\ref{1ss})$ it is possible to study the structure of the spin-$s$ electromagnetic coupling. Let us stress that there are only two truly different terms, one with $F^{\m\n}$ contracted with the polarization tensors and the other with $A^\m$ contracted with a momentum, so that we expect that the lower derivative terms arise only because of the massive equations of motion for the external states, while the true off-shell couplings are the one with maximum number of derivative, that is Born-Infeld-like, and by the one with $2s-1$ derivatives, that deforms the abelian gauge symmetry, as already stressed in \cite{Boulanger:2008tg}. This structure, as can be seen from $(\ref{1ss})$, is very general, and in fact if the external states are massive one can see that each of these two couplings splits into multiple contributions with lower number of derivatives. At any rate, in the massless limit only the higher derivative terms survive and on-shell for massless fields one is left with
\begin{gather}
\begin{split}
\cA^{\pm}&_{1-s-s}\,=\,a\,\left(\pm\sqrt{\frac{\a'}{2}}\right)^{2s-1}\ \left[\,k_{\,23}^{s-1}\cdot H^{\,(1)}_{\m}{H^{\,(3)}}_{\n}\cdot k_{\,12}^{s-1}\ F^{\,(2)\m\n}\right.\\
&\left.+\,2k_{\,23}^{s-1}\cdot H_\a^{(1)}\, H^{(3)\a}\cdot k_{\,12}^{s-1}\,A\cdot k_{\,31}
\right]\vphantom{\left(\pm\sqrt{\frac{\a'}{2}}\right)^{2s-1}}\\
&+\left(\pm\sqrt{\frac{\a'}{2}}\right)^{2s+1}\ k_{\,23}^{s}\cdot H^{\,(1)}{H^{\,(3)}}\cdot k_{\,12}^{s}\ A\cdot k_{\,31}\ ,
\end{split}
\end{gather}
that displays only the two expected gauge invariant contributions weighted by a relative coefficient $a$.


\scss{$2-s-s$ Couplings}
Here we consider the $2-s-s$ coupling that codifies the spin-$2$ interaction of a spin-$s$ current. As in the previous case, eq. $(\ref{expl})$ can be computed via the binomial expansion, with the result
\begin{small}
\begin{gather}
\begin{split}
\cA&^{\pm}_{2-s-s}\,=\,\left[\sum_{k\,=\,0}^{s-2}\,\frac{(s!)^2}{k!(s-k)!(s-k-2)!}\, \left(\pm\sqrt{\frac{\a'}{2}}\right)^{2s-2k-2}\right.\\
&\left.\times k_{\,23}^{s-k-2}\cdot H^{(1)}_{\a_{s-k-1}\cdots\a_{s-2}\m_1\m_2}\,{H^{(3)\a_{s-k-1}\cdots\a_{s-2}}}_{\n_1\n_2}\cdot k_{\,12}^{s-k-2}\vphantom{\sqrt{\frac{\a'}{2}}}\vphantom{\left(\pm\sqrt{\frac{\a'}{2}}\right)^{2s-2k-2}}\right]\\
&\times\Big(\phi^{\n_1\n_2}k_{\,23}^{\m_1}k_{\,23}^{\m_2}+\phi^{\m_1\m_2}k_{\,12}^{\n_1}k_{\,12}^{\n_2}\Big)
\vphantom{\left(\pm\sqrt{\frac{\a'}{2}}\right)^{2s-2k-2}}\\
&+\left[\sum_{k\,=\,0}^{s-1}\frac{2(s!)^2}{k!(s-k)!(s-k-1)!}\left(\pm\sqrt{\frac{\a'}{2}}\right)^{2s-2k}\right.\\
&\left.\vphantom{\left(\pm\sqrt{\frac{\a'}{2}}\right)^{2s-2k-2}}\times k_{\,23}^{s-k-1}\cdot H^{(1)}_{\a_{s-k}\cdots\a_{s-1}\m_1}\ {H^{(3)\a_{s-k}\cdots\a_{s-1}}}_{\n_1}\cdot k_{\,12}^{s-k-1}\vphantom{\sqrt{\frac{\a'}{2}}}\right]\\
&\vphantom{\left(\pm\sqrt{\frac{\a'}{2}}\right)^{2s-2k-2}}\times\left(k_{\,31}\cdot\phi^{\n_1}k_{\,23}^{\m_1}+
k_{\,31}\cdot\phi^{\m_1}k_{\,12}^{\n_1}\right)+\\
&+\left[\sum_{k\,=\,0}^s\frac{(s!)^2}{k!(s-k)!^2}\left(\pm\sqrt{\frac{\a'}{2}}\right)^{2s-2k+2}\right.\\
&\left.\vphantom{\left(\pm\sqrt{\frac{\a'}{2}}\right)^{2s-2k-2}}\times k_{\,23}^{s-k}\cdot H^{(1)}_{\a_{s-k+1}\cdots\a_{s}}\,{H^{(3)\a_{s-k+1}\cdots\a_{s}}}\cdot k_{\,12}^{s-k}\vphantom{\sqrt{\frac{\a'}{2}}}\right]\times k_{\,31}^2\cdot \phi\\
&+\left[\sum_{k\,=\,1}^s\frac{2k(s!)^2}{k!(s-k)!^2}\left(\pm\sqrt{\frac{\a'}{2}}\right)^{2s-2k}\right.\\
&\left.\vphantom{\left(\pm\sqrt{\frac{\a'}{2}}\right)^{2s-2k-2}}\times k_{\,23}^{s-k}\cdot H^{(1)}_{\a_{s-k+1}\cdots\a_{s-1}\m_1}\,{H^{(3)\a_{s-k+1}\cdots\a_{s-1}}}_{\n_1}\cdot k_{\,12}^{s-k}\vphantom{\sqrt{\frac{\a'}{2}}}\right]\times \,\phi^{\m_1\n_1}\ ,\label{2ss}
\end{split}
\end{gather}
\end{small}where, as before, $H^{(1)}$, $H^{(3)}$ and $\phi_{\m\n}$ are respectively the two spin-$s$ totally symmetric irreducible polarization tensors and the spin-$2$ irreducible symmetric polarization tensor. It is very instructive to consider in detail the simple $2-3-3$ example, already analyzed in \cite{Boulanger:2008tg}, so that $(\ref{2ss})$ turns into
\begin{small}
\begin{gather}
\begin{split}
\cA^{\pm}_{2-3-3}&\sim\\\sim\left[\vphantom{k^{(2)}}\right.&\left.H^{(1)}_{\a_1\a_2\m_1}\,{H^{(3)\a_1\a_2}}_{\n_1}\ \phi^{\m_1\n_1}\right]\\
+&\left(\pm\sqrt{\frac{\a'}{2}}\right)^2\left[\frac{1}{2}\,H^{(1)}_{\a_1\m_1\m_2}\,{H^{(3)\a_1}}_{\n_1\n_2}\ \left(\phi^{\n_1\n_2}k_{\,23}^{\m_1}k_{\,23}^{\m_2}+\phi^{\m_1\m_2}k_{\,12}^{\n_1}k_{\,12}^{\n_2}\right)\right.\\
&\vphantom{\left(\pm\sqrt{\frac{\a'}{2}}\right)^2}+H^{(1)}_{\a_1\a_2\m_1} {H^{(3)\a_1\a_2}}_{\n_1}\ \left(k_{\,31}\cdot\phi^{\n_1}\,k_{\,23}^{\m_1}+k_{\,31}\cdot\phi^{\m_1}\,k_{\,12}^{\n_1}\right)+ \frac{1}{6}\,H^{(1)}\cdot H^{(3)}\ k_{\,31}^2\cdot\phi\\ \vphantom{\left(\pm\sqrt{\frac{\a'}{2}}\right)^2}&\left.+2k_{\,23}\cdot H^{(1)}_{\a_1\m_1}\,{H^{(3)\a_1}}_{\n_1}\cdot k_{\,12}\,\phi^{\m_1\n_1}\vphantom{\frac{1}{2}}\right]+\\
+&\left(\pm\sqrt{\frac{\a'}{2}}\right)^4\left[\frac{1}{6}\,k_{\,23}\cdot H^{(1)}_{\m_1\m_2}\ {H^{(3)}}_{\n_1\n_2}\cdot k_{\,12}\ \left(\phi^{\n_1\n_2}\,k_{\,23}^{\m_1}k_{\,23}^{\m_2}+\phi^{\m_1\m_2}\,k_{\,12}^{\n_1}k_{\,12}^{\n_2}\right)\right.
\\\vphantom{\left(\pm\sqrt{\frac{\a'}{2}}\right)^2}&+k_{\,23}\cdot H^{(1)}_{\a_1\m_1}\, {H^{(3)\a_1}}_{\n_1}\cdot k_{\,12}\ \left(k_{\,31}\cdot\phi^{\n_1}\,k_{\,23}^{\m_1}+k_{\,31}\cdot\phi^{\m_1}\,k_{\,12}^{\n_1}\right)\\
&\vphantom{\left(\pm\sqrt{\frac{\a'}{2}}\right)^2}\left.+\frac{1}{2}\,k_{\,23}\cdot H^{(1)}_{\a_1\a_2}\,H^{(3)\a_1\a_2}\cdot k_{\,12}\ k_{\,31}^{\,2}\cdot\phi+\frac{1}{2}\,k_{\,23}^{\,2}\cdot H^{(1)}_{\m_1}\,H^{(3)}_{\n_1}\cdot k_{\,12}^{\,2}\ \phi^{\m_1\n_1}\right]\\
+&\left(\pm\sqrt{\frac{\a'}{2}}\right)^6\left[\frac{1}{6}\,k_{\,23}^{\,2}\cdot H^{(1)}_{\m_1}\ H^{(3)}_{\n_1}\cdot k_{\,12}^{\,2}\ \left(k_{\,31}\cdot\phi^{\n_1}\,k_{\,23}^{\m_1}+k_{\,31}\cdot\phi^{\m_1}\,k_{\,12}^{\n_1}\right)\right.\\ &\vphantom{\left(\pm\sqrt{\frac{\a'}{2}}\right)^2}\left.+\frac{1}{4}\,k_{\,23}^{\,2}\cdot H^{(1)}_{\a_1}\,H^{(3)\a_1}\cdot k_{\,12}^{\,2}\ k_{\,31}^{\,2}\cdot \phi\right]\\
+&\left(\pm\sqrt{\frac{\a'}{2}}\right)^8\left[\frac{1}{36}\,k_{\,23}^3\cdot H^{(1)}\ H^{(3)}\cdot k_{\,12}^3\ k_{\,31}^2\cdot\phi\right]\ ,\label{233}
\end{split}
\end{gather}
\end{small}where we have grouped together terms with the same number of derivatives.

As before $(\ref{2ss})$ and $(\ref{233})$ show the general structure of such couplings with contributions involving different number of derivatives. At any rate, the lower derivative terms are expected to be there only because of the massive equations of motion for the external states, as in the previous case. Analyzing these couplings, one can see that there are only four different structures. The one multiplied with
\begin{gather}
\Big(\phi^{\n_1\n_2}k_{\,23}^{\m_1}k_{\,23}^{\m_2}+\phi^{\m_1\m_2}k_{\,12}^{\n_1}k_{\,12}^{\n_2}\Big)\ ,
\end{gather}
the one with the factor
\begin{gather}
\phi^{\m_1\n_1}\ ,
\end{gather}
the one with the factor
\begin{gather}
\Big(k_{\,31}\cdot\phi^{\n_1}k_{\,23}^{\m_1}+k_{\,31}\cdot\phi^{\m_1}k_{\,12}^{\n_1}\Big)\ ,
\end{gather}
and finally the one with the factor
\begin{gather}
k_{\,31}^2\cdot \phi\ .
\end{gather}
In the massive case all these structures split into a tower of contributions with different number of derivatives, while in the massless case all lower derivative terms must cancel because the contributions proportional to $k_{\,i}^{\,2}$ are identically zero. As a result, in the massless case one is left only with three higher derivative contributions weighted by some relative coefficients $a$ and $b$:

\begin{gather}
\begin{split}
\cA&^{\pm}_{2-s-s}\,=\,a\,\left(\pm\sqrt{\frac{\a'}{2}}\right)^{2s-2}\ \left[k_{\,23}^{s-2}\cdot H^{(1)}_{\m_1\m_2}\,{H^{(3)}}_{\n_1\n_2}\cdot k_{\,12}^{s-2}\right.\\\vphantom{\left(\pm\sqrt{\frac{\a'}{2}}\right)^{2s-2}}
&\times\left((\phi^{\n_1\n_2}\,k_{\,23}^{\m_1}k_{\,23}^{\m_2}+
\phi^{\m_1\m_2}\,k_{\,12}^{\n_1}k_{\,12}^{\n_2})+2\,\phi^{\m_1\n_1}\,k_{\,23}^{\m_2}k_{\,12}^{\n_2}\right)\\ \vphantom{\left(\pm\sqrt{\frac{\a'}{2}}\right)^{2s-2}}
&+2k_{\,23}^{s-2}\cdot H_{\a\m_1}^{(1)}\, {H^{(3)\a}}_{\n_1}\cdot k_{\,12}^{s-2}\,\left(k_{\,31}\cdot \phi^{\n_1}k_{\,23}^{\m_1}+k_{\,31}\cdot \phi^{\m_1}k_{\,12}^{\n_1}\right)\\
\vphantom{\left(\pm\sqrt{\frac{\a'}{2}}\right)^{2s-2}}
&\left.+\,k_{\,23}^{s-2}\cdot H^{(1)}_{\a_1\a_2}\, H^{(3)\a_1\a_2}\cdot k_{\,12}^{s-2}\ k_{\,31}^2\cdot\phi
\right]\\
\vphantom{\left(\pm\sqrt{\frac{\a'}{2}}\right)^{2s-2}}
&+b\left(\pm\sqrt{\frac{\a'}{2}}\right)^{2s}\ \left[\,k_{\,23}^{s-1}\cdot H^{(1)}_{\m_1}\,{H^{(3)}}_{\n_1}\cdot k_{\,12}^{s-1}\ \left(k_{\,31}\cdot\phi^{\n_1}\,k_{\,23}^{\m_1}+k_{\,31}\cdot\phi^{\m_1}\,k_{\,12}^{\n_1}\right)\right.\\
\vphantom{\left(\pm\sqrt{\frac{\a'}{2}}\right)^{2s-2}}
&\left.+\,k_{\,23}^{s-1}\cdot H^{(1)}_\a\,H^{(3)\a}\cdot k_{\,12}^{s-1}\ k_{\,31}^2\cdot \phi
\right]\\
\vphantom{\left(\pm\sqrt{\frac{\a'}{2}}\right)^{2s-2}}
&+\left(\pm\sqrt{\frac{\a'}{2}}\right)^{2s+2}\ k_{\,23}^{s}\cdot H^{(1)}\,{H^{(3)}}\cdot k_{\,12}^{s}\ k_{\,31}^{\,2}\cdot \phi\ .
\end{split}
\end{gather}
In fact, as already observed in \cite{Boulanger:2008tg}, in this case the off-shell amplitude is expected to be composed of a Born-Infeld-like term carrying the maximum number of derivatives, a coupling with $2s$ derivatives, that can exist only in space-time dimensions $d>4$, and by only another coupling with $2s-2$ derivatives that should deform the abelian gauge symmetry to a non-abelian one.


\scs{Field Theory Scattering Amplitudes}

Putting together the results of the previous paragraphs one can compute some interesting processes involving higher-spin exchanges. We shall do this in the relatively simple case of massless higher-spin particles in $d\,=\,4$, in order to study some properties of the resulting amplitudes.

Let us begin by recalling that the propagator generating function is given by
\begin{multline}
\hat{\cP}\,=\,-\frac{1}{k^{\,2}}\left[a\left(\frac{1}{2}\,p\cdot q+\frac{1}{2}\sqrt{(p\cdot q)^2-p^{\,2}\,q^{\,2}}\right)\right.\\\left.+a\left(\frac{1}{2}\,p\cdot q-\frac{1}{2}\sqrt{(p\cdot q)^2-p^{\,2}\,q^{\,2}}\right)-a_0\right]\ ,
\end{multline}
while the conserved currents that we want to study are expressed in terms of generating functions as
\begin{gather}
\cJ^{\pm}_1(k')\,=\,-4\partial_\l\left(\partial_q\cdot\partial_q\right)
\exp\left(\pm(1+\l)\sqrt{\frac{\a'}{2}}[k_{\,12\m}+F_{\m\a}\,q^\a]\,k^{'\m}\right)\Bigg|_{q\,=\,\l\,=\,0}\ ,\label{J40}
\end{gather}
and
\begin{gather}
\cJ^{\pm}_2(k')\,=\,\exp\left(\pm\sqrt{\frac{\a'}{2}}\ k_{\,12}\cdot k'\right)\,H_1\left(k_{\,1},\pm\sqrt{2\a'}\ k_{\,2}\right)\,H_{2}\left(k_{\,2},\mp\sqrt{2\a'}\ k_{\,1}\right)\ .\label{J30}
\end{gather}
The last expression is exactly the higher derivative part of the current $(\ref{Curr})$, conserved for any spin, that gives rise to $(\ref{J10})$ in the case of spin one and to $(\ref{J20})$ in the case of spin zero.

The general expression for the amplitude can be written, in terms of contractions among generating functions, as
\begin{gather}
\cA\,=\,\cJ(p)\cdot\hat{\cP}\cdot\cJ(q)+(t-\text{channel})+(u-\text{channel})\ ,
\end{gather}
where $s$, $t$ and $u$ are the Mandelstam variables defined, in our conventions, by
\begin{gather}
\begin{split}
s\,=\,-(k_{\,1}+k_{\,2})^2\,=\,-2k_{\,1}\cdot k_{\,2}\ ,\\
t\,=\,-(k_{\,1}+k_{\,3})^2\,=\,-2k_{\,1}\cdot k_{\,3}\ ,\\
u\,=\,-(k_{\,1}+k_{\,4})^2\,=\,-2k_{\,1}\cdot k_{\,4}\ .
\end{split}
\end{gather}
Let us also stress that given
\begin{gather}
\begin{split}
A\,=\,\sum_{n\,=\,0}^\infty\frac{1}{n!}\ A_{\m_1\cdots\m_n}\,p^{\,\m_1}\cdots p^{\,\m_n}\ ,\\
B\,=\,\sum_{n\,=\,0}^\infty\frac{1}{n!}\ B_{\m_1\cdots\m_n}\,p^{\,\m_1}\cdots p^{\,\m_n}\ ,
\end{split}
\end{gather}
the contraction has been defined in Chapter $4$ by
\begin{gather}
A\cdot B\,=\,\sum_n \frac{1}{n!}\ A_{\m_1\cdots\m_n}\,B^{\m_1\cdots \m_n}\ .
\end{gather}
For simplicity one can compute only the $s$-channel contribution, since the others can be obtained from it by redefinitions dictated by crossing symmetry.

Considering the amplitude with two $\cJ^{\pm}_2$ one can compute the $s$-channel contribution observing that the current is exponential in $p$ and $q$. The contraction thus amounts to affecting in the propagator generating function the substitution
\begin{gather}
\begin{split}
p\cdot q\ \lra&\ \frac{\a'}{2}\,k_{\,12}\cdot k_{\,34}\,=\,\frac{\a'}{2}\,(u-t)\ ,\\
p^{\,2}\ \lra&\ \frac{\a'}{2}\,k_{\,12}\cdot k_{\,12}\,=\,\frac{\a'}{2}\,s\,=\,-\frac{\a'}{2}\,(u+t)\ ,\\
q^{\,2}\ \lra&\ \frac{\a'}{2}\,k_{\,34}\cdot k_{\,34}\,=\,\frac{\a'}{2}\,s\,=\,-\frac{\a'}{2}\,(u+t)\ ,
\end{split}
\end{gather}
and, as a result, the $s$-channel amplitude becomes
\begin{small}
\begin{multline}
\cA^{(s)}\,=\,-\frac{1}{\a's}\left[a\left(\frac{\a'}{4}\,(u-t)+\frac{\a'}{2}\sqrt{-ut}\right)
+a\left(\frac{\a'}{4}\,(u-t)-\frac{\a'}{2}\sqrt{-ut}\right)\,-\,a_0\right]\\\times H_1\left(k_{\,1},\pm\sqrt{2\a'}\ k_{\,2}\right)\,H_{2}\left(k_{\,2},\mp\sqrt{2\a'}\ k_{\,1}\right)\\H_3\left(k_{\,1},\pm\sqrt{2\a'}\ k_{\,4}\right)\,H_{4}\left(k_{\,2},\mp\sqrt{2\a'}\ k_{\,3}\right)\ .
\end{multline}
\end{small}In this case the behavior is the same for all external states, and is determined by the factor
\begin{gather}
a\left(\frac{\a'}{4}\,(u-t)+\frac{\a'}{2}\sqrt{-ut}\right)+
a\left(\frac{\a'}{4}\,(u-t)-\frac{\a'}{2}\sqrt{-ut}\right)\,-\,a_0\ .
\end{gather}
This result can be extended in closed form to even space-time dimensions $d>4$, for $a(z)\,=\,e^z$, with the result
\begin{gather}
\begin{split}
\cA^{(s)}\,=\,&-\frac{1}{\a' s}\left[e^{\frac{\a'}{4}\,(u-t)+\frac{\a'}{2}\sqrt{-ut}}\sum_{k\,=\,0}^{\a-1}\binom{\a-1}{k}\ g_-^{(k)}\left(\frac{\a'}{4}\,(u-t)+\frac{\a'}{2}\sqrt{-ut}\right)\right.\\
&\left.+e^{\frac{\a'}{4}\,(u-t)-\frac{\a'}{2}\sqrt{-ut}}\sum_{k\,=\,0}^{\a-1}\binom{\a-1}{k}\ g_+^{(k)}\left(\frac{\a'}{4}(u-t)-\frac{\a'}{2}\sqrt{-ut}\right)\right]\\&\times H_1\left(k_{\,1},\pm\sqrt{2\a'}\ k_{\,2}\right)\,H_{2}\left(k_{\,2},\mp\sqrt{2\a'}\ k_{\,1}\right)\\&\times H_3\left(k_{\,1},\pm\sqrt{2\a'}\ k_{\,4}\right)\,H_{4}\left(k_{\,2},\mp\sqrt{2\a'}\ k_{\,3}\right)\ ,
\end{split}
\end{gather}
where
\begin{gather}
g_\pm(z)\,=\,\left(\frac{z}{z-\frac{\a'}{4}(u-t)\mp\frac{\a'}{2}\sqrt{-ut}}\right)^\a\ .
\end{gather}
A similar manipulation can be applied to the amplitude with two $\cJ^{\pm}_1$, and
the end result are the following substitutions in the propagator generating function
\begin{gather}
\begin{split}
p\cdot q\ \lra&\ \frac{\a'}{2}\,(1+\l_1)(1+\l_2)[k_{\,12\m}+F_{\m\a}q_1^\a][k_{\,34\m}+F_{\m\a}\,q_2^\a]\ ,\\
p^{\,2}\ \lra&\ \frac{\a'}{2}\,(1+\l_1)^2[k_{\,12\m}+F_{\m\a}\,q_1^\a]^2\ ,\\
q^{\,2}\ \lra&\ \frac{\a'}{2}\,(1+\l_2)^2[k_{\,34\m}+F_{\m\a}\,q_2^\a]^2\ .
\end{split}
\end{gather}
In four dimensions one thus arrives at the following generalization of the result of \cite{Bekaert:2009ud}:
\begin{gather}
\begin{split}
{\cA}\,=\,-\frac{\hat{A}}{\a' s}\Bigg[&a\Bigg(\frac{\a'}{4}\,(1+\l_1)(1+\l_2)[k_{\,12\m}+F_{\m\a}\,q_1^\a] [k_{\,34\m}+F_{\m\a}\,q_2^\a]\\+
&\frac{\a'}{4}\Big[\Big((1+\l_1)(1+\l_2)[k_{\,12\m}+F_{\m\a}\,q_1^\a][k_{\,34\m}+F_{\m\a}\,q_2^\a]\Big)^2
\\-&(1+\l_1)^2[k_{\,12\m}+F_{\m\a}\,q_1^\a]^2(1+\l_2)^2[k_{\,34\m}+F_{\m\a}\,q_2^\a]^2\Big]^{1/2}\Bigg)\\
+&a\Bigg(\frac{\a'}{4}(1+\l_1)(1+\l_2)[k_{\,12\m}+F_{\m\a}\,q_1^\a][k_{\,34\m}+F_{\m\a}\,q_2^\a]\\
-&\frac{\a'}{4}\Big[\Big((1+\l_1)(1+\l_2)[k_{\,12\m}+F_{\m\a}\,q_1^\a][k_{\,34\m}+F_{\m\a}\,q_2^\a]\Big)^2
\\-&(1+\l_1)^2[k_{\,12\m}+F_{\m\a}\,q_1^\a]^2(1+\l_2)^2[k_{\,34\m}+F_{\m\a}\,q_2^\a]^2\Big]^{1/2}\Bigg)-a_0\Bigg]\ ,
\end{split}
\end{gather}
where $\hat{A}$ is the differential operator
\begin{gather}
\hat{A}\,=\,\partial_{\l_1}\partial_{\l_2}(\partial_{q_1}\cdot\partial_{q_1})
(\partial_{q_2}\cdot\partial_{q_2})\Big|_{\l_i\,=\,q_{\,i}\,=\,0}\ .
\end{gather}
Another possibility is to consider the mixed amplitude, with at one end the current $\cJ^{\pm}_1$ and at the other end the current $\cJ^{\pm}_2$.
In this case the result can be obtained making the substitutions
\begin{gather}
\begin{split}
p\cdot q\ \lra&\ \frac{\a'}{2}\,(1+\l_2)k_{\,12}^{\m}\cdot[k_{\,34\m}+F_{\m\a}\,q_2^\a]\ ,\\
p^{\,2}\ \lra&\ \frac{\a'}{2}\,k_{\,12}\cdot k_{\,12}\,=\,-\frac{\a'}{2}(u+t)\ ,\\
q^{\,2}\ \lra&\ \frac{\a'}{2}\,(1+\l_2)^2[k_{\,34\m}+F_{\m\a}\,q_2^\a]^2\ ,
\end{split}
\end{gather}
so that the amplitude is finally
\begin{gather}
\begin{split}
\cA\,=\,-\frac{\hat{B}}{\a' s}\Bigg[&a\Bigg(\frac{\a'}{4}(1+\l_2)k_{\,12\m}[k_{\,34\m}+F_{\m\a}\,q_2^\a]\\
+&\frac{\a'}{4}\Big[\Big((1+\l_2)k_{\,12\m}[k_{\,34\m}+F_{\m\a}\,q_2^\a]\Big)^2\\
-&(u+t)^2(1+\l_2)^2[k_{\,34\m}+F_{\m\a}\,q_2^\a]^2\Big]^{1/2}\Bigg)\\
+&a\Bigg(\frac{\a'}{4}(1+\l_2)k_{\,12\m}[k_{\,34\m}+F_{\m\a}\,q_2^\a]\\
-&\frac{\a'}{4}\Big[\Big((1+\l_2)k_{\,12\m}[k_{\,34\m}+F_{\m\a}\,q_2^\a]\Big)^2\\
-&(u+t)^2(1+\l_2)^2[k_{\,34\m}+F_{\m\a}\,q_2^\a]^2\Big]^{1/2}\Bigg)-a_0\Bigg]\\
\times&H_1\Big(k_{\,1},\pm\sqrt{2\a'}k_{\,2}\Big)H_{2}\Big(k_{\,2},\mp\sqrt{2\a'}k_{\,1}\Big)\ ,
\end{split}
\end{gather}
with $\hat{B}$ defined as
\begin{gather}
\hat{B}\,=\,\partial_{\l_2}(\partial_{q_2}\cdot\partial_{q_2})\Big|_{\l_2\,=\,q_2\,=\,0}\ .
\end{gather}
Let us stress that the dominant behavior, apart from the tensorial structure, is basically determined by the zeros of $a(z)$ and its derivatives, so that each of these amplitudes can be extremely soft at high energies. A more general discussion of the ultraviolet behavior of these amplitudes will be given in the Discussion.


\chapter{Discussion}
In this Thesis we have studied $3$-point and $4$-point scattering amplitudes for open bosonic string states in the first Regge trajectory obtaining from them, for the first time, handy explicit forms for a number of cubic string couplings involving higher-spin particles. The computation provides some evidence of how an effective theory for the first Regge trajectory of open strings is naturally embedded in an enlarged space with auxiliary coordinates $k_{\,i}'$, and thus points onto a well definite framework in which the higher-spin systematics could be better understood. The crucial observation, recently made in \cite{Bekaert:2009ud}, is that in this framework the Wigner-Weyl calculus becomes a natural tool to construct consistent non-abelian deformations of the gauge symmetry. As a result, it is possible to argue that the group of gauge symmetries of these unconstrained metric-like theories is isomorphic, to lowest order, to the group of unitary operators defined on $\mathbb{R}^n$ \cite{Bekaert:2007mi,Bekaert:2009ud}.

The amplitudes obtained from String Theory have shown several beautiful properties over the years, and here we have seen how to recover from them some gauge invariant couplings. This has made it possible to compute a number of higher-spin four-point scattering amplitudes from the Quantum Field Theory side. The amplitudes thus obtained can be studied in the high energy regime and, as already pointed out in \cite{Bekaert:2009ud}, their behavior depends crucially on the coupling function defined in the previous chapter,
\begin{gather}
a(z)\,=\,\sum_{n\,=\,0}^{\infty}\frac{a_n}{n!}\ z^n\ ,\ \ \ a_n\geq 0\ .\label{couplings}
\end{gather}
Let us stress that the total amplitude is generally a sum of contributions from different channels, that contain terms in which the argument of $a(z)$ goes both to $+\infty$ and to $-\infty$, in the high-energy fixed-angle limit. As a result, the amplitudes can be well behaved at high energies only if the function $a(z)$ has zeros both at $z\,=\,+\infty$ and $z\,=\,-\infty$. In String Theory, the coupling function is $a(z)\,=\,e^z$ and $z\,=\,+\infty$ is not a zero, which makes the sub-leading Regge-trajectories necessary to improve the high-energy behavior of the theory. Moreover, it seems also necessary, in general, to add non-local quartic terms in order to cancel the non-polar parts arising from higher-spin propagators. It should be appreciated that, these non-polar contributions are not present in the Veneziano formula.

Non-local quartic terms, however, are very natural in this kind of framework. Moreover, allowing non-local Lagrangians opens up new interesting possibilities, as observed in \cite{Moeller:2002vx}, and highlights an interesting similarity with the ``unfolding'' procedure (for a review see \cite{Bekaert:2005vh}), a powerful first-order formulation underlying Vasiliev's construction of higher-spin dynamics in $(A)dS$. 
The point is that there is a substantial difference between a differential equation involving an infinite number of space-time derivatives and one involving a large but finite number of them, since an infinite number of derivatives requires an infinite number of initial conditions. Assuming analiticity, this leads to reconstruct the whole function locally from its Taylor expansion, while the solution appears completely determined independently of the differential equation. As a result, a differential equation involving an infinite number of derivatives puts strong constraints on the possible initial value configurations. In this picture, dynamics is completely specified by the solution of the initial value problem while the differential equation turns out to be only a set of constraints. This argument points to several connections with the ``unfolding'' procedure. In fact, even in this case the dynamical content of the system is encoded in a set of constraints that allow to reconstruct the Taylor expansion of the solution and, by analiticity, determine it completely at least locally. The basic difference with String Field Theory is that the unfolding procedure can be rather regarded as an analogue of the Hamiltonian approach, since as we stressed it is a first order formulation of the dynamics. 

From this point of view a very interesting problem, and also one of the motivations for this Thesis, was to try and dissect the Veneziano amplitude, considering separately the exchanges of all sub-leading Regge trajectories. The aim is, eventually, to understand how String Theory is connected to such field theory constructions. To address fully this question, the key laking information are the couplings chosen by String Theory for the sub-leading states. In the simple case of tachyon scattering with trivial Chan-Paton factors, however, starting from the result obtained here it can be possible to construct other amplitudes in which different numbers of Regge trajectories are exchanged.

Another amusing possibility, that in some sense tries to go beyond String Theory, it is to consider, following \cite{Bekaert:2009ud}, different coupling functions $a(z)$ having the correct zeros at $z\,=\,\pm\infty$.
In this case the series $(\ref{couplings})$ must have a finite radius of convergence, and can thus be defined by analytic continuation at $z\,=\,\pm\infty$. One simple example was already exhibited in \cite{Bekaert:2009ud} and it is $a_n\,=\,n!$ with $a(z)\,=\,\frac{1}{1-z}$. Anyway any power series with positive coefficients and with finite radius of convergence $R$ has a pole for $z\,=\,R$, so that it would seem problematic in defining a consistent $S$-matrix. Further developments along these lines are currently under investigation.

Summarizing, scattering amplitudes resulting from the exchange of an infinite tower of higher spin particles can be extremely soft in the ultraviolet regime. In fact, the idea behind this phenomenon is very simple: although a single spin-$m$ $t$-channel exchange grows as $\frac{(-s)^m}{t}$ in the Regge limit, the whole sum of these contributions, with coefficients $(m!)^{-1}$, is $e^{-s}$, that it is far smaller then any of the individual contributions. It may well be that allowing an infinite number of Regge trajectories could improve further the behavior of the resulting amplitude even making choices that are different from those realized in String Theory.
Moreover, even in Field Theory one is led to consider seriously the emerging non-local structures, whose systematic investigation may well prove a most important step towards a deeper understanding of String Theory leading nonetheless to a well-defined $S$-matrix.\\

Let us conclude by outlining some expected future developments along the lines of this Thesis. First of all, a systematic study of the higher-spin currents that we have computed, with the aim of extracting off-shell couplings and to study their massless limit. This systematic study could also lead to the conclusion that the off-shell string currents have a non-singular massless limit that is exactly, as suggested by the simple examples analyzed in this Thesis, that found in \cite{Berends:1985xx}. Moreover, this kind of analysis could help us to understand explicitly how the conjectured string mass generation takes place, giving an encouraging starting point for a higher-spin systematics.\\
A second interesting possibility could be to try and extend the results obtained in \cite{Zinoviev:2008jz,Porrati:2009bs} to higher-spins. The particular form of the spin-$1$ string couplings found in this Thesis can be used in principle to construct consistent Lagrangians, at least in the case of constant electromagnetic background fields.

Another beautiful perspective on this problem is to go beyond the cubic level so far considered, trying to gain a better understanding of the non-linear structure of higher-spin interactions. This problem appears to be the most difficult one of those listed here, and in fact no consistent quartic couplings for higher-spin fields has yet been constructed explicitly. The reader can perhaps appreciate the source of difficulties, considering that, at the quartic level, the consistency for a gauge theory depends crucially on the Jacobi identity. Analogously, a consistent quartic coupling requires the complete non linear structure of the theory with its deformed gauge symmetry, and is expected to couple together infinitely many higher-spin fields. String Theory can give, and in fact has already given, several fundamental hints to face this very complicated challenge.


\end{document}